\documentclass[letterpaper]{article} %
\usepackage{aaai24}  %
\usepackage{times}  %
\usepackage{helvet}  %
\usepackage{courier}  %
\usepackage[hyphens]{url}  %
\usepackage{graphicx} %
\usepackage{natbib}  %
\usepackage{caption} %
\usepackage{algorithm}
\usepackage{algorithmic}

\usepackage{newfloat}
\usepackage{listings}
\DeclareCaptionStyle{ruled}{labelfont=normalfont,labelsep=colon,strut=off} %
\floatstyle{ruled}
\newfloat{listing}{tb}{lst}{}
\floatname{listing}{Listing}
\usepackage{tabularx}
\usepackage{multirow}
\usepackage{enumitem}
\usepackage{csquotes}
\usepackage{booktabs}
\usepackage{makecell}
\usepackage{array}
\usepackage{longtable}
\usepackage{epsfig}
\usepackage{hyperref}
\usepackage[capitalize,noabbrev]{cleveref}

    \newcolumntype{s}{>{\hsize=.25\hsize}X}
    \newcolumntype{Y}{>{\centering\arraybackslash}X} %

\title{How Do AI Companies \enquote{Fine-Tune} Policy? Examining Regulatory Capture in AI Governance}
\author{
    Kevin Wei\textsuperscript{\rm 1, \rm 2},
    Carson Ezell\textsuperscript{\rm 3},
    Nick Gabrieli\textsuperscript{\rm 3},
    Chinmay Deshpande\textsuperscript{\rm 3}
}
\affiliations{

    \textsuperscript{\rm 1}Technology and Security Policy Fellow at RAND 
    \textsuperscript{\rm 2}Harvard Law School 
    \textsuperscript{\rm 3}Harvard University \\

    kwei@rand.org
}

\defcitealias{AIA}{EU AI Act}

\begin{document}

\pagenumbering{roman}

\maketitle

\section*{Executive Summary} \label{sec:Exec_Summary}

Industry actors in the United States have gained extensive influence in conversations about the regulation of general-purpose artificial intelligence (AI) systems. examines the ways in which industry influence in AI policy can result in policy outcomes that are detrimental to the public interest, i.e., scenarios of ``regulatory capture.'' First, we provide a framework for understanding regulatory capture. Then, we report the results from 17 expert interviews identifying what policy outcomes could constitute capture in AI policy and how industry actors (e.g., AI companies, trade associations) currently influence AI policy. We conclude with suggestions for how capture might be mitigated or prevented.

In accordance with prior work, we define ``regulatory capture'' as situations in which:

\begin{enumerate}
    \item \textbf{A policy \textit{outcome} contravenes the public interest.} These policy outcomes are characterized by regulatory regimes that prioritize private over public welfare and that could hinder such regulatory goals as ensuring the safety, fairness, beneficence, transparency, or innovation of general-purpose AI systems. Potential outcomes can include changes to policy, enforcement of policy, or governance structures that develop or enforce policy. 
    \item \textbf{Industry actors exert influence on policymakers through particular \textit{mechanisms} to achieve that policy outcome.} We identify 15 mechanisms through which industry actors can influence policy. These mechanisms include advocacy, revolving door (employees shuttling between industry and government), agenda-setting, cultural capture, and other mechanisms as defined in Table \ref{tab:Exec_Summary_Mechanisms}. Policy outcomes that arise absent industry influence---even those which may benefit industry---do not reflect capture.
\end{enumerate}

To contextualize these outcomes and mechanisms to AI policy, we interview 17 AI policy experts across academia, government, and civil society. We seek to identify possible outcomes of capture in AI policy as well as the ways that AI industry actors are currently exerting influence to achieve those outcomes. 

With respect to potential captured outcomes in AI policy, experts were primarily concerned with capture leading to a lack of AI regulation, weak regulation, or regulation that over-emphasizes certain policy goals above others. 

Experts most commonly identified that AI industry actors use the following mechanisms to exert policy influence:
\begin{itemize}
    \item \textbf{Agenda-setting (15 of 17 interviews):} Interviewees expressed that industry actors advance anti-regulation narratives and are able to steer policy conversations toward or away from particular problems posed by AI. These actors, including AI companies, are also able to set default standards, measurement metrics, and regulatory approaches that fail to reflect public interest goals.
    \item \textbf{Advocacy (13):} Interviewees were concerned with AI companies' and trade associations' advocacy activities targeted at legislators.
    \item \textbf{Academic capture (10):} Interviewees identified ways that industry actors can direct research agendas or promote particular researchers, which could in turn influence policymakers.
    \item \textbf{Information management (9):} Interviewees indicated that industry actors have large information asymmetries over government actors and are able to shape policy narratives by strategically controlling or releasing specific types of information.
\end{itemize}

To conclude, we explore potential measures to mitigate capture. Systemic changes are needed to protect the AI governance ecosystem from undue industry influence---building technical capacity within governments and civil society (e.g., promoting access requirements, providing funding independent of industry, and creating public AI infrastructure) could be a first step towards building resilience to capture. Procedural and institutional safeguards may also be effective against many different types of capture; examples include building regulatory capacity in government, empowering watchdogs, conducting independent review of regulatory rules, and forming advisory boards or public advocates. Other mitigation measures that are specific to different types of industry influence are outlined in Table \ref{tab:Exec_Summary_Mechanisms}.

Although additional research is needed to identify more concrete solutions to regulatory capture, we hope that this article provides a starting point and common framework for productive discussions about industry influence in AI policy.

\begingroup

\renewcommand{\arraystretch}{1.25} %
\setcounter{table}{-1}
\begin{table*}[!p]
    \centering
    \small

    \begin{tabularx}{\textwidth}{ l X X }
        \toprule

        \multicolumn{1}{l}{\textbf{Mechanism}} &
        \multicolumn{1}{l}{\textbf{Definition}} &
        \multicolumn{1}{l}{\textbf{Potential Mitigations}} 
        \\ 
        
        \midrule

        \multicolumn{3}{l}{\textit{Personal Engagement}} 
        \\
        
        \multicolumn{1}{l}{\multirow[t]{4}{*}{Advocacy}} &
        Industry actors directly participate in formal policy making processes---e.g., interacting directly with policymakers or regulators to provide information or convince them of a particular point of view. &
        \begin{itemize}[after=\vspace{-\baselineskip}, topsep=0pt]
            \vspace{-0.75em} %
            \item Increase transparency requirements
            \item Build robust civil society institutions
         \end{itemize}
        \\

        \multicolumn{1}{l}{\multirow[t]{2}{*}{Procedural obstruction}} &
        Industry actors intentionally impede policymaker or regulator action through procedural interference. &
        \begin{itemize}[after=\vspace{-\baselineskip}, topsep=0pt]
            \vspace{-0.75em}
            \item N/A; additional research needed
        \end{itemize}
        
        \\ 
        
        \midrule
        
        \multicolumn{3}{l}{\textit{Incentive Shaping}}
        \\
        
        \multicolumn{1}{l}{\multirow[t]{3}{*}{Donations, gifts, and bribes}} &
        Industry actors make financial contributions to elected officials’ campaigns, or give personal gifts to elected officials (or their staff). &
        \begin{itemize}
            \vspace{-1.15em}
            \item Increase transparency requirements
            \item Build robust civil society institutions
        \end{itemize}
        \\
        
        \multicolumn{1}{l}{\multirow[t]{4}{*}{Private threats}} &
        Industry actors make explicit or implicit threats of litigatory, reputational, or other negative consequences to prevent policy enactment or enforcement. &
        \begin{itemize}[after=\vspace{-\baselineskip}, topsep=0pt]
            \vspace{-0.75em}
            \item N/A to the United States
        \end{itemize}
        
        \\ 

        \multicolumn{1}{l}{\multirow[t]{7}{*}{Revolving door}} &
        Industry actors hire government officials, regulators, policymakers, or their staff; or employees of industry actors leave to work for government officials, regulators, or policymakers.
        
        \,
        
        \,
        
        \,
        
        &
        \begin{itemize}[after=\vspace{-\baselineskip}, topsep=0pt]
            \vspace{-0.75em}
            \item Strengthen and enforce government ethics policies such as conflict-of-interest reviews
            \item Fund and provide AI-specific training for government ethics offices
            \item Invest in regulator salaries, work environments, and professional development to make government careers more desirable
        \end{itemize}
        \\
        
        \midrule

        \multicolumn{3}{l}{\textit{Information Capture}} 
        \\
        
        \multicolumn{1}{l}{\multirow[t]{6}{*}{Agenda-setting}} &
        Industry actors emphasize or de-emphasize particular perspectives or data, set priorities in policy conversations, or frame regulatory problems in ways that favor industry actors. Regulators may then adopt such views---e.g., goals, norms, practices, activities, and models of risks and markets. &
        
        \multirow[t]{12.5}{=}{
        \begin{itemize}[after=\vspace{-\baselineskip}, topsep=0pt]
            \vspace{-1.0em}
            \item Increase access for non-industry stakeholders to policy processes, particularly at early stages of policy development (to address all information capture mechanisms)
            \item Consider consumer empowerment programs to enable civic participation (to address all information capture mechanisms)
            \item Institute reporting and monitoring requirements to raise regulatory visibility and verify industry information (to address all information capture mechanisms)
        \end{itemize}
        }
        
        \\

        \multicolumn{1}{l}{\multirow[t]{3}{*}{Information management}} &
        Industry actors selectively share, control access to, withhold, or provide misleading or false information to policymakers or regulators. &
        \\

        \multicolumn{1}{l}{\multirow[t]{5}{*}{Information overload}} &
        Industry actors inundate policymakers or regulators with similar information or communications supporting their points of view, which challenges the ability of regulators to process and interpret the information. &
        \\

        \midrule
    \end{tabularx}
    \caption{Mechanisms of industry influence in the policy process (cont'd on next page)}
    \label{tab:Exec_Summary_Mechanisms}
\end{table*}

\setcounter{table}{-1}
\begin{table*}[!p]
    \centering
    \small

    \begin{tabularx}{\textwidth}{ l X X }
        \toprule

        \multicolumn{1}{l}{\textbf{Mechanism}} &
        \multicolumn{1}{l}{\textbf{Definition}} &
        \multicolumn{1}{l}{\textbf{Potential Mitigations}} 
        \\ 
        
        \midrule
        
        \multicolumn{3}{l}{\textit{Cultural Capture}} 
        \\
                
        \multicolumn{1}{l}{\multirow[t]{3}{*}{Group identity}} &
        Policymakers or regulators may be ``more likely to adopt positions advanced by people whom they perceive as being their in-group'' \cite{kwak_cultural_2013}. &
        \multirow[t]{9}{=}{
        \begin{itemize}[after=\vspace{-\baselineskip}, topsep=0pt]
            \vspace{-1em}
            \item N/A; additional research needed
        \end{itemize}
        }
        \\

        \multicolumn{1}{l}{\multirow[t]{3}{*}{Relationship networks}} &
        Policymakers or regulators may be ``more likely to adopt positions advanced by people who are in their social networks'' \cite{kwak_cultural_2013}. &
        \\

        \multicolumn{1}{l}{\multirow[t]{4}{*}{Status}} &
        Policymakers or regulators may be ``more likely to adopt positions advanced by people whom they perceive to be of higher status in social, economic, intellectual, or other terms'' \cite{kwak_cultural_2013}. &
        \\ 
        
        \midrule

        \multicolumn{3}{l}{\textit{Indirect Capture}} 
        \\
        
        \multicolumn{1}{l}{\multirow[t]{5}{*}{Academic capture}} &
        Industry actors influence academic actors, who may then influence policymakers or regulators. 
                
        \,
        
        \,

        \,
        
        &
        \begin{itemize}[after=\vspace{-\baselineskip}, topsep=0pt]
            \vspace{-0.75em}
            \item Provide funding sources independent of industry
            \item Increase non-industry career opportunities
            \item Ensure academic access to compute and data resources
        \end{itemize}
        \\

        \multicolumn{1}{l}{\multirow[t]{3}{*}{Private regulator capture}} &
        Industry actors influence private organizations that serve regulatory functions---e.g., auditors or standards-setting bodies. &
        \begin{itemize}[after=\vspace{-\baselineskip}, topsep=0pt]
            \vspace{-0.75em}
            \item Provide funding sources independent of industry
        \end{itemize}
        \\

        \multicolumn{1}{l}{\multirow[t]{3}{*}{Public relations}} &
        Industry actors engage in direct public communications, which may then influence policymakers or regulators directly or via shaping public opinion. &
        \begin{itemize}[after=\vspace{-\baselineskip}, topsep=0pt]
            \vspace{-0.75em}
            \item Increase transparency requirements
            \item Build robust civil society institutions
        \end{itemize}
        \\

        \multicolumn{1}{l}{\multirow[t]{4}{*}{Media capture}} &
        Industry actors influence journalists or outputs from media channels, which may then influence policymakers or regulators directly or via shaping public opinion. &
        \begin{itemize}[after=\vspace{-\baselineskip}, topsep=0pt]
            \vspace{-0.75em}
            \item Increase transparency requirements
            \item Build robust civil society institutions
        \end{itemize}
        \\

        \bottomrule
    \end{tabularx}
    \caption{Mechanisms of industry influence in the policy process (cont'd)}
\end{table*}

\endgroup

\clearpage

\onecolumn
\tableofcontents
\clearpage

\listoffigures
\listoftables

\clearpage

\section*{Readers' Guide\footnote{Inspired by \citet{weidinger_ethical_2021} and \citet{weidinger_sociotechnical_2023}.}} \label{sec:Readers_Guide}

We recommend the following reading strategies for different types of readers:

\begin{itemize}
    \item \textbf{2-minute read}: Read Table \ref{tab:Exec_Summary_Mechanisms} in full. Browse the interview results in Figure \ref{fig:Results_Goals} (interview responses to the goals of AI regulation), Figure \ref{fig:Appendix_Results_Actors} (interview responses to actors influencing AI policy), and Figure \ref{fig:Results_Mechanisms} (interview responses to the mechanisms of industry influence in AI policy).
    \item \textbf{10-minute read}: Start with the \nameref{sec:Exec_Summary}. Skip Table \ref{tab:Exec_Summary_Mechanisms} but read through the definitions in Table \ref{tab:Appendix_Mechanisms_Definitions}. Finish with Section \ref{sec:Prevention}, on preventing and mitigating capture.
    \item \textbf{Policymakers}: Start with the \nameref{sec:Exec_Summary}, particularly Table \ref{tab:Exec_Summary_Mechanisms}. Browse the interview results in Figure \ref{fig:Results_Goals} (interview responses to the goals of AI regulation), Figure \ref{fig:Appendix_Results_Actors} (interview responses to actors influencing AI policy), and Figure \ref{fig:Results_Mechanisms} (interview responses to the mechanisms of industry influence in AI policy). Then read Section \ref{sec:Prevention} on preventing and mitigating capture.
    \item \textbf{AI governance researchers}: Start with Section \ref{sec:Definition} (defining regulatory capture) before reading the results in Sections \ref{sec:Outcomes} and \ref{sec:Mechanisms}. Finish with Section \ref{sec:Prevention} on preventing and mitigating capture.
    \item \textbf{AI developers/researchers}: Start with Section \ref{sec:Definition} (defining regulatory capture), then read the discussions on academic capture in Section \ref{subsec:Mechanisms_Indirect_Capture} and in the penultimate paragraph of Section \ref{sec:Prevention}. Optionally, read in full the interview results and suggested capture mitigation measures in Sections \ref{sec:Outcomes}, \ref{sec:Mechanisms}, and \ref{sec:Prevention}.
\end{itemize}

\clearpage

\twocolumn

\pagenumbering{arabic}

\section{Introduction} \label{sec:Intro}

Jurisdictions around the world are proposing or implementing regulations of general-purpose artificial intelligence (AI) systems \cite{iapp_global_2024, lexis_artificial_2024}. As AI policy develops, industry players---including AI developers; AI deployers; trade associations; cloud companies; and providers of tools, services, or hardware used in AI development or deployment---have gained widespread influence in AI policy. In the European Union (EU), industry actors have fought against the regulation of general-purpose AI and foundation models in the EU AI Act \cite{ceo_big_2022, ceo_byte_2023, bertuzzi_eus_2023, perrigo_exclusive_2023, lomas_frances_2023, neerven_will_2023}. In the United States, the number of lobbyists working on AI-related issues in 2023 increased by 120\% relative to 2022, with 85\% of lobbyists hired by industry organizations \cite{cheng_artificial_2024}.

Industry participation is essential in the policy process, but it can also lead to regulatory capture: situations in which industry co-opts regulatory regimes to prioritize private over public welfare. In the context of AI governance, industry influence that leads to capture can impede effective AI regulation and harm public interests \cite{chomanski_missing_2021, dal_bo_regulatory_2006} because AI companies’ goals and incentives are not always aligned with those of the public \cite{khanal_why_2024}. Commentators have warned that regulatory capture could result in general-purpose AI policies and enforcement practices that are ineffective, unsafe, or unjust---or even no regulation at all \cite{roberts_achieving_2021, guha_ai_2024, goodman_ai_2024, mollman_bill_2023, von_thun_monopoly_2023}. Understanding what capture is, how it could occur, and how it might be mitigated is therefore critical to ensuring that AI policy promotes the public interest.

This article aims to fill a gap in the literature on industry influence and regulatory capture in AI policy. Our main contributions are to outline what types of policy outcomes could constitute regulatory capture, explain the main channels through which industry actors are currently influencing US AI policy and attempting to achieve those outcomes, and discuss how different forms of influence and capture can be addressed. To answer these questions, we conducted 17 semi-structured expert interviews and a survey of observational data. We focus our discussion on corporate influence in US policy for general-purpose AI systems,\footnote{See Appendix \ref{sec:Appendix_Interview_Protocol} for a definition of ``general-purpose AI systems.'' See also \citetalias[Art. 3(63)]{AIA}; \citet{gutierrez_proposal_2023, triguero_general_2024, xia_ai_2024}.} and we hope to provide a framework for productive discussions of capture in AI governance.

\section{Defining \enquote{Regulatory Capture}} \label{sec:Definition}

Investigations of regulatory capture in general-purpose AI policy have been limited, even though concerns about capture have permeated AI governance research \cite{hendrycks_overview_2023, hadfield_regulatory_2023, whyman_ai_2023, thierer_problem_2023}, legislative testimony \cite{lawrence_large_2023}, media coverage \cite{herrman_how_2023, thornhill_ai_2023, davis_ai_2023}, and public discourse \cite{goanta_regulation_2023, all-in_podcast_all-summit_2023, marcus_mustafasuleymn_2023, lecun_tegmark_2023, tegmark_classic_2023, renieris_concerned_2023, delangue_im_2023}. These discussions have been incomplete: Commentators often only discuss capture in passing, do not focus on \textit{how} capture occurs, or consider only the classical model of capture in which monopolists attempt to raise barriers to entry through over-regulation \citetext{see Appendix \ref{sec:Appendix_Review}}. The sincerity of some discourse is also questionable, particularly when corporate actors levy accusations of capture that seem to serve their own interests \citetext{\citealp{yackee_regulatory_2022}; e.g., \citealp{andreessen_horowitz_andreessen_2023, andreessen_techno-optimist_2023, altman_oversight_2023, nquiringminds_written_2023, openuk_written_2023, delangue_im_2023, lecun_tegmark_2023}}.

In the social science literature, the phenomenon of regulatory capture has been well-theorized since \citet{stigler_theory_1971}. Definitions of regulatory capture vary widely \cite{mitnick_capturing_2011}, partly because of the heterogeneity of forms of capture \cite{carpenter_detecting_2013}, but the consensus is that capture requires industry influence in the policy process \cite{dal_bo_regulatory_2006}. Adapting from \citet{carpenter_detecting_2013}, \citet{wren-lewis_regulatory_2011}, and \citet{dal_bo_regulatory_2006}, we define ``regulatory capture'' as consisting of (1) a policy \textit{outcome} contravening the public interest that (2) results from industry influence exerted on policymakers through particular \textit{mechanisms}.\footnote{In AI governance, most commentators use the term ``regulatory capture'' to refer to the model of capture presented in \citet{stigler_theory_1971}, in which market incumbents support stringent regulations (such as licensing requirements) to block new market entrants. Our definition is aligned with the modern academic literature on regulatory capture: That definition is more expansive and would also include, e.g., companies lobbying to \textit{weaken} instead of strengthen regulations.} 

We emphasize that not all industry influence reflects regulatory capture. To the contrary, some industry participation in policy processes is both inevitable and desirable, especially when it provides regulators with greater visibility and technical expertise \cite{slayton_beyond_2018, kaminski_regulating_2022, thaw_enlightened_2014, wansley_virtuous_2015}---both of which are particularly important in AI \cite{anderljung_frontier_2023, scherer_regulating_2016}. Capture occurs only when corporate influence leads to regulation that unjustly prioritizes private interests over public ones \cite{wren-lewis_regulatory_2011}. 

We developed and present in Sections \ref{subsec:Def_Outcomes} and \ref{subsec:Def_Mechanisms} a framework that encapsulates the range of models of regulatory capture.

\subsection{Outcomes of Regulatory Capture} \label{subsec:Def_Outcomes}

A necessary condition for regulatory capture is a policy \textit{outcome} contravening the public interest. Capture would not occur, e.g., if industry actors successfully influenced policymakers to implement policies beneficial to the public or if a policy beneficial to the public were to be implemented despite industry opposition. Although the public interest is often ill-defined, the social science literature cites catastrophic safety incidents, financial downturns, and monopolistic market structures as outcomes in contravention of the public interest.

The classic model of regulatory capture emphasizes industry influence on a regulator leading to price-fixing or anti-competitive conditions that protected market incumbents, usually via over-regulation of prospective market entrants in industries with natural monopolies \cite{stigler_theory_1971, dal_bo_regulatory_2006, rex_agency_2022}. However, the literature has since developed broader models that account for \textit{under-regulation} \cite{carpenter_corrosive_2013} or for influence across various governmental entities \cite{magill_courts_2013, anderson_court_2018, rex_agency_2022, rex_anatomy_2020}. We use ``regulatory capture'' as an umbrella term for all these models of capture.

Resolving the question of which outcomes are ``properly'' considered capture of AI policy is beyond the scope of this article because the goals of general-purpose AI regulation are contested and sometimes conflicting. However, discourse about capture in AI policy can benefit from understanding how these models of capture operate. We categorize the ways in which policy outcomes can diverge from the public interest as affecting the content of government policies, the enforcement of government policies, or the institutional structures of regulation. Examples of outcomes in each category are outlined in Table \ref{tab:Outcomes}. 

\begin{table}[t]
    \centering
    \small
    \setlist[itemize]{leftmargin=*} %

    \renewcommand{\tabularxcolumn}[1]{m{#1}} %
    
    \begin{tabularx}{\linewidth}{ l Y } 
        \toprule 
        
        \textbf{Category} &  
        \textbf{Examples}
        \\ 
        
        \midrule  
         
        \hfil \rotatebox[origin=c]{90}{\makecell{Policy \\ content}} \hfill &  
        \begin{itemize}[after=\vspace{-\baselineskip}, topsep=0pt]
            \item Regulations are too weak to protect the public (or are nonexistent) \cite{carpenter_corrosive_2013, papyshev_limitation_2022}.
            \item Regulation creates high barriers to entry that protect market incumbents \cite{stigler_theory_1971, barrett_bus_2004}.
            \item Product standards are set to favor (particular) industry players \cite{berman_industry_2017}.
            \item Regulations make suboptimal value trade-offs between, e.g., safety and justice \cite{guha_ai_2024, wong_ai_2023}.
         \end{itemize}
         \\ 
         \midrule 

         \hfil \rotatebox[origin=c]{90}{\makecell{Policy \\ enforcement}} \hfill &  
         \begin{itemize}[after=\vspace{-\baselineskip}, topsep=0pt]
            \item Policies are not enforced, or exceptions are created, for particular companies \cite{teachout_market_2014}.
            \item Enforcement is biased toward a subset of firms \cite{mariniello_antitrust_2015}.
         \end{itemize}
         \\ 
         
         \midrule  

         \hfil \rotatebox[origin=c]{90}{\makecell{Governance \\ structures}} \hfill &  
         \begin{itemize}[after=\vspace{-\baselineskip}, topsep=0pt]
             \item Agencies lack funding to enact or enforce policies \cite{shapiro_complexity_2012, neudert_regulatory_2023}.
             \item A lack of uniform rules allows regulated entities to engage in regulatory arbitrage \cite{etzioni_capture_2009}.
             \item An agency's dual mandate results in the agency achieving only one mandate \cite{carrigan_captured_2013, rex_anatomy_2020}.
             \item Suboptimal federal policies preempt state policies \cite{carpenter_corrosive_2013}.
         \end{itemize}
         \\ 
         
         \bottomrule
    \end{tabularx}
    \caption{Examples of outcomes of regulatory capture}
    \label{tab:Outcomes}
\end{table}

\subsection{Mechanisms of Industry Influence} \label{subsec:Def_Mechanisms}

Industry actors can seek the outcomes outlined above by exerting influence on the policy process; capture occurs when these actors succeed. Therefore, policy failures absent corporate involvement are not capture, and neither is corporate influence that does not result in an outcome contravening the public interest. We distinguish between direct and indirect \textit{mechanisms} of capture, and definitions for all mechanisms are provided in Table \ref{tab:Appendix_Mechanisms_Definitions}.

\textbf{Direct mechanisms} are channels of influence immediately exerted on policymakers---including legislatures, regulatory agencies, courts, White House offices, and individual staffers or officials. We introduce four categories of direct mechanisms:

\begin{itemize}
    \item \textit{Personal engagement}: Industry actors directly participate in formal policy making processes. Mechanisms in this category are \textit{advocacy} \cite{barkow_insulating_2010, etzioni_capture_2009,  de_figueiredo_advancing_2014, godwin_lobbying_2013} and \textit{procedural obstruction} \citetext{R16\footnote{These citations indicate expert interviews. See Table \ref{tab:Interviewees}.}}.
    
    \item \textit{Incentive shaping}: Industry actors offer positive or negative incentives to shape policymakers' decisions. Mechanisms in this category are \textit{donations and gifts} \cite{aggarwal_corporate_2012, kuntze_lobbying_2023},\footnote{Bribes also fall within this category but are illegal. We consider bribes to be less relevant to the United States---though they are discussed in the regulatory capture literature in non-US contexts \citetext{see \citealp{dal_bo_regulatory_2006}}---and thus exclude bribes from our discussion below.} \textit{private threats} \cite{dal_bo_regulatory_2006, brezis_vulnerability_2011},\footnote{Private threats include coercion \cite{dal_bo_regulatory_2006}. We consider private threats to be less relevant in the United States, so we exclude this mechanism from our discussion below.} and \textit{revolving door} \cite{rex_agency_2022, de_chiara_dynamic_2021, tabakovic_revolving_2018}.
    
    \item \textit{Information capture}: Industry actors shape policymakers' information environment. Mechanisms in this category are \textit{agenda-setting} \cite{bachrach_two_1962, li_regulatory_2023, walters_capturing_2019, baxter_capture_2011, shapiro_complexity_2012, king_effects_2018, rilinger_who_2023}, \textit{information management} \cite{mitnick_developing_2015, rilinger_who_2023, shapiro_complexity_2012, dal_bo_regulatory_2006}, and \textit{information overload} \cite{wagner_administrative_2009, heims_mechanisms_2023}.
    
    \item \textit{Cultural capture}: Capture is made more likely by policymakers' and industry actors' shared underlying assumptions or backgrounds, including \textit{group identity} \cite{kwak_cultural_2013, caprio_regulatory_2013},  \textit{relationship networks} \cite{kwak_cultural_2013, caprio_regulatory_2013, rex_anatomy_2020}, and \textit{status} \cite{goanta_regulation_2023,  saltelli_science_2022, bode_constructing_2023, perlman_for_2020, kwak_cultural_2013}. Unlike the previous types of influence, cultural capture is less clear-cut and can consist of industry actors creating new or taking advantage of pre-existing social dynamics.
\end{itemize}

\textbf{Indirect mechanisms} are channels of influence exerted on intermediaries---academics and think tanks \cite{abdalla_grey_2021}, the media \cite{schiffrin_media_2021}, private regulators \cite{laux_taming_2021, berman_industry_2017, tartaro_assessing_2023, terzis_law_2024, casper_black-box_2024}, or the public---who in turn influence policymakers. Corporate influence exerted on these intermediaries can occur through the mechanisms outlined above, as well as other mechanisms specific to the relevant intermediaries. Note that indirect capture requires that influenced intermediaries in turn influence policy to create the types of outcomes described in Section \ref{sec:Outcomes}.

\begingroup

\renewcommand{\arraystretch}{1.25}

\begin{table*}[!p]
    \centering
    \small

    \begin{tabularx}{\textwidth}{ l X X }
        \toprule

        \multicolumn{1}{l}{\textbf{Mechanism}} &
        \multicolumn{1}{l}{\textbf{Definition}} &
        \multicolumn{1}{l}{\textbf{Examples}} 
        \\ 
        
        \midrule

        \multicolumn{3}{l}{\textit{Personal Engagement}} 
        \\
        
        \multicolumn{1}{l}{\multirow[t]{3}{*}{Advocacy}} &
        Industry actors directly participate in formal policy making processes---e.g., interacting directly with policymakers or regulators to provide information or convince them of a particular point of view. &
        Activities include lobbying, private meetings, speaking events, public hearings, constituent engagement, court filings (\textit{amicus} briefs).
        \\

        \multicolumn{1}{l}{\multirow[t]{6}{*}{Procedural obstruction}} &
        Industry actors intentionally impede policymaker or regulator action through procedural interference. &
        An industry actor participates in a standards-setting committee and repeatedly stalls the conversation; an industry actor files multiple lawsuits against a regulatory agency to prevent enforcement; an industry actor files many requests for reconsideration or to otherwise slow down policy enforcement. 
        \\ 
        
        \midrule

        \multicolumn{3}{l}{\textit{Incentive Shaping}} 
        \\
        
        \multicolumn{1}{l}{\multirow[t]{3}{*}{Donations, gifts, and bribes}} &
        Industry actors make financial contributions to elected officials’ campaigns, or give personal gifts to elected officials (or their staff). &
        A company donates money to a Congress member's campaign; a company gives free vacations to an agency staff member. 
        \\

        \multicolumn{1}{l}{\multirow[t]{6}{*}{Private threats}} &
        Industry actors make explicit or implicit threats of litigatory, reputational, or other negative consequences to prevent policy enactment or enforcement. &
        Regulators decline to investigate a company because they believe that the company would sue; a policymaker stops advocating for a policy because they are wary of negative press coverage; a company threatens to release material that would portray a policymaker in a negative light.
        \\ 

        \multicolumn{1}{l}{\multirow[t]{4}{*}{Revolving door}} &
        Industry actors hire government officials, regulators, policymakers, or their staff; or employees of industry actors leave to work for government officials, regulators, or policymakers.  &
        A company hires a legislator's chief of staff; a general counsel of a company is nominated for a political appointment.
        \\
        
        \midrule

        \multicolumn{3}{l}{\textit{Information Capture}} 
        \\
        
        \multicolumn{1}{l}{\multirow[t]{11}{*}{Agenda-Setting}} &
        Industry actors emphasize or de-emphasize particular perspectives or data, set priorities in policy conversations, or frame regulatory problems in ways that favor industry actors. Regulators may then adopt such views---e.g., goals, norms, practices, activities, and models of risks and markets. &
        A company frames industry regulation as a question of regulating downstream users or upstream producers (but not of the company itself); a company tells policymakers or regulators that particular policy goals are more important than other ones when thinking about regulating the industry, which then results in industry-biased regulation; many companies coordinate to repeat the same message to policymakers so that policymakers perceive a ``united front'' of industry voices on a particular issue.
        \\

        \multicolumn{1}{l}{\multirow[t]{6}{*}{Information management}} &
        Industry actors selectively share, control access to, withhold, or provide misleading or false information to policymakers or regulators. &
        A company fails to report important information about its product or business practices to regulators/policymakers; a company makes a presentation to a policymaker/regulator in which they highlight the benefits of their technology but fail to discuss its risks.
        \\

        \multicolumn{1}{l}{\multirow[t]{6}{*}{Information overload}} &
        Industry actors inundate policymakers or regulators with similar information or communications supporting their points of view, which challenges the ability of regulators to process and interpret the information. &
        Industry actors organize a comment submission drive and overwhelm the notice-and-comment process with comments favorable to the industry position; industry actors send (or organize) a barrage of phone calls, letters, or other communications to a policymaker to create the illusion of support for their position.
        \\

        \midrule
    \end{tabularx}
    \caption{Mechanisms of industry influence in the policy process (cont'd on next page)}
    \label{tab:Appendix_Mechanisms_Definitions}
\end{table*}

\setcounter{table}{1}
\begin{table*}[!p]
    \centering
    \small

    \begin{tabularx}{\textwidth}{ l X X }
        \toprule

        \multicolumn{1}{l}{\textbf{Mechanism}} &
        \multicolumn{1}{l}{\textbf{Definition}} &
        \multicolumn{1}{l}{\textbf{Examples}} 
        \\ 
        
        \midrule
        
        \multicolumn{3}{l}{\textit{Cultural Capture}}
        \\
        
        \multicolumn{1}{l}{\multirow[t]{6}{*}{Group identity}} &
        Policymakers or regulators may be ``more likely to adopt positions advanced by people whom they perceive as being their in-group'' \cite{kwak_cultural_2013}. &
        A regulator was formerly a business executive and identifies with employees in that industry as people of the same trade; a company sends a lobbyist of the same gender and ethnic background as a legislator to speak to them, in the hopes that the legislator would be more sympathetic.
        \\

        \multicolumn{1}{l}{\multirow[t]{7}{*}{Relationship networks}} &
        Policymakers or regulators may be ``more likely to adopt positions advanced by people who are in their social networks'' \cite{kwak_cultural_2013}. &
        A legislator has a relative who works in a particular industry, and the legislator adopts their relative's views about regulating that industry after speaking to them; a regulator regularly plays golf with trade association executives, and the regulator begins to adopt industry-friendly views after discussing policy issues with those executives.
        \\

        \multicolumn{1}{l}{\multirow[t]{6}{*}{Status}} &
        Policymakers or regulators may be ``more likely to adopt positions advanced by people whom they perceive to be of higher status in social, economic, intellectual, or other terms'' \cite{kwak_cultural_2013}. &
        A policymaker adopts the views of a someone testifying at a hearing because of their status as a technical expert; a legislator wishes to associate with CEOs in an industry that many people consider ``hot'' and ``the next big thing'', and adopts industry-friendly views as a result.
        \\ 
        
        \midrule

        \multicolumn{3}{l}{\textit{Indirect Capture}}
        \\
        
        \multicolumn{1}{l}{\multirow[t]{3}{*}{Academic Capture}} &
        Industry actors influence academic actors, who may then influence policymakers or regulators. &
        A company funds an academic's research; a company donates a large sum to a think tank; a company donates to a university to set up a research lab.
        \\

        \multicolumn{1}{l}{\multirow[t]{4}{*}{Private regulator capture}} &
        Industry actors influence private organizations that serve regulatory functions---e.g., auditors or standards-setting bodies. &
        A company actively participates in standards-setting, and the standards are then adopted by regulators; a company develops a close relationship with its auditors, leading to ineffective audits. 
        \\

        \multicolumn{1}{l}{\multirow[t]{3}{*}{Public relations}} &
        Industry actors engage in direct public communications, which may then influence policymakers or regulators directly or via shaping public opinion. &
        A company puts out a press release or runs an advertising campaign supporting or opposing a regulation.
        \\

        \multicolumn{1}{l}{\multirow[t]{4}{*}{Media Capture}} &
        Industry actors influence journalists or outputs from media channels, which may then influence policymakers or regulators directly or via shaping public opinion. &
        A company puts out paid media pieces advocating for its policy stances.
        \\

        \bottomrule
    \end{tabularx}
    \caption{Mechanisms of industry influence in the policy process (cont'd)}
\end{table*}

\endgroup

\clearpage

\section{Methods} \label{sec:Methods}

We conducted 17 semi-structured interviews with AI policy experts.\footnote{We conducted one group interview, so we spoke with more than 17 experts in total. Group interviewees were not assigned distinct IDs in Table \ref{tab:Interviewees}.} An expert interview method is appropriate for our research questions because most information about industry influence is non-public. Additionally, political influence---and regulatory capture in particular---is difficult to measure quantitatively, and our data captures many informal interactions and processes \cite{soest_why_2023}. Interviews were anonymous to enable more in-depth conversations about the policy process. The study protocol was approved by the Human Subjects Protection Committee at RAND.

We also examined observational data on industry influence in AI governance: media sources, academic research, online discourse, and other public artifacts. This data can help corroborate and triangulate our findings \cite{beyers_lets_2014}; observational data was gathered through targeted Google searches. Finally, we conducted a scoping review of the AI governance literature mentioning regulatory capture, with details and results in Appendix \ref{sec:Appendix_Review}. This review helped inform our definitions in Section \ref{sec:Definition} and generate examples discussed in interviews.

\subsection{Interviewee Selection} \label{subsec:Methods_Sampling}

We used expert and convenience sampling methods to identify interviewees, who were selected based on their academic or professional expertise in AI policy, as identified by: employment at a relevant government or civil society institution \cite{atih_responsible_2024, hicock_ai_nodate}, membership on lists of AI policy experts \cite{thebridge_thebridge_2024, oecd_oecd_nodate, sourcelist_sourcelist_nodate, waie_open_2024}, Google search, and referrals from individuals known to the authors. The first author contacted the interviewees and collected consent forms via email.

We primarily recruited experts located in the United States but also included some based in the United Kingdom and the European Union. To maintain sample diversity and reduce bias, we purposively invited experts with diverse demographic, organizational, and ideological backgrounds. We did not contact any experts currently affiliated with companies that develop general-purpose AI models. Experts' backgrounds are described in Table \ref{tab:Interviewees}.\footnote{Given the small size of our population of interest,  demographics are not reported to protect interviewee confidentiality.}

\begin{table*}[!ht]
    \centering
    \small
    \begin{tabular}{ l l l }
         \toprule
         
         \textbf{ID} & \textbf{Type} & \textbf{Role} 
         \\ 
         
         \midrule
         
         R1 &
         Government &
         A congressional staffer 
         \\
         
         R2 &
         Government & 
         A former congressional staffer 
         \\ 
         
         \midrule
         
         R3 &  
         Academia/research & 
         An academic in a university studying technology policy 
         \\
         
         R4 &  
         Academia/research & 
         An academic who has worked on AI in different sectors 
         \\
         
         R5 &  
         Academia/research & 
         An AI ethics researcher 
         \\
         
         R6 & 
         Academia/research & 
         A policy analyst working on technology issues 
         \\
         
         R7 & 
         Academia/research & 
         An expert at a technical research organization 
         \\ 
         
         \midrule
         
         R8 & 
         Civil society & 
         An executive at a US advocacy group that works on technology issues 
         \\
         
         R9 & 
         Civil society & 
         An executive at a US think tank working on technology policy issues 
         \\
         
         R10 & 
         Civil society & 
         An expert at a US think tank 
         \\
         
         R11 & 
         Civil society & 
         An economist at a US think tank 
         \\
         
         R12 & 
         Civil society & 
         A leader at a US think tank working on technology policy issues 
         \\
         
         R13 & 
         Civil society & 
         Grantmakers at a philanthropic foundation focused on technology 
         \\ 
         
         \midrule

         R14 & 
         EU civil society & 
         A researcher at a technology policy think tank 
         \\
         
         R15 & 
         EU civil society & 
         A policy executive at a think tank 
         \\
         
         R16 & 
         EU civil society & 
         An employee of an EU think tank working on technology policy 
         \\
         
         \midrule
         
         R17 & 
         UK civil society & 
         A policy expert at a UK think tank 
         \\
         
         \bottomrule
    \end{tabular}
    \caption{Overview of interviews, with interview IDs and descriptions of experts' roles}
    \label{tab:Interviewees}
\end{table*}

Although a sample size of $n = 17$ is relatively small, saturation \cite{fusch_are_2015, saunders_saturation_2018} in qualitative interview studies has been achieved with comparable or even smaller sample sizes \cite{hennink_sample_2022, guest_how_2006}. The nascency of AI regulation, especially in the United States, and the correspondingly small number of professionals in the field further justifies a smaller sample size \cite{baker_how_2012}.

\subsection{Interview Protocol} \label{subsec:Methods_Protocol}

All interviews were conducted online by the first author in January--February 2024 and lasted 40--60 minutes each. At the beginning of each interview, the first author described to interviewees the goals of this study and detailed our practices for protecting interviewee confidentiality. They then verbally re-obtained consent to record the interviews, to use the descriptions reported in Table \ref{tab:Interviewees}, and to use transcribed quotes.\footnote{When interviewees did not consent to recording, analysis was performed with the first author's contemporaneous interview notes.}

They identified to interviewees the focus of these interviews as being government policy related to AI. They defined ``AI systems'' as ``general-purpose AI systems that have a wide variety of use cases, rather than narrow or domain-specific AI systems'' and ``AI policy'' as ``government policies intended to regulate, restrict, or promote the development and deployment of general-purpose AI systems.''

The interview protocol was designed to elicit information about industry actors' preferred policy outcomes that could constitute capture, as well as what mechanisms of influence industry actors are currently using to facilitate those outcomes. Interviewees were first asked about the public interest goals of AI regulation, the types of industry actors involved in AI policy, and the policy goals of those actors. The first author then presented interviewees with a table of influence mechanisms (an early version of Table \ref{tab:Appendix_Mechanisms_Definitions}), asked interviewees to list any additional mechanisms of influence, and asked which mechanisms were currently most relevant to AI policy. Where interviewees indicated that industry actors used a specific mechanism to influence policy, they asked follow-up questions asking for examples of such influence, what mitigation measures were in place to curb it, whether and why those measures were effective, and whether similar dynamics existed in other industries. The full interview protocol is contained in Appendix \ref{sec:Appendix_Interview_Protocol}.

\subsection{Data Analysis} \label{subsec:Analysis}

The first author transcribed the interviews with the assistance of a private OpenAI Whisper instance, then de-identified the interview transcripts following \citet{saunders_anonymising_2015} and \citet{stam_qualitative_2023}. In preliminary analysis, the first and second authors deductively developed codes for the goals of AI regulation, the types of actors involved in AI regulation, and the mechanisms through which industry actors influenced the policy process. The third and fourth authors then independently coded the de-identified transcripts, and the first and second authors subsequently refined the codes and adjudicated any disagreements in coding using an open discussion method \cite{chinh_ways_2019}. The final coding manual for influence mechanisms was substantially similar to the table presented in Table \ref{tab:Appendix_Mechanisms_Definitions}. Agreement between coders was very high.\footnote{We do not provide quantitative inter-rater reliability metrics, as ease of coding was relatively high \cite{mcdonald_reliability_2019} and as there were few disagreements between coders.}

\section{Outcomes of Regulatory Capture in US AI Policy} \label{sec:Outcomes}

In this section, we present our findings on what policy outcomes could constitute regulatory capture in US AI policy. 

\begin{figure*}[!htb]
    \centering
    \includegraphics[width=0.60\textwidth]{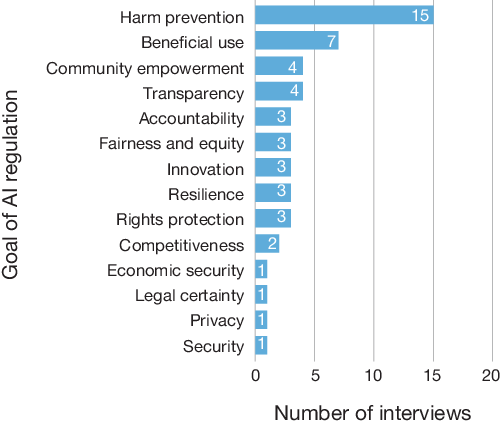} 
    \caption{Goals of AI regulation suggested by interviewees}
    \label{fig:Results_Goals}
\end{figure*}

Defining the public interest goals of general-purpose AI regulation---which is necessary to diagnose capture---is controversial. When we asked interviewees to identify goals that would be in the public interest when regulating general-purpose AI systems, responses included 14 distinct goals. Although experts indicated in 15 of 17 interviews that regulation should attempt to prevent harms from AI, they diverged on the specific possible goals of regulation such as community empowerment, innovation, competitiveness, and legal certainty (Figure \ref{fig:Results_Goals}). These results are perhaps unsurprising given the variety of stakeholders and interests that could be affected by general-purpose AI systems, because ``AI is . . . coming for every sector and industry'' \citetext{R13}.

\begin{figure*}[!htb]
    \centering
    \includegraphics[width=0.60\textwidth]{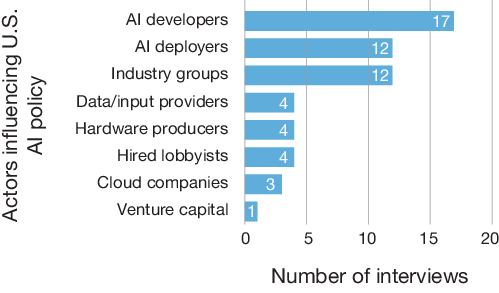} 
    \caption{Industry actors mentioned by interviewees as exerting policy influence}
    \label{fig:Appendix_Results_Actors}
\end{figure*}

Similarly, interviewees identified many actors in the AI industry who are participating in the policy process. Figure \ref{fig:Appendix_Results_Actors} presents a frequency chart of the number of interviews that mentioned each type of actor as participating in AI policy. Any mentions of specific companies or industry actors were re-coded into one of the categories below. Actor categories, definitions, and examples are presented in Table \ref{tab:Appendix_Actors_Definitions}.

\begin{table}[!ht]
    \centering
    \small

    \begin{tabularx}{\linewidth}{ l X }
        \toprule

        \textbf{Actor} &
        \textbf{Definition}
        \\ 
        
        \midrule

        AI deployers &
        Companies that are deploying AI products or services
        \\

        AI developers &
        Companies that are building AI models or products
        \\

        Cloud companies &
        Companies that provide virtual computing services but do not primarily manufacture hardware
        \\

        Data/input providers &
        Companies that provide datasets, data labeling services, or technical infrastructure or services to AI developers 
        \\

        Hardware producers &
        Companies that produce computing hardware
        \\

        Hired lobbyists &
        External lobbyists paid to advocate on behalf of industry actors listed elsewhere in this table
        \\

        Industry groups &
        Membership organizations consisting of industry actors listed elsewhere in this table
        \\

        Venture capital &
        Financial firms that provide funding to early-stage AI developers or deployers
        \\

        \bottomrule
    \end{tabularx}
    \caption{Definitions for industry actors participating in the policy process}
    \label{tab:Appendix_Actors_Definitions}
\end{table}

Interviewees noted that industry actors had many incentives and that these incentives sometimes differed. Some incentives that interviewees attributed to these actors when they influenced policy included: ``to look good'' \citetext{R7}, ``keeping regulatory burden . . . low'' \citetext{R16}, ``to maximize their profit'' \citetext{R6}, ``regulatory certainty, that regulations don't just fluctuate every second'' \citetext{R10}, and ``to preserve the ability to research and deploy AI systems'' \citetext{R12}. Moreover, although ``it is easy to think about industry or industry actors as monolithic . . . there are a variety of . . . industry actors who engage with policy conversation[s] with different motivations'' \citetext{R12}. While different industry players desire different and sometimes opposing policy outcomes, many of these outcomes conflict with some of the policy goals in Figure \ref{fig:Results_Goals}.

Notably, most interviewees did not believe that the current state of general-purpose AI policy in the United States reflects capture, but many expressed concern with existing, high levels of industry influence. We caution against conclusions that AI policy has already been ``captured''; accusations of regulatory capture in the United States while regulatory debates are ongoing can be premature and thus confusing \cite{wexler_which_2011} or counterproductive \citetext{e.g., \citealp{adler_framing_2021}}. Overall, however, our interviews do suggest that it is important to remain vigilant about industry influence resulting in policy outcomes detrimental to the public interest.\footnote{Given the experiences of other industries \cite{yeoh_capture_2019} and on such issues as data privacy and content moderation \cite{hilden_politics_2019, neudert_regulatory_2023, thierer_history_2013}, current levels of corporate influence in AI governance appear to raise real risks of capture.}

Of the types of outcomes resulting from regulatory capture, interviewees most often described outcomes related to the content of government policy, with substantially less discussion of policy enforcement and governance structures. We discuss below each of these outcome categories in turn.

\subsection{Policy Content} \label{subsec:Outcomes_Policy_Content}

Unlike much of the public discourse and the original literature on capture---suggesting that incumbents may seek over-regulation to protect their market advantages---the majority of interviewees expressed concerns that industry capture could result in regulation that is \textit{too weak} (or in no regulation at all). Examples of such outcomes included: ``weak requirements baked into legally binding requirements'' \citetext{R14}, ``creat[ing] an exemption for [particular companies]'' \citetext{R2}, ``remov[ing] requirements on those who are developing general-purpose AI models and hav[ing] requirements on [deployers]'' \citetext{R15}, and limiting regulation to industry self-regulation \citetext{R8}. Interviewees referenced specific industry attempts to weaken the EU AI Act by expansively defining an open source exemption and by raising compute thresholds to narrow the definition of advanced models \citetext{R14; \citealp{coulter_eus_2023}}.

Interviewees noted that corporate actors occasionally do advocate for regulation. However, interviewees offered disparate explanations for when and why companies would do so. One expert stated that some companies could ``very much [be] on board with the idea that protection of public health and safety should be a priority'' \citetext{R6}. Other interviewees were more skeptical: R12 conveyed that although some industry actors may genuinely support regulations, their policy positions can be distorted and ``be accidentally or purposely co-opted'' by actors with more self-interested goals. R13 said that certain large companies may support regulation for political advantage when competitors are engaging in politically unpopular behavior---e.g., around consumer privacy. And for R2, large companies may wish to simply codify their existing practices into regulation so that they can avoid increased investments in compliance while forcing ``competitors . . . to be operating at that level.'' 

On the other hand, only three interviewees discussed industry actors' use of AI policy to affect market competition. Of these, R9 noted that developing competitive moats through regulation is not currently a goal of AI companies because AI developers have yet to successfully commercialize their products, while R11 expressed that building regulatory barriers to entry ``just simply doesn't apply when we're talking about billion-dollar training runs.''

Finally, general-purpose AI regulations could embody suboptimal trade-offs among public interest goals: ``There's a really active sort of battle for the soul of AI regulation right now'' \citetext{R13}. Specific goals may be over- or under-emphasized in AI regulation, which could neglect particular rights or interests \citetext{R5}.

\subsection{Policy Enforcement} \label{subsec:Outcomes_Policy_Enforcement}

Another potential outcome of regulatory capture is uneven enforcement of government policy, on either an ad hoc or systematic basis. Although interviewees discussed these outcomes relatively less frequently---likely because no general-purpose AI regulations are currently being enforced in the United States, United Kingdom, or European Union---they raised concerns that historically lax enforcement practices on the technology industry could be repeated in AI regulation. Describing Facebook's enablement of racially targeted housing advertisements on its platform, R13 explained that:

\begin{displayquote}

``all sorts of rules have been allowed to be broken for sometimes like 10 years . . . because the enforcement community didn't understand what was happening and didn't have the capacity to go and extend those historic rules [to the technology industry] . . . every year that goes by that historic rules are not enforced in the digital environment normalizes the lack of enforcement, the lack of protections . . . So the [technology] systems have been built now in ways that protections we used to have don't exist anymore.''

\end{displayquote}

Relatedly, AI companies could also adopt ``the kind of Uber model of `let's just disrupt the taxi commissions and then they have to legalize us'\,'' \citetext{R11}. 

\subsection{Governance Structures} \label{subsec:Outcomes_Governance_Structures}

No experts suggested that industry actors are currently seeking to facilitate capture through governance structures. 

\section{Mechanisms of Industry Influence in US AI Policy} \label{sec:Mechanisms}

In this section, we present findings from interviews and from observational data about which mechanisms of influence (from Section \ref{subsec:Def_Mechanisms}) industry actors are currently employing to facilitate captured outcomes (from Section \ref{sec:Outcomes}).

\begin{figure*}[!htb]
    \centering
    \includegraphics[width=0.60\textwidth]{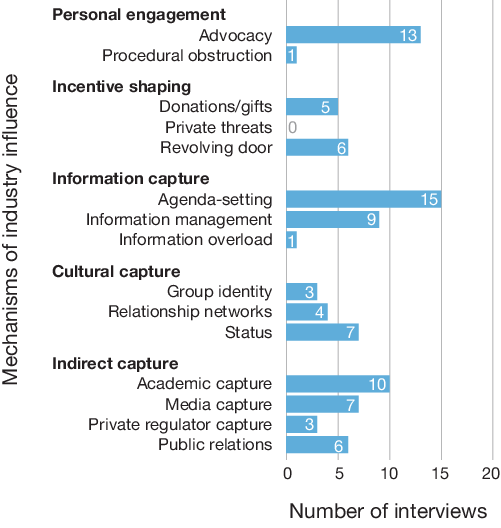} 
    \caption{Number of interviews that discussed each mechanism of industry influence in AI policy}
    \label{fig:Results_Mechanisms}
\end{figure*}

Figure \ref{fig:Results_Mechanisms} displays the influence mechanisms from Section \ref{subsec:Def_Mechanisms}, along with the number of interviews in which experts indicated each mechanism to be important in AI policy. Extended definitions and examples are in Table \ref{tab:Appendix_Mechanisms_Definitions}. Our discussion proceeds according to these categories of mechanisms. 

\subsection{Personal Engagement} \label{subsec:Mechanisms_Personal_Engagement}

\subsubsection{Advocacy.} 

Thirteen interviewees listed advocacy as one of the most important influence mechanisms in AI policy. Advocacy activities include, e.g., salon dinners and other social events \citetext{R5; R9; R10}, private meetings \citetext{R2; R12; R15}, and informational events or panels \citetext{R16}. These interactions are facilitated by social norms because ``elected officials kind of have to meet with [lobbyists] . . . There's a perception that it's a snub if they don't meet with the lobbyists'' \citetext{R7}.

Publicly available records provide an idea of the scale of lobbying by AI industry actors; R1 reported that lobbying by industry actors is much greater than lobbying by non-industry actors, and \citet{cheng_artificial_2024} finds that 85\% of DC lobbyists working on AI-related issues in 2023 were hired by industry organizations. In addition, R15 mentioned that in 2023, 77\% of registered meetings on AI taken by high-level European Commission members were with AI industry players \citetext{\citealp{ceo_meetings_2024}; see \citealp{ceo_byte_2023} and \citealp{field_ai_2024}}. Disclosures reveal that lobbying on AI issues in the United States increased significantly in 2023 at both the federal level \cite{ratanpal_federal_2024, oprysko_ai_2024} and the state level \cite{bordelon_as_2023, williams_amazon_2023}. Large AI developers have registered to lobby the federal government \cite{oprysko_openai_2023} and hired former congressional staff who were involved in crafting key AI policies as lobbyists (see Section \ref{subsec:Mechanisms_Incentive_Shaping}). Notably, many industry players who are supportive of general-purpose AI regulation in public are often much less supportive in private \cite{henshaw_theres_2024, perrigo_exclusive_2023}. 

Trade associations also engage in advocacy \citetext{R15; R16}, generally lobbying against any general-purpose AI regulations \citetext{R1}. Moreover, larger corporations may have more control over trade associations' decision-making, thus expanding their influence on policymakers through these associations \citetext{R15; R16; see \citealp{johnson_access_2020}}. Advocacy by trade associations may give policymakers the impression that particular policies represent an industry consensus. ``By far the biggest channel of influence is industry consensus,'' and perceptions of industry consensus can create an agenda-setting effect because ``governments are often very deferential to industry'' \citetext{R7}.

\subsubsection{Procedural Obstruction.}

According to our interviews, this mechanism is not widely used by industry to influence general-purpose AI regulation. Only R16 mentioned procedural obstruction, suggesting that some companies could be intentionally delaying or disrupting the work of AI standards-setting bodies.

\subsection{Incentive Shaping} \label{subsec:Mechanisms_Incentive_Shaping}

\subsubsection{Donations and Gifts.}

Four experts indicated that donations and gifts are currently an important mechanism of industry influence in AI policy,\footnote{A variation on the mechanism of donations and gifts is industry investment in legislators' districts \citetext{see, e.g., \citealp{de_figueiredo_economic_2022}}. These investments were not discussed in our interviews.} and another noted that this mechanism was not currently dominant but was likely to occur in the future. One interviewee specifically highlighted venture capital firm Andreessen Horowitz's 2023 announcement that it would donate to ``support like-minded candidates [who are against `misguided regulatory policy'] and oppose candidates who aim to kill America’s advanced technological future'' \cite{horowitz_politics_2023}. That expert reported that many elected officials perceived this statement as opposition to any general-purpose AI regulation.

In addition to monetary donations, gifts can also come in intangible forms, including status or social reputation, such as speaking engagements or other ``opportunities that fall below the thresholds of . . . anything that would be captured by [ethics] policies'' \citetext{R4}.

\subsubsection{Private Threats.}

No experts suggested that industry actors are currently using private threats to influence policy. 

\subsubsection{Revolving Door.}

The ``revolving door'' occurs when government staff take on industry employment and industry staff take on government roles. Experts noted that revolving doors are currently pervasive in Congress. Many AI companies have hired lobbying firms with former congressional technology policy staffers to lobby the US government on AI issues \citetext{R2; R10; \citealp{giorno_bottom_2024, scarcella_openai_2024}}. This dynamic is likely to intensify once regulations are implemented: Staffers currently working on AI policy would be particularly desirable hires for AI companies because ``whoever wrote the first [AI rules] has a lot of networks, has a lot of connections, [and] is highly sought after [by industry]'' \citetext{R10}.

Firms can also hire former staffers as in-house lobbyists. Two experts independently brought up the case of French AI startup Mistral and Cédric O, a former high-level French official known to be critical of the technology industry but who, after co-founding Mistral, lobbied for a provision in the EU AI Act that would exempt Mistral from regulatory requirements \cite{bergen_former_2023}. Another interviewee recalled that the top aide to Senate Majority Leader Chuck Schumer had become a public policy executive with Microsoft \cite{evers-hillstrom_lobbying_2023}, which then became the only company to have two representatives attend the first AI Insight Forum organized by that legislator \cite{miller_us_2023}.

Experts diverged on the extent to which revolving doors are relevant to AI policy, with some experts expressing that it was not currently a major issue \citetext{R9; R11}. Other interviewees indicated that revolving doors could become an issue once general-purpose AI regulations are enacted. R8 compared the AI industry to the nascent nuclear energy industry in the 1950s, indicating that revolving doors are particularly likely to occur because there are few credible leaders on AI policy who are unaffiliated with industry and who would not seek industry employment after working in government. Similarly, because ``the regulatory stakes are so high where a company could . . . be above water or underwater just based [on] how one line of law is interpreted under the agency . . . the revolving door will increase in a way potentially reaching a scale that we see in the defense sector'' \citetext{R10}. Head-hunting agencies specializing in hiring former government officials into industry may intensify these dynamics \citetext{R4}.

Aside from direct employment relationships, industry actors can directly fund roles in government: ``[T]here's a number of roles within government that there either is no funding for or requires third party funding to staff'' \citetext{R4}. Companies can help ``fill the gap'' in funding or even fund new roles \citetext{R4; \citealp{bordelon_key_2023, thompson_ex-google_2022, thompson_google_2022}}.

One example of such roles is industry-funded positions for staffers with technical backgrounds. Companies may genuinely wish to ensure that regulation is functional because badly designed regulation is bad for business \citetext{R7} or because they view impending regulation as inevitable. However, adding more technical staff within key decision-making bodies can create agenda-setting effects (see Section \ref{subsec:Mechanisms_Information_Capture}) or militate against stricter regulations: ``Congressional offices having additional technical expertise . . . does sometimes at least make it less likely that they will do the most extreme . . . forms of regulation'' \citetext{R2}.

\subsection{Information Capture} \label{subsec:Mechanisms_Information_Capture}

\subsubsection{Agenda-Setting.}

Experts expressed widespread concern about agenda-setting by AI companies, with one interviewee concluding that ``the battle is being fought on the front of agenda-setting'' \citetext{R10}. 

AI companies are advancing anti-regulation narratives based on national security or economic competitiveness. In conversations with policymakers, corporate actors regularly claim that general-purpose AI regulation must be avoided so that the United States can win an ``AI arms race'' with China \citetext{R13; \citealp{bordelon_dc_2024}}, argue that AI developers have a special economic or political role that is too important to be interfered with \citetext{R8}, create dichotomies between safety and innovation \citetext{R8; R16}, or oppose regulation of open-source models because ``open source is an engine of economic growth'' \citetext{R9}. These narratives can frequently be extensions of broader techno-libertarian rhetoric that has existed as far back as the 1990s \citetext{R5; R11}, such as ``the permissionless innovation framework'' \citetext{R11}.

Industry may also steer policy discussions toward particular problems---e.g., by hosting workshops or presentations with policymakers to ``frame the question of `what are the salient harms and risks from AI influence?'\,'' \citetext{R17}. AI developers can direct attention toward specific use cases and away from others: for R5, an outsized focus on text or image generation allows other uses---e.g., law enforcement, military, or surveillance---to go ignored. Even when large language models are deployed for these functions,

\begin{displayquote}
    ``[AI companies] sort of toss up their hands and say, `this [problem or type of use] has nothing to do with us.' And I think that accomplishes two things . . . They avoid a lot of the critiques and also the idea that policymakers would regulate them for that. And I think the dual-use narrative also happens a lot here. That means that, oh, if you're interested in this topic, you should regulate it away from us because that's not our thing.'' \citetext{R5}
\end{displayquote}

In addition, industry can set the policy agenda by promoting default standards or regulatory approaches \citetext{R7; R8}, which can fail to reflect broader public interest goals \cite{hacker_sustainable_2023}. Companies have promulgated self-governance practices, such as the Responsible Scaling Policy from \citet{anthropic_anthropics_2023} and the voluntary commitments on AI from the \citet{white_house_ensuring_2023}. Experts were concerned that these announcements ``essentially set[] the agenda for regulation'' \citetext{R9} and endeavor to prove to policymakers that ``there is some sense of . . . thoughtfulness going into how the technology is being built'' \citetext{R12}.\footnote{We do not mean to imply that such agenda-setting is always harmful or that companies are not being thoughtful by joining voluntary commitments. Self-regulation can often be beneficial.} Companies have also pushed for regulation of individual AI use cases rather than direct regulation of general-purpose AI systems \citetext{R3; R6; R15}.

Another form of agenda-setting occurs when industry actors determine metrics for measuring harm prevention, equity, or system safety. Companies frequently ``[cast] demographic diversity in the text and images in the generated content . . . as the holy grail of [AI] ethics, and it's . . . not even close . . . It's not a sliver of the problems being created'' \citetext{R5}. Metrics used are often misleading proxies for the harms they sought to address, which can detract from broader concerns about AI's effects on labor markets, economic inequality, democratic governance of AI systems, and the goal of ``shifting power to marginalized communities'' \citetext{R5}.

Moreover, some companies attempt to steer regulatory discussions toward technical or engineering details at the expense of broader policy discussions. Although governments require technical expertise to make effective policy, it may be undesirable for discussions to be overly limited to technical details because

\begin{displayquote}
    ``companies prefer to have fights in the . . . technical weeds . . . [One analogy is] you would be trying to argue with the military on their own terms where you're saying, like, oh, this strategy is going to lead you to capture or not capture this number of things or have this effect. But the military knows way more about all those stats and has way more ability to produce those stats . . . If you're arguing with them about their stats and on their terms, it feels like you're just losing.'' \citetext{R2}
\end{displayquote}

This framing may help advance more-moderate regulations or make enacting regulation more difficult, and it could also exacerbate cultural or informational effects, forcing governments to defer to industry information \cite{hakenes_regulatory_2014}.\footnote{On the other hand, technical discussions can benefit the regulatory process, and it may be the case that more moderate regulations are sometimes desirable. It is important, however, to ensure that this framing does not derail public interest goals such as preventing harms from AI technologies.} It can also sideline civil society groups by reducing ``a political policy issue that many groups in society might have a stake in . . . and turning it into that sort of technocratic discipline'' \citetext{R17}.

\subsubsection{Information Management.}

AI companies possess large information asymmetries over government actors, who have limited expertise and visibility into advanced AI systems \cite{taeihagh_governance_2021, lupo_risky_2023, anderljung_frontier_2023}. Companies are capitalizing on these asymmetries to shape the information environment in their favor \citetext{R1; R2; R3; R11; R14}, usually by withholding important information about general-purpose AI systems or about AI development processes. Interviewees gave examples of such ``information management'', including the lack of information publicly disclosed about OpenAI's GPT-4 compared with previous models, even when contemporary industry norms were to release certain data \citetext{R14}; the lack of transparency about firms' data collection policies and the contents of training data \citetext{R1; R11; see \citealp{gallifant_peer_2024}}; and companies' claims that they were not in possession of information about the labor practices of data suppliers \citetext{R5}. AI developers could further engage in ``stonewalling or slow-walking . . . like lack of forthcomingness to inquiries that make it difficult for government actors to have a full picture of what's going on in an official capacity'' \citetext{R12}.

The flip side of refusing to disclose key information is releasing information that could serve industry goals---e.g., cherry-picked, positive use cases of AI technologies. Although promoting beneficial use of AI is a possible goal of regulation (Figure \ref{fig:Results_Goals}), ``industry [is] selectively sharing examples of proactive, responsible use or responsible approaches, which both are intended to demonstrate a reduced need for regulatory intervention or to shape the kind of regulatory imagination of what good practices look like'' \citetext{R12}.

\subsubsection{Information Overload.}

Only one expert discussed information overload, which occurs when actors inundate policymakers with more information than they can assess to lower scrutiny of policy outcomes \cite{wagner_administrative_2009}. R6 indicated that many industry actors submitted comments to the National Institute of Standards and Technology (NIST) on the AI Risk Management Framework \cite{nist_ai_2023}, which could be a limited example of this mechanism at play.

\subsection{Cultural Capture} \label{subsec:Mechanisms_Cultural_Capture}

\subsubsection{Group Identity.}

Interviewees indicated that influence exerted through group identity is limited and largely occurs along partisan lines---e.g., when corporate lobbyists lean on partisan affiliations to push policymakers toward particular policies \citetext{R9}. At least in Europe, lobbyists are also appealing to national interests in an attempt to obtain more-favorable regulations for local AI companies \citetext{R14; R15; \citealp{volpicelli_power_2023}}. No interviewees suggested that regulators may identify with the AI industry, as \citet{kwak_cultural_2013} describes having occurred in the financial industry.

\subsubsection{Relationship Networks.}

Relationship networks are an important channel of influence for professional lobbyists \citetext{R1; R8}. Corporate executives also leverage personal relationships to advance their policy goals: One particular executive at a major AI company was described as ``ha[ving] so many relationships and [is] so extroverted and so personable and so forth, [which] is a key input into why they have more voice on a lot of these debates'' \citetext{R11}.

\subsubsection{Status.}

Seven experts expressed concerns with the more-subtle influence of intellectual status (technical expertise) and social status on policymaking. The epistemic construction of technical expertise in particular has been a battleground for AI companies in seeking to influence policy \citetext{R4; \citealp{bode_constructing_2023}}. Company executives who oppose general-purpose AI regulation frequently disparage non-industry experts as ``old'' and outdated \citetext{R7}. On the other hand, these industry players tend to promote ``the builders [who these players claim] understand what it takes to actually build stuff'' and who they claim are ``not worried about [risks from AI] because [the builders] really understand how AI works. Once you're a technical engineer, and you really understand this stuff, you'd understand that you shouldn't be afraid of [AI]'' \citetext{R7}. Furthermore, anchoring expertise to industry employment furthermore allows companies to promote the views of ``big-wig influencer[s] within an AI sphere'' who may not have technical understanding of AI systems \citetext{R5; \citealp{goanta_regulation_2023}}.\footnote{Industry experts often \textit{do} have real expertise. R2 described a congressional hearing at which ``we had some people come from [a cryptocurrency exchange], and they hired a bunch of former senior [executive branch] officials who worked on preventing cybercrime . . . And they are giving you a mix of genuinely helpful information that they are gaining as someone with their genuine expertise, mixed in with things that fit the company's goals.''}

Executives of AI companies are also viewed as having high social status, which makes policymakers more favorable to their views \citetext{R16}. Specifically, elected officials may seek to mirror industry executives’ views because they seek proximity ``with the cool tech CEO'' \citetext{R16} or because they see the ``zeitgeist of AI being important right now'' \citetext{R1}.

\subsection{Indirect Capture} \label{subsec:Mechanisms_Indirect_Capture}

\subsubsection{Academia.}

Experts most frequently discussed academia as a target of indirect capture. Industry funding of academic labs or research centers can shape research agendas and directions \citetext{R5; R17; \citealp{jiang_ai_2023, tafani_whats_2022, cath_your_2022,himmelreich_artificial_2022, weinkle_experts_2019}}. Direct institutional funding can help industry actors generate positive publicity, shape conversations, and influence decisions at universities and at major academic AI conferences \cite{abdalla_grey_2021}. Academic researchers are often dependent on collaborations with industry because academic institutions lack the large-scale computing resources needed for research into large models \cite{ahmed_-democratization_2020, whittaker_steep_2021}, which can result in ``research that would be useful to the player that has that infrastructure rather than purely exploratory research'' \citetext{R12; \citealp{widder_open_2023, abdalla_elephant_2023}}. R13 noted that these dynamics combine to create academic and ``employment structures [that] incentivize a certain kind of belief system'' and prevent academic institutions from effectively training technologists to consider downstream harms of AI.

Additionally, industry organizations can support ``people who genuinely have ideological commitments that happen to line up with theirs, rather than paying them to change their view'' \citetext{R2}. For instance, companies can promote academic researchers who share their views by inviting them to serve on advisory boards, endowing professorships, funding teaching buyouts, or starting partnerships with universities \citetext{R4; R9; R17}. Providing these incentives can advantage industry views even where the independence of academic research is not compromised. 

Industry ties are pervasive even among researchers who study the ethics and social effects of AI \cite{cath_your_2022, young_confronting_2022}. R5 recalled:
\begin{displayquote}
   ``I was at FAccT [the Association of Computing Machinery (ACM) Conference on Fairness, Accountability, and Transparency] once, and I was in this circle of people . . . [who are] very critical of corporate influence, and I asked this small circle I was standing in how many of them had ever been affiliated with DeepMind, and it was [a large percentage] of people in the circle.''
\end{displayquote}
Interviewees also indicated that similar dynamics could occur in think tanks as in academia \citetext{R4; R7; R17}. 

\subsubsection{Media Capture.}

Many interviewees expressed that media capture is occurring \citetext{see also \citealp{brennen_industry-led_2018, barakat_interrogating_2024}}. Companies often pay for sponsored content \citetext{R12; e.g., \citealp{wang_sponsored_2023}} or co-host high-profile media events \citetext{R14; e.g., \citealp{politico_davos_2024}}. Media staff may also become culturally captured by industry through relationship networks or status \citetext{R4; R5; R14}. These concerns, however, were tempered by the observation that many outlets can be ``super anti-tech'' \citetext{R1; R2}. 

\subsubsection{Private Regulator Capture.}

Private bodies that serve regulatory functions can be captured in much the same way as government institutions. The structure of the current auditor and model evaluator landscape raises questions about the independence of AI assessments \cite{raji_change_2023}. Auditors may not be sufficiently independent or could have perverse incentives that impair the integrity or validity of assessment results \citetext{R3; \citealp{manheim_necessity_2024}}. AI auditors are also vulnerable to cultural capture---as was common in the financial industry during the run-up to 2008 \cite{kwak_cultural_2013}---because ``they're being invited [to] or speaking at some of the same events, AI policy, AI standards, events [where] industry actors . . . are. So certainly I think a bit of indirect influence is happening'' \citetext{R6}.

Private standards-setting bodies can be subject to capture as well \citetext{R3; R6}. Industry actors are generally well resourced and better positioned to engage in standards-setting processes \citetext{R6; \citealp{sri_discerning_2023}}, and over-representation of industry on standards-setting bodies could help facilitate capture \citetext{R14}.\footnote{Similarly, industry over-representation on advisory councils could facilitate capture \citetext{R14; \citealp{vassei_ethical_2019}; \citealp{schaferling_case_2023}; see also, e.g., \citealp{dhs_over_2024}}. See \citet{westgarth_understanding_2022} for a different view.} Companies can also hinder standards-setting work by engaging in procedural obstruction \citetext{R16}. 
Overall, these interactions can result in the creation of standards that are particularly partial to industry \cite{gornet_european_2023, tartaro_regulating_2023, tartaro_towards_2023, ebers_standardizing_2022}.

\subsubsection{Public Relations.}

Industry players attempt to influence policymakers by shifting public opinion. Though interviewees were unable to specify the extent to which these methods were effective, several discussed specific public relations campaigns. One expert mentioned a trade association’s \$25 million expenditure on an ad campaign detailing positive AI use cases \citetext{R10; \citealp{technet_technets_2024}}, and another was concerned that AI companies may adopt practices from other parts of the technology industry of sending ``direct calls to action to customers, . . . like Uber, sending notifications directly to users'' \citetext{R11; \citealp{stempeck_are_2015}}. 

\section{Mitigating or Preventing Regulatory Capture in AI Policy} \label{sec:Prevention}

Drawing on our interviews and on the social science literature, we preliminarily discuss below mitigation measures for different influence mechanisms. Overall, systemic changes are needed to build an AI governance ecosystem that aligns AI development and deployment with the public interest. One urgent need is to build governance capacity for AI within government and civil society \citetext{see \citealp{reuel_position_2024}}---ensuring that government, academia, private regulators, and other organizations receive sufficient funding and talent will be crucial both to resisting capture in the overall ecosystem as well as to combating informational capture. For instance, R12 pointed at ``public [AI] infrastructure and investments'' a significant intervention, and full funding for NIST could help build technical capacity in government \cite{gastfriend_leading_2024} and thereby reduce reliance on industry expertise. 

Procedural and institutional safeguards may also contribute to aligning AI with the public interest. The literature frequently discusses the importance of such mechanisms as watchdogs \cite{caprio_regulatory_2013}, independent review of regulatory rules \cite{livermore_can_2013}, and appointed advisory boards or public advocates \cite{wagner_administrative_2009, cuellar_coalitions_2013, schlanger_offices_2014, baxter_understanding_2012}. Interviewees were also supportive of participatory processes such as notice-and-comment,\footnote{Notice and comment is an administrative procedure in the United States used by federal agencies when setting regulation. During the notice-and-comment process, an agency announces a potential regulation to the public, solicits public comments on a draft rule, and then promulgates a final rule and responds to those comments \cite{garvey_brief_2017}.} which could help increase policy access for individual and civil society actors. 

Specific mechanisms of industry influence may also be addressed via targeted measures, and we discuss below some of these interventions. Increased transparency and robust civil society institutions may be generally useful for combating mechanisms such as advocacy, donations and gifts, and indirect capture through public relations or media capture \citetext{R6; R9; \citealp{baxter_understanding_2012, anderljung_towards_2023, hirsch_policy-development_2018}}. R16 suggested, for instance, that requiring disclosures of lobbyist meetings with legislators and policymakers may be slightly but not very effective. R9 described having attended ``dinners where gifting limits were stretched to their broadest definition,'' and forcing disclosure of these types of events and expenditures could increase accountability for both corporate and government actors. Most experts were pessimistic about the sufficiency of transparency requirements in curbing industry influence, though R10 pointed at the Lobbying Disclosure Act and the Foreign Agents Registration Act as examples of effective, strictly enforced transparency requirements. Generally, however, the ability to capture policy through advocacy, donations and gifts, or indirect capture may reflect systemic weaknesses in US societal structures \cite{mitnick_capturing_2011};  targeted interventions to address these mechanisms may run into legal barriers \cite{killion_free_2023}. 

Revolving doors can be addressed through more robust (enforcement of) ethics requirements limiting post-government employment \cite{kwak_cultural_2013}. Government ``conflicts [of interest] reviews that are more straightforward in other legacy fields [are] not as rigorous for both AI and tech more generally'' because ethics offices often do not have AI ``industry knowledge and [familiarity with] the variety of roles that can exist in different types of companies'' \citetext{R4}. Offices managing these types of reviews might benefit from additional funding and from AI industry-specific training. Interventions such as these, however, which limit personnel transfers between government and industry, must be balanced with governments' needs to be aware of industry developments and to possess top technical expertise. Another intervention for revolving doors could be to make government careers relatively more desirable by investing in regulator salaries, work environments, and professional development \cite{mitnick_developing_2015, hempling_regulatory_2014}. 

Information capture can be addressed through giving non-industry stakeholders greater access to the policy process. In particular, ``the very early stage'' before legislative proposals have been drawn up ``is potentially the most important because it fixes what [a draft] law looks like. So the framing is quite important, . . . and the most powerful industry stakeholders could be more involved in that process'' \citetext{R15; see \citealp{khanal_why_2024}}. To address this issue, consumer empowerment programs could help enable civic participation in AI policy \cite{schwarcz_preventing_2013}. Reporting and monitoring requirements could also mitigate agenda-setting and information management by increasing regulatory visibility and verifying industry information \cite{rubinstein_reiss_benefits_2012}. At the same time, however, stakeholder engagement can also slow down the policy process. Given the pace of AI development, additional research may be needed to find new methods to enable participation without unduly slowing down AI policy.

In contrast with other mechanisms of influence, many interviewees suggested that targeted interventions for cultural capture were largely intractable \citetext{e.g., R11}. Some potential solutions are ensuring that policymakers come from diverse backgrounds \cite{kwak_cultural_2013}, changing agency cultures \cite{hempling_regulatory_2014}, and increasing policymaker awareness of cultural capture mechanisms so that policymakers can be on guard against these mechanisms \citetext{R6}. 

Finally, academic capture has been relatively well-studied in the context of AI \cite{abdalla_elephant_2023, abdalla_grey_2021, rikap_dynamics_2024}. Analogous to revolving doors in government, evening the playing field for non-industry academics would mitigate academic capture---e.g., through increasing non-industry career opportunities and ensuring access to compute and data resources \cite{zingales_preventing_2013, egan_action_2023, besiroglu_compute_2024}. Independent funding sources can also alleviate industry control over universities and think tanks \citetext{R17}, as well as over private regulators such as auditors \citetext{R6}.

\section{Limitations} \label{sec:Limitations}

This work is highly contextual and subject to a number of limitations. Methodologically, our interview sampling method was not random, which could affect sample representativeness and result validity. Our instrument was also descriptive and targeted toward uncovering \textit{current} industry practices in influencing general-purpose AI policy; influence dynamics may change over time as regulations are implemented or updated.

Our scope is also limited: We examine only US policy, general-purpose AI regulation, and adverse policy influence exerted by corporate actors. Our results may not generalize outside these contexts; regulatory capture in international institutions or other jurisdictions \citetext{see, e.g., \citealp{lall_why_2009, young_confronting_2022}}, or relating to other types of AI systems, may take on very different forms. Furthermore, while corporate actors tend to dominate the US policymaking process, other special interests can also exert policy influence in ways distinct from industry \citetext{e.g., \citealp{ahmed_building_2023}}. We also excluded consideration of ways in which industry influence could be net beneficial for the public.

Finally, our discussion of solutions to regulatory capture is preliminary and intended only to suggest directions for future research. Models of capture differ widely depending on the actors involved and the types of government policy at hand. Further research is needed to identify targeted institutional changes that would address context-specific models of capture in general-purpose AI regulation. 

\section{Conclusion} \label{sec:Conclusion}

Industry influence in AI governance can harm consumers and decrease social welfare through regulatory capture. Capture occurs when companies steer policy processes or dynamics to create an industry-leaning regulatory environment---causing, e.g., policymakers to under-regulate unsafe, opaque, or inequitable AI systems. Our findings suggest that AI developers, deployers, trade associations, and other industry actors are attempting to achieve captured outcomes most often through mechanisms of agenda-setting, advocacy, academic capture, and information management. 

These channels of influence operate in heterogeneous ways and can lead to a variety of undesirable outcomes. Researchers should understand the various models of capture and the goals of different actors, and additional work is needed to identify workable solutions to these different models. Although not all industry participation in AI policy is problematic, policymakers must be on guard against both more-conspicuous and subtler forms of corporate influence in order to prevent capture.

\clearpage

\section*{Ethical Considerations Statement}

This study was reviewed by the Human Subjects Protection Committee at RAND (HSPC ID: 2023-N0581) and was determined to be exempt from further review. The interviewers obtained written informed consent forms from all interviewees prior to their participation in this study, as well as verbal consent to record interviews prior to the start of each interview (interviews were not recorded when interviewees declined to be recorded). The authors conducted interviews via secure platforms, took steps to securely store and transmit study data, and generally followed best practices in protecting the confidentiality of interviewees and in de-identifying interview data \cite{saunders_anonymising_2015, stam_qualitative_2023}. AI assistance was used in the form of a private OpenAI Whisper instance hosted by RAND to transcribe interview recordings where applicable; at no point were interview recordings or other identifiable data input to other AI systems. At the time of this writing, the authors have deleted all identifiable project data.

\section*{Adverse Impacts Statement}

The broader impact of this research is generally expected to be positive, but a project on industry influence and regulatory capture in AI policy also has risks. 

First, there is a risk that by focusing on regulatory capture, this article overly emphasizes in the policy conversation the pitfalls of industry participation in the policy process. We emphasize that industry players have an important role in regulation and in AI policy, and it is important for policymakers to balance that role with the potential of regulatory capture. 

Second, there is a risk that readers may conclude that AI policy has already been captured in the United States, thus justifying undesirable policy corrections or other measures. We do not claim in this article that US AI policy has already been captured, and we caution against calls to this effect because of the difficulty of measuring capture and the heterogeneity of possible policy goals for AI regulation. 

Third, there is a risk that readers may conclude that capture is inevitable, and perhaps relatedly that all AI regulation is doomed to fail. We do not claim in this article that capture is inevitable in US AI policy, nor do we believe that all AI regulation will inexorably be entangled by capture dynamics. We believe that it would be unreasonable to infer such claims from this article.

Fourth, there is a risk that our work may be interpreted as predictive of regulatory capture dynamics in the future---e.g., after the regulatory and governance environments have matured. We emphasize that our research is meant to identify trends in influence and to outline possibilities of capture at the current point in time in the United States; we do not make predictive claims about influence and capture dynamics in the future or after significant changes in the regulatory and governance landscapes.

Finally, there is a risk that by thoroughly examining the ways in which industry actors can influence policy, this article paradoxically \textit{increases} the risk of capture by highlighting for AI industry actors some strategies of influencing policy that have been used to great effect in other industries. Although this risk may be real, we hope that by documenting these strategies, we are able to provide clarity and raise awareness among policymakers and researchers such that future AI policy may be designed to guard against undue or harmful industry influence.

\section*{Acknowledgments}

KW was the primary organizer and writer of this article. KW and CE were responsible for data collection and conceptualization. NG and CD assisted with data analysis and with review and editing. The authors are grateful for conversations with and feedback from (in random order): Brodi Kotila, Keller Scholl, Janet Egan, Rashida Richardson, Markus Anderljung, Anthony Barrett, Karson Elmgren, Risto Uuk, Diane Bernabei, Stephen Casper, Matt Davies, Anton Schenk, Gaurav Sett, Jon Menaster, Michelle Shevin, Everett Smith, Michael Aird, Shaun Ee, and Ashwin Acharya. The authors also thank Sandra Petitjean for graphics design, David Adamson for editing support, Nora Spiering and Rebecca Fowler for copy-editing support, and three anonymous AIES reviewers for comments and suggestions. 

This work was undertaken by the Technology and Security Policy Center within RAND Global and Emerging Risks, and funding for this work was provided by gifts from RAND supporters. Global and Emerging Risks is a division of RAND that delivers rigorous and objective public policy research on the most consequential challenges to civilization and global security. The Technology and Security Policy Center explores how high-consequence, dual-use technologies change the global competition and threat environment, then develops policy and technology options to advance the security of the United States, its allies and partners, and the world. For more information, contact tasp@rand.org.

\clearpage

\bibliography{references_AAAI_RC}

\begin{thebibliography}{420}
\providecommand{\natexlab}[1]{#1}

\bibitem[{Abdalla and Abdalla(2021)}]{abdalla_grey_2021}
Abdalla, M.; and Abdalla, M. 2021.
\newblock The {Grey} {Hoodie} {Project}: {Big} {Tobacco}, {Big} {Tech}, and the {Threat} on {Academic} {Integrity}.
\newblock In \emph{Proceedings of the 2021 {AAAI}/{ACM} {Conference} on {AI}, {Ethics}, and {Society}}, {AIES} '21, 287--297. New York, NY, USA: Association for Computing Machinery.
\newblock ISBN 978-1-4503-8473-5.

\bibitem[{Abdalla et~al.(2023)Abdalla, Wahle, Ruas, Névéol, Ducel, Mohammad, and Fort}]{abdalla_elephant_2023}
Abdalla, M.; Wahle, J.~P.; Ruas, T.; Névéol, A.; Ducel, F.; Mohammad, S.; and Fort, K. 2023.
\newblock The {Elephant} in the {Room}: {Analyzing} the {Presence} of {Big} {Tech} in {Natural} {Language} {Processing} {Research}.
\newblock In Rogers, A.; Boyd-Graber, J.; and Okazaki, N., eds., \emph{Proceedings of the 61st {Annual} {Meeting} of the {Association} for {Computational} {Linguistics} ({Volume} 1: {Long} {Papers})}, 13141--13160. Toronto, Canada: Association for Computational Linguistics.

\bibitem[{Abdu, Pasquetto, and Jacobs(2023)}]{abdu_empirical_2023}
Abdu, A.~A.; Pasquetto, I.~V.; and Jacobs, A.~Z. 2023.
\newblock An {Empirical} {Analysis} of {Racial} {Categories} in the {Algorithmic} {Fairness} {Literature}.
\newblock In \emph{Proceedings of the 2023 {ACM} {Conference} on {Fairness}, {Accountability}, and {Transparency}}, {FAccT} '23, 1324--1333. New York, NY, USA: Association for Computing Machinery.
\newblock ISBN 9798400701924.

\bibitem[{Abebe et~al.(2022)Abebe, Hardt, Jin, Miller, Schmidt, and Wexler}]{abebe_adversarial_2022}
Abebe, R.; Hardt, M.; Jin, A.; Miller, J.; Schmidt, L.; and Wexler, R. 2022.
\newblock Adversarial {Scrutiny} of {Evidentiary} {Statistical} {Software}.
\newblock In \emph{Proceedings of the 2022 {ACM} {Conference} on {Fairness}, {Accountability}, and {Transparency}}, {FAccT} '22, 1733--1746. New York, NY, USA: Association for Computing Machinery.
\newblock ISBN 978-1-4503-9352-2.

\bibitem[{Abou-Zeid, Bayingana, and Amazouz(2022)}]{abou-zeid_au_2022}
Abou-Zeid, A.; Bayingana, M.; and Amazouz, S. 2022.
\newblock {AU} {Data} {Policy} {Framework}.
\newblock Technical report, African Union.

\bibitem[{Adler(2021)}]{adler_framing_2021}
Adler, L. 2021.
\newblock Framing disruption: how a regulatory capture frame legitimized the deregulation of {Boston}’s ride-for-hire industry.
\newblock \emph{Socio-Economic Review}, 19(4): 1421--1450.

\bibitem[{Aggarwal, Meschke, and Wang(2012)}]{aggarwal_corporate_2012}
Aggarwal, R.~K.; Meschke, F.; and Wang, T.~Y. 2012.
\newblock Corporate {Political} {Donations}: {Investment} or {Agency}?
\newblock \emph{Business and Politics}, 14(1): 1--38.

\bibitem[{Ahmed and Wahed(2020)}]{ahmed_-democratization_2020}
Ahmed, N.; and Wahed, M. 2020.
\newblock The {De}-democratization of {AI}: {Deep} {Learning} and the {Compute} {Divide} in {Artificial} {Intelligence} {Research}.
\newblock arXiv:2010.15581.

\bibitem[{Ahmed et~al.(2023)Ahmed, Jaźwińska, Ahlawat, Winecoff, and Wang}]{ahmed_building_2023}
Ahmed, S.; Jaźwińska, K.; Ahlawat, A.; Winecoff, A.; and Wang, M. 2023.
\newblock Building the {Epistemic} {Community} of {AI} {Safety}.
\newblock SSRN:4641526.

\bibitem[{{AI Governance Alliance}(2024)}]{ai_governance_alliance_generative_2024}
{AI Governance Alliance}. 2024.
\newblock Generative {AI} {Governance}: {Shaping} a {Collective} {Global} {Future}.
\newblock Technical Report~3, World Economic Forum.

\bibitem[{Ajena et~al.(2022)Ajena, Bossard, Clément, Hibeck, Tiselli, and Oehen}]{ajena_agroecology_2022}
Ajena, F.; Bossard, N.; Clément, C.; Hibeck, A.; Tiselli, E.; and Oehen, B. 2022.
\newblock Agroecology \& {Digitalisation}: traps and opportunities to transform the food system.
\newblock Working paper, IFOAM Organics Europe.

\bibitem[{Alaga and Schuett(2023)}]{alaga_coordinated_2023}
Alaga, J.; and Schuett, J. 2023.
\newblock Coordinated pausing: {An} evaluation-based coordination scheme for frontier {AI} developers.
\newblock arXiv:2310.00374.

\bibitem[{{All-In Podcast}(2023)}]{all-in_podcast_all-summit_2023}
{All-In Podcast}. 2023.
\newblock All-{In} {Summit}: {Bill} {Gurley} presents 2,851 {Miles}.
\newblock \url{https://www.youtube.com/watch?v=F9cO3-MLHOM}.

\bibitem[{Allen(2019)}]{allen_regulating_2019}
Allen, T.~C. 2019.
\newblock Regulating {Artificial} {Intelligence} for a {Successful} {Pathology} {Future}.
\newblock \emph{Archives of Pathology \& Laboratory Medicine}, 143(10): 1175--1179.

\bibitem[{Almada and Petit(2023)}]{almada_eu_2023}
Almada, M.; and Petit, N. 2023.
\newblock The {EU} {AI} {Act}: a medley of product safety and fundamental rights?
\newblock SSRN:4308072.

\bibitem[{Altman, Marcus, and Montgomery(2023)}]{altman_oversight_2023}
Altman, S.; Marcus, G.; and Montgomery, C. 2023.
\newblock Oversight of {A}.{I}.: {Rules} for {Artificial} {Intelligence}.
\newblock \url{https://www.c-span.org/video/?528117-1/openai-ceo-testifies-artificial-intelligence}.

\bibitem[{Anderljung et~al.(2023{\natexlab{a}})Anderljung, Barnhart, Korinek, Leung, O'Keefe, Whittlestone, Avin, Brundage, Bullock, Cass-Beggs, Chang, Collins, Fist, Hadfield, Hayes, Ho, Hooker, Horvitz, Kolt, Schuett, Shavit, Siddarth, Trager, and Wolf}]{anderljung_frontier_2023}
Anderljung, M.; Barnhart, J.; Korinek, A.; Leung, J.; O'Keefe, C.; Whittlestone, J.; Avin, S.; Brundage, M.; Bullock, J.; Cass-Beggs, D.; Chang, B.; Collins, T.; Fist, T.; Hadfield, G.; Hayes, A.; Ho, L.; Hooker, S.; Horvitz, E.; Kolt, N.; Schuett, J.; Shavit, Y.; Siddarth, D.; Trager, R.; and Wolf, K. 2023{\natexlab{a}}.
\newblock Frontier {AI} {Regulation}: {Managing} {Emerging} {Risks} to {Public} {Safety}.
\newblock arXiv:2307.03718.

\bibitem[{Anderljung et~al.(2023{\natexlab{b}})Anderljung, Smith, O'Brien, Soder, Bucknall, Bluemke, Schuett, Trager, Strahm, and Chowdhury}]{anderljung_towards_2023}
Anderljung, M.; Smith, E.~T.; O'Brien, J.; Soder, L.; Bucknall, B.; Bluemke, E.; Schuett, J.; Trager, R.; Strahm, L.; and Chowdhury, R. 2023{\natexlab{b}}.
\newblock Towards {Publicly} {Accountable} {Frontier} {LLMs}: {Building} an {External} {Scrutiny} {Ecosystem} under the {ASPIRE} {Framework}.
\newblock arXiv:2311.14711.

\bibitem[{Anderson(2018)}]{anderson_court_2018}
Anderson, J.~J. 2018.
\newblock Court {Capture}.
\newblock \emph{Boston College Law Review}, 59: 1543.

\bibitem[{Andreessen(2023)}]{andreessen_techno-optimist_2023}
Andreessen, M. 2023.
\newblock The {Techno}-{Optimist} {Manifesto}.
\newblock \url{https://perma.cc/6SNE-AP98}.
\newblock Accessed: 2023-12-16.

\bibitem[{{Andreessen Horowitz}(2023)}]{andreessen_horowitz_andreessen_2023}
{Andreessen Horowitz}. 2023.
\newblock Andreessen {Horowitz}—written evidence ({LLM0114}) to the {House} of {Lords} {Communications} and {Digital} {Select} {Committee} inquiry: {Large} language models.
\newblock \textit{House of Lords Communications and Digital Select Committee}. \url{https://perma.cc/UF3M-JL2Y}.

\bibitem[{{Anthropic}(2023)}]{anthropic_anthropics_2023}
{Anthropic}. 2023.
\newblock Anthropic's {Responsible} {Scaling} {Policy}.
\newblock \textit{Anthropic}. \url{https://perma.cc/D9HQ-96HW}.
\newblock Accessed: 2023-09-30.

\bibitem[{{ATIH}(2024)}]{atih_responsible_2024}
{ATIH}. 2024.
\newblock Responsible {Tech} {Organizations}.
\newblock \url{https://perma.cc/C745-KD8F}.
\newblock Accessed: 2024-01-26.

\bibitem[{Attard-Frost and Widder(2023)}]{attard-frost_ethics_2023}
Attard-Frost, B.; and Widder, D.~G. 2023.
\newblock The {Ethics} of {AI} {Value} {Chains}.
\newblock arXiv:2307.16787.

\bibitem[{Bachrach and Baratz(1962)}]{bachrach_two_1962}
Bachrach, P.; and Baratz, M.~S. 1962.
\newblock Two {Faces} of {Power}.
\newblock \emph{The American Political Science Review}, 56(4): 947--952.

\bibitem[{Badran(2021)}]{badran_thoughts_2021}
Badran, A. 2021.
\newblock Thoughts and {Reflections} on the {Case} of {Qatar}: {Should} {Artificial} {Intelligence} {Be} {Regulated}?
\newblock In Azar, E.; and Haddad, A.~N., eds., \emph{Artificial {Intelligence} in the {Gulf}: {Challenges} and {Opportunities}}, 69--92. Singapore: Springer.
\newblock ISBN 9789811607714.

\bibitem[{Bajohr(2023)}]{bajohr_whoever_2023}
Bajohr, H. 2023.
\newblock Whoever {Controls} {Language} {Models} {Controls} {Politics}.
\newblock \textit{Neue Zürcher Zeitung}. \url{https://perma.cc/KQL4-B3K5}.

\bibitem[{Baker, Edwards, and Doidge(2012)}]{baker_how_2012}
Baker, S.; Edwards, R.; and Doidge, M. 2012.
\newblock How many qualitative interviews is enough?: expert voices and early career reflections on sampling and cases in qualitative research.
\newblock \textit{National Centre for Research Methods}. \url{https://perma.cc/5JFN-A8EL}.

\bibitem[{Bannerman et~al.(2020)Bannerman, Baade, Bivens, Regan~Shade, Shepherd, and Zeffiro}]{bannerman_platforms_2020}
Bannerman, S.; Baade, C.; Bivens, R.; Regan~Shade, L.; Shepherd, T.; and Zeffiro, A. 2020.
\newblock Platforms and {Power}: {A} {Panel} {Discussion}.
\newblock \emph{Canadian Journal of Communication}, 45(3): 473--490.

\bibitem[{Barabas(2023)}]{barabas_uninventing_2023}
Barabas, C.~M. 2023.
\newblock \emph{Uninventing {Carceral} {Technology}: {Four} {Experiments} in {Imagining} the {World} {More} {Rigorously}}.
\newblock Ph.{D}. diss., Massachusetts Institute of Technology.

\bibitem[{Barakat(2024)}]{barakat_interrogating_2024}
Barakat, H. 2024.
\newblock Interrogating {Mainstream} {Reporting} on {Artificial} {Intelligence}.
\newblock \textit{Tech Policy Press}. \url{https://techpolicy.press/interrogating-mainstream-reporting-on-artificial-intelligence}.
\newblock Accessed: 2024-08-12.

\bibitem[{Barkow(2010)}]{barkow_insulating_2010}
Barkow, R.~E. 2010.
\newblock Insulating {Agencies}: {Avoiding} {Capture} through {Institutional} {Design}.
\newblock \emph{Texas Law Review}, 89(1): 15--80.

\bibitem[{Barrett(2004)}]{barrett_bus_2004}
Barrett, S. 2004.
\newblock Bus {Competition} in {Ireland} — {The} {Case} for {Market} {Forces}.
\newblock \emph{Quarterly Economic Commentary}.

\bibitem[{Baumberger(2023)}]{baumberger_unveiling_2023}
Baumberger, J. 2023.
\newblock Unveiling {AI}’s {Existential} {Threats} and {Societal} {Responsibilities}.
\newblock \emph{Filozofia i Nauka}, 1(11): 65--80.

\bibitem[{Baxter(2012)}]{baxter_understanding_2012}
Baxter, L. 2012.
\newblock Understanding {Regulatory} {Capture}: {An} {Academic} {Perspective} from the {United} {States}.
\newblock In Pagliari, S., ed., \emph{The {Making} of {Good} {Financial} {Regulation}: {Towards} a {Policy} {Response} to {Regulatory} {Capture}}, 31--39. International Centre for Financial Regulation.
\newblock ISBN 978-1-78148-548-4.

\bibitem[{Baxter(2011)}]{baxter_capture_2011}
Baxter, L.~G. 2011.
\newblock Capture in {Financial} {Regulation}: {Can} {We} {Channel} {It} toward the {Common} {Good} {Essay}.
\newblock \emph{Cornell Journal of Law and Public Policy}, 21(1): 175--200.

\bibitem[{Bechara et~al.(2021)Bechara, Bossu, Liu, and Rossi}]{bechara_impact_2021}
Bechara, M.~M.; Bossu, W.; Liu, M.~Y.; and Rossi, A. 2021.
\newblock \emph{The {Impact} of {Fintech} on {Central} {Bank} {Governance}: {Key} {Legal} {Issues}}.
\newblock International Monetary Fund.
\newblock ISBN 978-1-5135-9247-3.

\bibitem[{Bedford et~al.(2022)Bedford, Mann, Foth, and Walters}]{bedford_post-capitalocentric_2022}
Bedford, L.; Mann, M.; Foth, M.; and Walters, R. 2022.
\newblock A {Post}-{Capitalocentric} {Critique} of {Digital} {Technology} and {Environmental} {Harm}: {New} {Directions} at the {Intersection} of {Digital} and {Green} {Criminology}.
\newblock \emph{International Journal for Crime, Justice and Social Democracy}, 11(1): 167--181.

\bibitem[{Bender and Grimsson~II(2024)}]{bender_power_2024}
Bender, E.~M.; and Grimsson~II, A. 2024.
\newblock Power {Shift}: {Toward} {Inclusive} {Natural} {Language} {Processing}.
\newblock In Hudley, A. H.~C., ed., \emph{Inclusion in {Linguistics}}, 199--221. Oxford University Press.
\newblock ISBN 978-0-19-775531-0.

\bibitem[{Bennett(2023)}]{bennett_transmuting_2023}
Bennett, S. 2023.
\newblock \emph{Transmuting {Values} in {Artificial} {Intelligence}:}.
\newblock Ph.{D}. diss., The University of Edinburgh.

\bibitem[{Bergen, Deutsch, and Berthelot(2023)}]{bergen_former_2023}
Bergen, M.; Deutsch, J.; and Berthelot, B. 2023.
\newblock Former {French} {Official} {Pushes} for {Looser} {AI} {Rules} {After} {Joining} {Startup}.
\newblock \textit{Bloomberg.com}. \url{https://perma.cc/8AUU-MZ23}.
\newblock Accessed: 2024-04-24.

\bibitem[{Berman(2017)}]{berman_industry_2017}
Berman, A. 2017.
\newblock Industry, {Regulatory} {Capture} and {Transnational} {Standard} {Setting}.
\newblock \emph{American Journal of International Law}, 111: 112--118.

\bibitem[{Berman, Goyal, and Madaio(2024)}]{berman_scoping_2024}
Berman, G.; Goyal, N.; and Madaio, M. 2024.
\newblock A {Scoping} {Study} of {Evaluation} {Practices} for {Responsible} {AI} {Tools}: {Steps} {Towards} {Effectiveness} {Evaluations}.
\newblock In \emph{Proceedings of the {CHI} {Conference} on {Human} {Factors} in {Computing} {Systems}}, {CHI} '24, 1--24. New York, NY, USA: Association for Computing Machinery.
\newblock ISBN 9798400703300.

\bibitem[{Bertuzzi(2023)}]{bertuzzi_eus_2023}
Bertuzzi, L. 2023.
\newblock {EU}’s {AI} {Act} negotiations hit the brakes over foundation models.
\newblock \textit{Euractiv}. \url{https://perma.cc/4KN6-UEV5}.
\newblock Accessed: 2023-12-16.

\bibitem[{Besiroglu et~al.(2024)Besiroglu, Bergerson, Michael, Heim, Luo, and Thompson}]{besiroglu_compute_2024}
Besiroglu, T.; Bergerson, S.~A.; Michael, A.; Heim, L.; Luo, X.; and Thompson, N. 2024.
\newblock The {Compute} {Divide} in {Machine} {Learning}: {A} {Threat} to {Academic} {Contribution} and {Scrutiny}?
\newblock arXiv:2401.02452.

\bibitem[{Beyers et~al.(2014)Beyers, Braun, Marshall, and De~Bruycker}]{beyers_lets_2014}
Beyers, J.; Braun, C.; Marshall, D.; and De~Bruycker, I. 2014.
\newblock Let’s talk! {On} the practice and method of interviewing policy experts.
\newblock \emph{Interest Groups \& Advocacy}, 3(2): 174--187.

\bibitem[{Bietti(2023)}]{bietti_genealogy_2023}
Bietti, E. 2023.
\newblock A {Genealogy} of {Digital} {Platform} {Regulation}.
\newblock \emph{Georgetown Law Technology Review}, 7: 1.

\bibitem[{Bode and Huelss(2023)}]{bode_constructing_2023}
Bode, I.; and Huelss, H. 2023.
\newblock Constructing expertise: the front- and back-door regulation of {AI}’s military applications in the {European} {Union}.
\newblock \emph{Journal of European Public Policy}, 30(7): 1230--1254.

\bibitem[{Boffel(2023)}]{boffel_influence_2023}
Boffel, L. 2023.
\newblock The {Influence} of {Artificial} {Intelligence} and {Emerging} {Technologies} on the {Regulation} of {Insurance} {Companies} in the {U}.{S}.: {An} {Exemplary} {Analysis} of {California}'s {Rate} {Making} {Law}.
\newblock \emph{Berkeley Business Law Journal}, 20(2): 254--315.

\bibitem[{Bordelon(2023{\natexlab{a}})}]{bordelon_as_2023}
Bordelon, B. 2023{\natexlab{a}}.
\newblock As states move on {AI}, tech lobbyists are swarming in.
\newblock \textit{Politico}. \url{https://perma.cc/BEE2-G9HJ}.
\newblock Accessed: 2024-01-29.

\bibitem[{Bordelon(2023{\natexlab{b}})}]{bordelon_key_2023}
Bordelon, B. 2023{\natexlab{b}}.
\newblock Key {Congress} staffers in {AI} debate are funded by tech giants like {Google} and {Microsoft}.
\newblock \textit{Politico}. \url{https://perma.cc/GR52-7P8G}.
\newblock Accessed: 2023-12-16.

\bibitem[{Bordelon(2024)}]{bordelon_dc_2024}
Bordelon, B. 2024.
\newblock In {DC}, a new wave of {AI} lobbyists gains the upper hand.
\newblock \textit{Politico}. \url{https://perma.cc/X9QY-SGRA}.
\newblock Accessed: 2024-05-12.

\bibitem[{Bova, Stefano, and Han(2024)}]{bova_both_2024}
Bova, P.; Stefano, A.~D.; and Han, T.~A. 2024.
\newblock Both eyes open: {Vigilant} {Incentives} help auditors improve {AI} safety.
\newblock \emph{Journal of Physics: Complexity}, 5(2): 025009.

\bibitem[{Brandusescu(2021)}]{brandusescu_artificial_2021}
Brandusescu, A. 2021.
\newblock Artificial {Intelligence} {Policy} and {Funding} in {Canada}: {Public} {Investments}, {Private} {Interests}.
\newblock SSRN:4089932.

\bibitem[{Brandusescu and Sieber(2023)}]{brandusescu_canadas_2023}
Brandusescu, A.; and Sieber, R. 2023.
\newblock Canada’s {Artificial} {Intelligence} and {Data} {Act}: a {Missed} {Opportunity} for {Shared} {Prosperity}.
\newblock SSRN:4602943.

\bibitem[{Bremmer and Suleyman(2023)}]{bremmer_ai_2023}
Bremmer, I.; and Suleyman, M. 2023.
\newblock The {AI} {Power} {Paradox}: {Can} {States} {Learn} to {Govern} {Artificial} {Intelligence} - before {It}'s {Too} {Late}?
\newblock \emph{Foreign Affairs}, 102: 26.

\bibitem[{Brennen, Howard, and Nielsen(2018)}]{brennen_industry-led_2018}
Brennen, J.~S.; Howard, P.~N.; and Nielsen, R.~K. 2018.
\newblock An {Industry}-{Led} {Debate}: {How} {UK} {Media} {Cover} {Artificial} {Intelligence}.
\newblock Technical report, Reuters Institute for the Study of Journalism.

\bibitem[{Brezis and Wiist(2011)}]{brezis_vulnerability_2011}
Brezis, M.; and Wiist, W.~H. 2011.
\newblock Vulnerability of health to market forces.
\newblock \emph{Medical Care}, 49(3): 232--239.

\bibitem[{Broughel(2023)}]{broughel_rules_2023}
Broughel, J. 2023.
\newblock Rules for {Robots}: {A} {Framework} for {Governance} of {AI}.
\newblock SSRN:4620277.

\bibitem[{Browne, Drage, and McInerney(2024)}]{browne_tech_2024}
Browne, J.; Drage, E.; and McInerney, K. 2024.
\newblock Tech workers’ perspectives on ethical issues in {AI} development: {Foregrounding} feminist approaches.
\newblock \emph{Big Data \& Society}, 11(1): 20539517231221780.

\bibitem[{Brownsword(2019)}]{brownsword_law_2019}
Brownsword, R. 2019.
\newblock Law {Disrupted}, {Law} {Re}-{Imagined}, {Law} {Re}-{Invented}.
\newblock \emph{Technology and Regulation}, 2019: 10--30.

\bibitem[{Brynjolfsson and Ng(2023)}]{brynjolfsson_big_2023}
Brynjolfsson, E.; and Ng, A. 2023.
\newblock Big {AI} {Can} {Centralize} {Decision}-{Making} and {Power}, {And} {That}'s a {Problem}.
\newblock In Prud’homme, B.; Régis, C.; and Farnadi, G., eds., \emph{Missing {Links} in {AI} {Governance}}, 65--88. Paris / Québec: UNESCO / MILA.
\newblock ISBN 978-92-3-100579-4.

\bibitem[{Bryson(2020)}]{bryson_artificial_2020}
Bryson, J.~J. 2020.
\newblock The {Artificial} {Intelligence} of the {Ethics} of {Artificial} {Intelligence}.
\newblock In Dubber, M.~D.; Pasquale, F.; and Das, S., eds., \emph{The {Oxford} handbook of ethics of {AI}}, Oxford handbooks series. New York, NY: Oxford University Press.
\newblock ISBN 978-0-19-006739-7.

\bibitem[{Bryson and Malikova(2021)}]{bryson_is_2021}
Bryson, J.~J.; and Malikova, H. 2021.
\newblock Is {There} an {AI} {Cold} {War}?
\newblock \emph{Global Perspectives}, 2(1): 24803.

\bibitem[{Caprio(2013)}]{caprio_regulatory_2013}
Caprio, G.~J. 2013.
\newblock Regulatory {Capture}: {Why} {It} {Occurs}, {How} to {Minimize} {It} ({Transcript}).
\newblock \emph{North Carolina Banking Institute}, 18(1): 39--50.

\bibitem[{Carlizzi and Quattrone(2023)}]{carlizzi_artificial_2023}
Carlizzi, D.~N.; and Quattrone, A. 2023.
\newblock Artificial {Intelligence} and {Data} {Governance} for {Precision} {ePolicy} {Cycle}.
\newblock In Marino, D.; and Monaca, M., eds., \emph{Artificial {Intelligence} and {Economics}: the {Key} to the {Future}}, Lecture {Notes} in {Networks} and {Systems}, 67--84. Cham: Springer International Publishing.
\newblock ISBN 978-3-031-14605-3.

\bibitem[{Carpenter(2013{\natexlab{a}})}]{carpenter_corrosive_2013}
Carpenter, D. 2013{\natexlab{a}}.
\newblock Corrosive {Capture}? {The} {Dueling} {Forces} of {Autonomy} and {Industry} {Influence} in {FDA} {Pharmaceutical} {Regulation}.
\newblock In Carpenter, D.; and Moss, D.~A., eds., \emph{Preventing {Regulatory} {Capture}: {Special} {Interest} {Influence} and {How} to {Limit} it}, 152--172. Cambridge: Cambridge University Press.
\newblock ISBN 978-1-107-03608-6.

\bibitem[{Carpenter(2013{\natexlab{b}})}]{carpenter_detecting_2013}
Carpenter, D. 2013{\natexlab{b}}.
\newblock Detecting and {Measuring} {Capture}.
\newblock In Carpenter, D.; and Moss, D.~A., eds., \emph{Preventing {Regulatory} {Capture}: {Special} {Interest} {Influence} and {How} to {Limit} it}, 57--68. Cambridge: Cambridge University Press.
\newblock ISBN 978-1-107-64670-4.

\bibitem[{Carrigan(2013)}]{carrigan_captured_2013}
Carrigan, C. 2013.
\newblock Captured by {Disaster}? {Reinterpreting} {Regulatory} {Behavior} in the {Shadow} of the {Gulf} {Oil} {Spill}.
\newblock In Carpenter, D.; and Moss, D.~A., eds., \emph{Preventing {Regulatory} {Capture}: {Special} {Interest} {Influence} and {How} to {Limit} it}, 239--291. Cambridge: Cambridge University Press.
\newblock ISBN 978-1-107-03608-6.

\bibitem[{Carter(2023)}]{carter_machine_2023}
Carter, R. 2023.
\newblock Machine {Visions}: {Mapping} {Depictions} of {Machine} {Vision} through {Critical} {Image} {Synthesis}.
\newblock \emph{Open Library of Humanities}, 9(2).

\bibitem[{Casper et~al.(2024)Casper, Ezell, Siegmann, Kolt, Curtis, Bucknall, Haupt, Wei, Scheurer, Hobbhahn, Sharkey, Krishna, Von~Hagen, Alberti, Chan, Sun, Gerovitch, Bau, Tegmark, Krueger, and Hadfield-Menell}]{casper_black-box_2024}
Casper, S.; Ezell, C.; Siegmann, C.; Kolt, N.; Curtis, T.~L.; Bucknall, B.; Haupt, A.; Wei, K.; Scheurer, J.; Hobbhahn, M.; Sharkey, L.; Krishna, S.; Von~Hagen, M.; Alberti, S.; Chan, A.; Sun, Q.; Gerovitch, M.; Bau, D.; Tegmark, M.; Krueger, D.; and Hadfield-Menell, D. 2024.
\newblock Black-{Box} {Access} is {Insufficient} for {Rigorous} {AI} {Audits}.
\newblock In \emph{Proceedings of the 2024 {ACM} {Conference} on {Fairness}, {Accountability}, and {Transparency}}.

\bibitem[{Cath and Keyes(2022)}]{cath_your_2022}
Cath, C.; and Keyes, O. 2022.
\newblock Your {Thoughts} for a {Penny}? {Capital}, {Complicity}, and {AI} {Ethics}.
\newblock In Phan, T.; Goldfein, J.; Kuch, D.; and Mann, M., eds., \emph{Economies of {Virtue}: {The} {Circulation} of ‘{Ethics}’ in {AI}}, number~46 in Theory on {Demand}, 24--39. Amsterdam: Institute of Network Cultures.
\newblock ISBN 978-94-92302-96-0.

\bibitem[{Cebulla(2023)}]{cebulla_future_2023}
Cebulla, A. 2023.
\newblock \emph{The {Future} of {Work} and {Technology}: {Global} {Trends}, {Challenges} and {Policies} with an {Australian} {Perspective}}.
\newblock CRC Press.
\newblock ISBN 978-1-00-382465-7.

\bibitem[{{CEO}(2022)}]{ceo_big_2022}
{CEO}. 2022.
\newblock Big {Tech}’s last minute attempt to tame {EU} tech rules: {Lobbying} in times of trilogues.
\newblock \url{https://perma.cc/4RMW-8K3U}.

\bibitem[{{CEO}(2023)}]{ceo_byte_2023}
{CEO}. 2023.
\newblock Byte by byte: {How} {Big} {Tech} undermined the {AI} {Act}.
\newblock \url{https://perma.cc/28NF-PSQY}.

\bibitem[{{CEO}(2024)}]{ceo_meetings_2024}
{CEO}. 2024.
\newblock Meetings {EU} officials on {AI} in 2023.
\newblock \url{https://perma.cc/C7R7-LBBW}.

\bibitem[{Chan, Bradley, and Rajkumar(2023)}]{chan_reclaiming_2023}
Chan, A.; Bradley, H.; and Rajkumar, N. 2023.
\newblock Reclaiming the {Digital} {Commons}: {A} {Public} {Data} {Trust} for {Training} {Data}.
\newblock In \emph{Proceedings of the 2023 {AAAI}/{ACM} {Conference} on {AI}, {Ethics}, and {Society}}, 855--868. Montr{\textbackslash}'\{e\}al QC Canada: ACM.
\newblock ISBN 9798400702310.

\bibitem[{Chan et~al.(2024)Chan, Ezell, Kaufmann, Wei, Hammond, Bradley, Bluemke, Rajkumar, Krueger, Kolt, Heim, and Anderljung}]{chan_visibility_2024}
Chan, A.; Ezell, C.; Kaufmann, M.; Wei, K.; Hammond, L.; Bradley, H.; Bluemke, E.; Rajkumar, N.; Krueger, D.; Kolt, N.; Heim, L.; and Anderljung, M. 2024.
\newblock Visibility into {AI} {Agents}.
\newblock arXiv:2401.13138.

\bibitem[{Chan, Papyshev, and Yarime(2022)}]{chan_balancing_2022}
Chan, K. J.~D.; Papyshev, G.; and Yarime, M. 2022.
\newblock Balancing the {Tradeoff} between {Regulation} and {Innovation} for {Artificial} {Intelligence}: {An} {Analysis} of {Top}-down {Command} and {Control} and {Bottom}-up {Self}-{Regulatory} {Approaches}.
\newblock SSRN:4223016.

\bibitem[{Charisi and Dignum(2024)}]{charisi_operationalizing_2024}
Charisi, V.; and Dignum, V. 2024.
\newblock Operationalizing {AI} {Regulatory} {Sandboxes} for {Children}'s {Rights} and {Well}-{Being}.
\newblock In \emph{Human-{Centered} {AI}}. Chapman and Hall/CRC.
\newblock ISBN 978-1-00-332079-1.

\bibitem[{Charlesworth et~al.(2023)Charlesworth, Fotheringham, Gavaghan, Sanchez-Graells, and Torrible}]{charlesworth_response_2023}
Charlesworth, A.; Fotheringham, K.; Gavaghan, C.; Sanchez-Graells, A.; and Torrible, C. 2023.
\newblock Response to the {UK}’s {March} 2023 {White} {Paper}.
\newblock SSRN:4477368.

\bibitem[{Charlesworth(2021)}]{charlesworth_regulating_2021}
Charlesworth, A.~J. 2021.
\newblock Regulating {Algorithmic} {Assemblages}: {Looking} {Beyond} {Corporatist} {AI} {Ethics}.
\newblock In Kohl, U.; and Eisler, J., eds., \emph{Data-{Driven} {Personalisation} in {Markets}, {Politics} and {Law}}, 243--262. Cambridge: Cambridge University Press.
\newblock ISBN 978-1-108-83569-5.

\bibitem[{Chauhan(2023)}]{chauhan_what_2023}
Chauhan, K. D.~S. 2023.
\newblock From ‘{What}’ and ‘{Why}’ to ‘{How}’: {An} {Imperative} {Driven} {Approach} to {Mechanics} of {AI} {Regulation}.
\newblock \emph{Global Jurist}, 23(2): 99--124.

\bibitem[{Cheng and Tanglis(2024)}]{cheng_artificial_2024}
Cheng, L.; and Tanglis, M. 2024.
\newblock Artificial {Intelligence} {Lobbyists} {Descend} on {Washington} {DC}.
\newblock Technical report, Public Citizen, Washington, D.C.

\bibitem[{Chesterman(2021{\natexlab{a}})}]{chesterman_we_2021}
Chesterman, S. 2021{\natexlab{a}}.
\newblock \emph{We, the {Robots}?}
\newblock Cambridge University Press.
\newblock ISBN 978-1-316-51768-0.

\bibitem[{Chesterman(2021{\natexlab{b}})}]{chesterman_weapons_2021}
Chesterman, S. 2021{\natexlab{b}}.
\newblock Weapons of mass disruption: artificial intelligence and international law.
\newblock \emph{Cambridge International Law Journal}, 10(2): 181--203.

\bibitem[{Chesterman(2023)}]{chesterman_tragedy_2023}
Chesterman, S. 2023.
\newblock The {Tragedy} of {AI} {Governance}.
\newblock SSRN:4600065.

\bibitem[{Chilson and Rinehart(2024)}]{chilson_public_2024}
Chilson, N.; and Rinehart, W. 2024.
\newblock Public {Interest} {Comment} on the {National} {Telecommunications} and {Information} {Administration} ({NTIA}) {AI} {Accountability} {Policy} {Request} for {Comment}.
\newblock \emph{The Center for Growth and Opportunity}.

\bibitem[{Chinen(2023)}]{chinen_international_2023}
Chinen, M. 2023.
\newblock \emph{The {International} {Governance} of {Artificial} {Intelligence}}.
\newblock Edward Elgar Publishing.
\newblock ISBN 978-1-80037-922-0.

\bibitem[{Chinh et~al.(2019)Chinh, Zade, Ganji, and Aragon}]{chinh_ways_2019}
Chinh, B.; Zade, H.; Ganji, A.; and Aragon, C. 2019.
\newblock Ways of {Qualitative} {Coding}: {A} {Case} {Study} of {Four} {Strategies} for {Resolving} {Disagreements}.
\newblock In \emph{Extended {Abstracts} of the 2019 {CHI} {Conference} on {Human} {Factors} in {Computing} {Systems}}, {CHI} {EA} '19, 1--6. New York, NY, USA: Association for Computing Machinery.
\newblock ISBN 978-1-4503-5971-9.

\bibitem[{Chomanski(2021)}]{chomanski_missing_2021}
Chomanski, B. 2021.
\newblock The {Missing} {Ingredient} in the {Case} for {Regulating} {Big} {Tech}.
\newblock \emph{Minds and Machines}, 31(2): 257--275.

\bibitem[{Cihon, Maas, and Kemp(2020{\natexlab{a}})}]{cihon_fragmentation_2020}
Cihon, P.; Maas, M.~M.; and Kemp, L. 2020{\natexlab{a}}.
\newblock Fragmentation and the {Future}: {Investigating} {Architectures} for {International} {AI} {Governance}.
\newblock \emph{Global Policy}, 11(5): 545--556.

\bibitem[{Cihon, Maas, and Kemp(2020{\natexlab{b}})}]{cihon_should_2020}
Cihon, P.; Maas, M.~M.; and Kemp, L. 2020{\natexlab{b}}.
\newblock Should {Artificial} {Intelligence} {Governance} be {Centralised}? {Design} {Lessons} from {History}.
\newblock In \emph{Proceedings of the {AAAI}/{ACM} {Conference} on {AI}, {Ethics}, and {Society}}, {AIES} '20, 228--234. New York, NY, USA: Association for Computing Machinery.
\newblock ISBN 978-1-4503-7110-0.

\bibitem[{Clarke and Whittlestone(2022)}]{clarke_survey_2022}
Clarke, S.; and Whittlestone, J. 2022.
\newblock A {Survey} of the {Potential} {Long}-term {Impacts} of {AI}.
\newblock In \emph{Proceedings of the 2022 {AAAI}/{ACM} {Conference} on {AI}, {Ethics}, and {Society}}, 192--202.

\bibitem[{Cohen and Jackson(2019)}]{cohen_rights_2019}
Cohen, A.~J.; and Jackson, J. 2019.
\newblock Rights as {Logistics}: {Notes} on the {Right} to {Food} and {Food} {Retail} {Liberalization} in {India}.
\newblock SSRN:3384351.

\bibitem[{Correa et~al.(2023)Correa, Danish, Ido, Mwangi, and Terán}]{correa_global_2023}
Correa, C.~M.; Danish; Ido, V. H.~P.; Mwangi, J.; and Terán, D.~U. 2023.
\newblock The {Global} {Digital} {Compact}: {Opportunities} and challenges for developing countries in a fragmented digital space.
\newblock Research {Report} 187, Research Paper.

\bibitem[{Couldry and Mejias(2019)}]{couldry_data_2019}
Couldry, N.; and Mejias, U.~A. 2019.
\newblock Data {Colonialism}: {Rethinking} {Big} {Data}’s {Relation} to the {Contemporary} {Subject}.
\newblock \emph{Television \& New Media}, 20(4): 336--349.

\bibitem[{Coulter et~al.(2023)Coulter, Mukherjee, Chee, Mukherjee, and Chee}]{coulter_eus_2023}
Coulter, M.; Mukherjee, S.; Chee, F.~Y.; Mukherjee, S.; and Chee, F.~Y. 2023.
\newblock {EU}'s {AI} {Act} could exclude open-source models from regulation.
\newblock \textit{Reuters}. \url{https://perma.cc/9U6T-HLSQ}.
\newblock Accessed: 2024-05-06.

\bibitem[{Critch and Russell(2023)}]{critch_tasra_2023}
Critch, A.; and Russell, S. 2023.
\newblock {TASRA}: a {Taxonomy} and {Analysis} of {Societal}-{Scale} {Risks} from {AI}.
\newblock arXiv:2306.06924.

\bibitem[{Cui et~al.(2024)Cui, Ho, Martin, and Joseph~O'Connell}]{cui_governing_2024}
Cui, I.; Ho, D.~E.; Martin, O.; and Joseph~O'Connell, A. 2024.
\newblock Governing by {Assignment}.
\newblock SSRN:4729432.

\bibitem[{Cuéllar(2013)}]{cuellar_coalitions_2013}
Cuéllar, M.-F. 2013.
\newblock Coalitions, {Autonomy}, and {Regulatory} {Bargains} in {Public} {Health} {Law}.
\newblock In Carpenter, D.; and Moss, D.~A., eds., \emph{Preventing {Regulatory} {Capture}: {Special} {Interest} {Influence} and {How} to {Limit} it}, 326--362. Cambridge: Cambridge University Press.
\newblock ISBN 978-1-107-64670-4.

\bibitem[{Cuéllar and Huq(2022)}]{cuellar_artificially_2022}
Cuéllar, M.-F.; and Huq, A.~Z. 2022.
\newblock Artificially {Intelligent} {Regulation}.
\newblock \emph{Daedalus}, 151(2): 335--347.

\bibitem[{Dal~Bó(2006)}]{dal_bo_regulatory_2006}
Dal~Bó, E. 2006.
\newblock Regulatory {Capture}: {A} {Review}.
\newblock \emph{Oxford Review of Economic Policy}, 22(2): 203--225.

\bibitem[{Dancy and Workman(2023)}]{dancy_integrating_2023}
Dancy, C.~L.; and Workman, D. 2023.
\newblock On {Integrating} {Generative} {Models} into {Cognitive} {Architectures} for {Improved} {Computational} {Sociocultural} {Representations}.
\newblock \emph{Proceedings of the AAAI Symposium Series}, 2(1): 256--261.

\bibitem[{Davis(2023)}]{davis_ai_2023}
Davis, W. 2023.
\newblock {AI} companies have all kinds of arguments against paying for copyrighted content.
\newblock \textit{The Verge}. \url{https://perma.cc/Y9YH-GPME}.
\newblock Accessed: 2023-12-16.

\bibitem[{De~Chiara and Schwarz(2021)}]{de_chiara_dynamic_2021}
De~Chiara, A.; and Schwarz, M.~A. 2021.
\newblock A {Dynamic} {Theory} of {Regulatory} {Capture}.
\newblock SSRN:3815456.

\bibitem[{de~Figueiredo and Raiha(2022)}]{de_figueiredo_economic_2022}
de~Figueiredo, J.~M.; and Raiha, D. 2022.
\newblock Economic influence activities and the strategic location of investment.
\newblock \emph{Business and Politics}, 24(3): 292--317.

\bibitem[{de~Figueiredo and Richter(2014)}]{de_figueiredo_advancing_2014}
de~Figueiredo, J.~M.; and Richter, B.~K. 2014.
\newblock Advancing the {Empirical} {Research} on {Lobbying}.
\newblock \emph{Annual Review of Political Science}, 17(1): 163--185.

\bibitem[{de~Laat(2021)}]{de_laat_companies_2021}
de~Laat, P.~B. 2021.
\newblock Companies {Committed} to {Responsible} {AI}: {From} {Principles} towards {Implementation} and {Regulation}?
\newblock \emph{Philosophy \& Technology}, 34(4): 1135--1193.

\bibitem[{Delangue(2023)}]{delangue_im_2023}
Delangue, C. 2023.
\newblock I’m in favor of more research on future catastrophic risks of {AI} but let’s make sure it doesn’t lead in the short-term to regulatory capture or blinds us from looking at current important challenges like biases, misinformation, lack of transparency, concentration of power, . . .
\newblock \textit{Twitter}. \url{https://perma.cc/U52M-Y4ZW}.

\bibitem[{Dempsey et~al.(2024)Dempsey, McBride, Haataja, and Bryson}]{dempsey_transnational_2024}
Dempsey, M.; McBride, K.; Haataja, M.; and Bryson, J.~J. 2024.
\newblock Transnational {Digital} {Governance} and {Its} {Impact} on {Artificial} {Intelligence}.
\newblock In Bullock, J.~B.; Chen, Y.-C.; Himmelreich, J.; Hudson, V.~M.; Korinek, A.; Young, M.~M.; and Zhang, B., eds., \emph{The {Oxford} {Handbook} of {AI} {Governance}}, 0. Oxford University Press.
\newblock ISBN 978-0-19-757932-9.

\bibitem[{Derczynski et~al.(2023)Derczynski, Kirk, Balachandran, Kumar, Tsvetkov, Leiser, and Mohammad}]{derczynski_assessing_2023}
Derczynski, L.; Kirk, H.~R.; Balachandran, V.; Kumar, S.; Tsvetkov, Y.; Leiser, M.~R.; and Mohammad, S. 2023.
\newblock Assessing {Language} {Model} {Deployment} with {Risk} {Cards}.
\newblock arXiv:2303.18190.

\bibitem[{{DHS}(2024)}]{dhs_over_2024}
{DHS}. 2024.
\newblock Over 20 {Technology} and {Critical} {Infrastructure} {Executives}, {Civil} {Rights} {Leaders}, {Academics}, and {Policymakers} {Join} {New} {DHS} {Artificial} {Intelligence} {Safety} and {Security} {Board} to {Advance} {AI}’s {Responsible} {Development} and {Deployment}.
\newblock \url{https://perma.cc/FDD4-Z4JU}.
\newblock Accessed: 2024-04-27.

\bibitem[{Dickens(2021)}]{dickens_right_2021}
Dickens, A.~D. 2021.
\newblock \emph{The right to health implications of data-driven health research partnerships}.
\newblock Ph.{D}. diss., University of Essex.

\bibitem[{Ebers(2022)}]{ebers_standardizing_2022}
Ebers, M. 2022.
\newblock Standardizing {AI}: {The} {Case} of the {European} {Commission}’s {Proposal} for an ‘{Artificial} {Intelligence} {Act}’.
\newblock In Poncibò, C.; DiMatteo, L.~A.; and Cannarsa, M., eds., \emph{The {Cambridge} {Handbook} of {Artificial} {Intelligence}: {Global} {Perspectives} on {Law} and {Ethics}}, Cambridge {Law} {Handbooks}, 321--344. Cambridge: Cambridge University Press.
\newblock ISBN 978-1-316-51280-7.

\bibitem[{Edwards(2022)}]{edwards_regulating_2022}
Edwards, L. 2022.
\newblock Regulating {AI} in {Europe}: four problems and four solutions.
\newblock Technical report, Ada Lovelace Institute.

\bibitem[{Egan and Heim(2023)}]{egan_oversight_2023}
Egan, J.; and Heim, L. 2023.
\newblock Oversight for {Frontier} {AI} through a {Know}-{Your}-{Customer} {Scheme} for {Compute} {Providers}.
\newblock arXiv:2310.13625.

\bibitem[{Egan and Milana(2023)}]{egan_action_2023}
Egan, J.; and Milana, D. 2023.
\newblock Action on {AI}: {Unpacking} the {Executive} {Order}’s {Security} {Implications} and the {Road} {Ahead}.
\newblock \url{https://perma.cc/CSE4-ZA32}.
\newblock Accessed: 2024-04-27.

\bibitem[{Eliot and Murakami~Wood(2022)}]{eliot_culling_2022}
Eliot, D.; and Murakami~Wood, D. 2022.
\newblock Culling the {FLoC}: {Market} forces, regulatory regimes and {Google}’s (mis)steps on the path away from targeted advertising 1.
\newblock \emph{Information Polity}, 27(2): 259--274.

\bibitem[{Erman and Furendal(2024)}]{erman_democratization_2024}
Erman, E.; and Furendal, M. 2024.
\newblock The democratization of global {AI} governance and the role of tech companies.
\newblock \emph{Nature Machine Intelligence}, 1--3.

\bibitem[{Etzioni(2009)}]{etzioni_capture_2009}
Etzioni, A. 2009.
\newblock The {Capture} {Theory} of {Regulations}—{Revisited}.
\newblock \emph{Society}, 46(4): 319--323.

\bibitem[{{European Parliament} and {Council of the European Union}(2024)}]{AIA}
{European Parliament}; and {Council of the European Union}. 2024.
\newblock Regulation ({EU}) 2024/1689 of the {European} {Parliament} and of the {Council} of 13 {June} 2024 laying down harmonised rules on artificial intelligence and amending {Regulations} ({EC}) {No} 300/2008, ({EU}) {No} 167/2013, ({EU}) {No} 168/2013, ({EU}) 2018/858, ({EU}) 2018/1139 and ({EU}) 2019/2144 and {Directives} 2014/90/{EU}, ({EU}) 2016/797 and ({EU}) 2020/1828 ({Artificial} {Intelligence} {Act}){Text} with {EEA} relevance.
\newblock \textit{Official Journal of the European Union}. \url{https://perma.cc/Z5UR-MUY4}.

\bibitem[{Evans et~al.(2021)Evans, Cotton-Barratt, Finnveden, Bales, Balwit, Wills, Righetti, and Saunders}]{evans_truthful_2021}
Evans, O.; Cotton-Barratt, O.; Finnveden, L.; Bales, A.; Balwit, A.; Wills, P.; Righetti, L.; and Saunders, W. 2021.
\newblock Truthful {AI}: {Developing} and governing {AI} that does not lie.
\newblock arXiv:2110.06674.

\bibitem[{Evers-Hillstrom(2023)}]{evers-hillstrom_lobbying_2023}
Evers-Hillstrom, K. 2023.
\newblock Lobbying {World}: {Top} {Schumer} aide heads to {Microsoft}.
\newblock \textit{The Hill}. \url{https://perma.cc/T964-7GJN}.
\newblock Accessed: 2024-04-24.

\bibitem[{Fagleman, Griffiths, and Mcateer(2023)}]{fagleman_ai_2023}
Fagleman, D.; Griffiths, J.; and Mcateer, M. 2023.
\newblock {AI} in {Financial} {Services}: {Avoiding} the {Big} {Risks}.
\newblock Technical report, Finnance Innovation Lab and Financial Inclusion Centre.

\bibitem[{Fahey(2022{\natexlab{a}})}]{fahey_eu_2022}
Fahey, E. 2022{\natexlab{a}}.
\newblock \emph{The {EU} as a {Global} {Digital} {Actor}}.
\newblock Bloomsbury Publishing.
\newblock ISBN 978-1-5099-5704-0.

\bibitem[{Fahey(2022{\natexlab{b}})}]{fahey_introduction_2022}
Fahey, E. 2022{\natexlab{b}}.
\newblock Introduction: {The} {Framework} of {Data} {Institutionalisation}.
\newblock In \emph{The {EU} as a {Global} {Digital} {Actor}: {Institutionalising} {Global} {Data} {Protection}, {Trade}, and {Cybersecurity}}. Hart Publishing.
\newblock ISBN 978-1-5099-5704-0 978-1-5099-5705-7 978-1-5099-5706-4 978-1-5099-5707-1.

\bibitem[{Fenwick and Vermeulen(2020{\natexlab{a}})}]{fenwick_banking_2020}
Fenwick, M.; and Vermeulen, E.~P. 2020{\natexlab{a}}.
\newblock Banking and regulatory responses to {FinTech} revisited- building the sustainable financial service 'ecosystems' of tomorrow.
\newblock \emph{Singapore Journal of Legal Studies}, 165--189.

\bibitem[{Fenwick and Vermeulen(2020{\natexlab{b}})}]{fenwick_fintech_2020}
Fenwick, M.; and Vermeulen, E. P.~M. 2020{\natexlab{b}}.
\newblock Fintech, {Overcoming} {Friction} and {New} {Models} of {Financial} {Regulation}.
\newblock In Fenwick, M.; Van~Uytsel, S.; and Ying, B., eds., \emph{Regulating {FinTech} in {Asia}: {Global} {Context}, {Local} {Perspectives}}, Perspectives in {Law}, {Business} and {Innovation}, 205--225. Singapore: Springer.
\newblock ISBN 9789811558191.

\bibitem[{Fenwick and Vermeulen(2021)}]{fenwick_future_2021}
Fenwick, M.; and Vermeulen, E. P.~M. 2021.
\newblock The {Future} of {Finance}.
\newblock In Liaw, K.~T., ed., \emph{The {Routledge} {Handbook} of {FinTech}}. Routledge.
\newblock ISBN 978-1-00-037570-1.

\bibitem[{Field(2024)}]{field_ai_2024}
Field, H. 2024.
\newblock {AI} lobbying spikes 185\% as calls for regulation surge.
\newblock \textit{CNBC}. \url{https://perma.cc/PPP9-H5YK}.
\newblock Accessed: 2024-02-09.

\bibitem[{Findlay et~al.(2022)Findlay, Ford, Seah, and Thampapillai}]{findlay_regulatory_2022}
Findlay, M.; Ford, J.; Seah, J.; and Thampapillai, D. 2022.
\newblock \emph{Regulatory {Insights} on {Artificial} {Intelligence}: {Research} for {Policy}}.
\newblock Edward Elgar Publishing.
\newblock ISBN 978-1-80088-078-8.

\bibitem[{Findlay and Seah(2020)}]{findlay_data_2020}
Findlay, M.; and Seah, J. 2020.
\newblock Data {Imperialism}: {Disrupting} {Secondary} {Data} in {Platform} {Economies} {Through} {Participatory} {Regulation}.
\newblock SSRN:3613562.

\bibitem[{Findlay, Seah, and Wong(2023)}]{findlay_ai_2023}
Findlay, M.; Seah, J.; and Wong, W. 2023.
\newblock \emph{{AI} and {Big} {Data}: {Disruptive} {Regulation}}.
\newblock Edward Elgar Publishing.
\newblock ISBN 978-1-80220-952-5.

\bibitem[{Ford and Clifford(2021)}]{ford_embracing_2021}
Ford, J.; and Clifford, D. 2021.
\newblock Embracing {Difference}: {Governance} of {Critical} {Technologies} in the {Indo}-{Pacific}.
\newblock Technical report, Australian National University.

\bibitem[{Fraser and Bello~y Villarino(2023)}]{fraser_acceptable_2023}
Fraser, H.; and Bello~y Villarino, J.-M. 2023.
\newblock Acceptable {Risks} in {Europe}’s {Proposed} {AI} {Act}: {Reasonableness} and {Other} {Principles} for {Deciding} {How} {Much} {Risk} {Management} {Is} {Enough}.
\newblock \emph{European Journal of Risk Regulation}, 1--16.

\bibitem[{Frazier(2023)}]{frazier_administrative_2023}
Frazier, K. 2023.
\newblock Administrative {X}-{Risk}: {Pinpointing} the {Flaws} of an {AI} {Regulatory} {Scheme} {Reliant} on {Administrative} {Action}.
\newblock SSRN:4582523.

\bibitem[{Friedman et~al.(2022)Friedman, Heydari, Isaacs, and Kinsey}]{friedman_policing_2022}
Friedman, B.; Heydari, F.; Isaacs, M.; and Kinsey, K. 2022.
\newblock Policing {Police} {Tech}: {A} {Soft} {Law} {Solution}.
\newblock \emph{Berkeley Technology Law Journal}, 37: 701.

\bibitem[{Fusch and Ness(2015)}]{fusch_are_2015}
Fusch, P.; and Ness, L. 2015.
\newblock Are {We} {There} {Yet}? {Data} {Saturation} in {Qualitative} {Research}.
\newblock \emph{The Qualitative Report}, 20(9): 1408--1416.

\bibitem[{Gallifant et~al.(2024)Gallifant, Fiske, Strekalova, Osorio-Valencia, Parke, Mwavu, Martinez, Gichoya, Ghassemi, Demner-Fushman, McCoy, Celi, and Pierce}]{gallifant_peer_2024}
Gallifant, J.; Fiske, A.; Strekalova, Y. A.~L.; Osorio-Valencia, J.~S.; Parke, R.; Mwavu, R.; Martinez, N.; Gichoya, J.~W.; Ghassemi, M.; Demner-Fushman, D.; McCoy, L.~G.; Celi, L.~A.; and Pierce, R. 2024.
\newblock Peer review of {GPT}-4 technical report and systems card.
\newblock \emph{PLOS Digital Health}, 3(1): e0000417.

\bibitem[{Gans(2024)}]{gans_how_2024}
Gans, J.~S. 2024.
\newblock How {Learning} {About} {Harms} {Impacts} the {Optimal} {Rate} of {Artificial} {Intelligence} {Adoption}.
\newblock \url{https://perma.cc/4NDW-N7GP}.

\bibitem[{Gantzias(2021)}]{gantzias_dynamics_2021}
Gantzias, G. 2021.
\newblock Dynamics of {Public} {Interest} in {Artificial} {Intelligence}: ‘{Business} {Intelligence} {Culture}’ and {Global} {Regulation} in the {Digital} {Era}.
\newblock In Park, S.~H.; Gonzalez-Perez, M.~A.; and Floriani, D.~E., eds., \emph{The {Palgrave} {Handbook} of {Corporate} {Sustainability} in the {Digital} {Era}}, 259--281. Cham: Springer International Publishing.
\newblock ISBN 978-3-030-42412-1.

\bibitem[{Garvey(2017)}]{garvey_brief_2017}
Garvey, T. 2017.
\newblock A {Brief} {Overview} of {Rulemaking} and {Judicial} {Review}.
\newblock Technical Report R41546, Congressional Research Service.

\bibitem[{Gaske(2023{\natexlab{a}})}]{gaske_operational_2023}
Gaske, M. 2023{\natexlab{a}}.
\newblock The {Operational} {Paradox} of {Centralized} {Artificial} {Intelligence} {Regulation}.
\newblock SSRN:4524342.

\bibitem[{Gaske(2023{\natexlab{b}})}]{gaske_regulation_2023}
Gaske, M. 2023{\natexlab{b}}.
\newblock Regulation {Priorities} for {Artificial} {Intelligence} {Foundation} {Models}.
\newblock SSRN:4417565.

\bibitem[{Gastfriend(2024)}]{gastfriend_leading_2024}
Gastfriend, E. 2024.
\newblock Leading {Technology} {Advocacy} {Organizations} {Urge} {Congress} to {Support} {NIST} {Funding} {Request} for {Responsible} {AI} {Innovation}.
\newblock \url{https://perma.cc/72A6-MPZJ}.

\bibitem[{Gazendam and Dawson(2023)}]{gazendam_mind_2023}
Gazendam, I.; and Dawson, P. 2023.
\newblock Mind the {Gap}: {The} {Challenges} of {Assurance} for {Artificial} {Intelligence}.
\newblock Technical report, Stanford Center for International Security and Cooperation.

\bibitem[{Gegenhuber et~al.(2022)Gegenhuber, Logue, Hinings, and Barrett}]{gegenhuber_institutional_2022}
Gegenhuber, T.; Logue, D.; Hinings, C.~B.; and Barrett, M. 2022.
\newblock Institutional {Perspectives} on {Digital} {Transformation}.
\newblock In Gegenhuber, T.; Logue, D.; (Bob)~Hinings, C.; and Barrett, M., eds., \emph{Digital {Transformation} and {Institutional} {Theory}}, volume~83 of \emph{Research in the {Sociology} of {Organizations}}, 1--32. Emerald Publishing Limited.
\newblock ISBN 978-1-80262-222-5 978-1-80262-221-8.

\bibitem[{Geiger et~al.(2023)Geiger, Tandon, Gakhokidze, Song, and Irani}]{geiger_making_2023}
Geiger, R.~S.; Tandon, U.; Gakhokidze, A.; Song, L.; and Irani, L. 2023.
\newblock Making {Algorithms} {Public}: {Reimagining} {Auditing} {From} {Matters} of {Fact} to {Matters} of {Concern}.
\newblock \emph{International Journal of Communication}, 18(0): 22.

\bibitem[{Gilbert(2021)}]{gilbert_modes_2021}
Gilbert, T. 2021.
\newblock \emph{Modes of {Deliberation} in {Machine} {Ethics}}.
\newblock Ph.{D}. diss., UC Berkeley.

\bibitem[{Gilbert et~al.(2022)Gilbert, Dean, Zick, and Lambert}]{gilbert_choices_2022}
Gilbert, T.~K.; Dean, S.; Zick, T.; and Lambert, N. 2022.
\newblock Choices, {Risks}, and {Reward} {Reports}: {Charting} {Public} {Policy} for {Reinforcement} {Learning} {Systems}.
\newblock arXiv:2202.05716.

\bibitem[{Giorno(2024)}]{giorno_bottom_2024}
Giorno, T. 2024.
\newblock Bottom {Line}: {Former} {Republican} chief signs on to guard nutrition programs.
\newblock \textit{The Hill}. \url{https://perma.cc/VBC5-67F5}.
\newblock Accessed: 2024-04-24.

\bibitem[{Giraudo, Fosch-Villaronga, and Malgieri(2023)}]{giraudo_competing_2023}
Giraudo, M.; Fosch-Villaronga, E.; and Malgieri, G. 2023.
\newblock Competing {Legal} {Futures}.
\newblock SSRN:4499785.

\bibitem[{Goanta et~al.(2023)Goanta, Aletras, Chalkidis, Ranchordás, and Spanakis}]{goanta_regulation_2023}
Goanta, C.; Aletras, N.; Chalkidis, I.; Ranchordás, S.; and Spanakis, G. 2023.
\newblock Regulation and {NLP} ({RegNLP}): {Taming} {Large} {Language} {Models}.
\newblock In Bouamor, H.; Pino, J.; and Bali, K., eds., \emph{Proceedings of the 2023 {Conference} on {Empirical} {Methods} in {Natural} {Language} {Processing}}, 8712--8724. Singapore: Association for Computational Linguistics.

\bibitem[{Godwin, Ainsworth, and Godwin(2013)}]{godwin_lobbying_2013}
Godwin, K.; Ainsworth, S.; and Godwin, E. 2013.
\newblock \emph{Lobbying and {Policymaking}: {The} {Public} {Pursuit} of {Private} {Interests}}.
\newblock Washington, DC: CQ Press.

\bibitem[{Goodlad(2023)}]{goodlad_editors_2023}
Goodlad, L. M.~E. 2023.
\newblock Editor's {Introduction}: {Humanities} in the {Loop}.
\newblock \emph{Critical AI}, 1(1-2).

\bibitem[{Goodman(2024)}]{goodman_ai_2024}
Goodman, E.~P. 2024.
\newblock {AI} {Accountability} {Policy} {Report}.
\newblock Technical report, U.S. National Telecommunications and Information Administration.

\bibitem[{Goodman, Gerstel, and Risberg(2019)}]{goodman_beyond_2019}
Goodman, M.~P.; Gerstel, D.; and Risberg, P. 2019.
\newblock Beyond the {Brink}: {Escalation} and {Conflict} in {U}.{S}.-{China} {Economic} {Relations}.
\newblock Technical report, Center for Strategic and International Studies (CSIS).

\bibitem[{Gornet(2023)}]{gornet_european_2023}
Gornet, M. 2023.
\newblock The {European} approach to regulating {AI} through technical standards.
\newblock \url{https://perma.cc/D6BT-TVE2}.

\bibitem[{Gottardo(2023)}]{gottardo_algorithmic_2023}
Gottardo, R. 2023.
\newblock Algorithmic {Decision}-{Making} and {Public} {Sector} {Accountability} in {Africa} –: {New} {Challenges} for {Law} and {Policy}.
\newblock In \emph{Comparative {Legal} {Metrics}}, 139--179. Brill Nijhoff.
\newblock ISBN 978-90-04-68094-4.

\bibitem[{Greenleaf, Clarke, and Lindsay(2019)}]{greenleaf_does_2019}
Greenleaf, G.; Clarke, R.; and Lindsay, D.~F. 2019.
\newblock Does {AI} {Need} {Governance}? – {The} {Potential} {Roles} of a ‘{Responsible} {Innovation} {Organisation}’ in {Australia} ({Submission} to the {Australian} {Human} {Rights} {Commissioner} on the {White} {Paper} {Artificial} {Intelligence}: {Governance} and {Leadership}).
\newblock SSRN:3346149.

\bibitem[{Guest, Bunce, and Johnson(2006)}]{guest_how_2006}
Guest, G.; Bunce, A.; and Johnson, L. 2006.
\newblock How {Many} {Interviews} {Are} {Enough}?: {An} {Experiment} with {Data} {Saturation} and {Variability}.
\newblock \emph{Field Methods}, 18(1): 59--82.

\bibitem[{Guha et~al.(2024)Guha, Lawrence, Gailmard, Rodolfa, Surani, Bommasani, Raji, Cuéllar, Honigsberg, Liang, and Ho}]{guha_ai_2024}
Guha, N.; Lawrence, C.; Gailmard, L.~A.; Rodolfa, K.; Surani, F.; Bommasani, R.; Raji, I.; Cuéllar, M.-F.; Honigsberg, C.; Liang, P.; and Ho, D.~E. 2024.
\newblock {AI} {Regulation} {Has} {Its} {Own} {Alignment} {Problem}: {The} {Technical} and {Institutional} {Feasibility} of {Disclosure}, {Registration}, {Licensing}, and {Auditing}.
\newblock \emph{George Washington Law Review}, 92.

\bibitem[{Guihot, Matthew, and Suzor(2017)}]{guihot_nudging_2017}
Guihot, M.; Matthew, A.~F.; and Suzor, N.~P. 2017.
\newblock Nudging {Robots}: {Innovative} {Solutions} to {Regulate} {Artificial} {Intelligence}.
\newblock \emph{Vanderbilt Journal of Entertainment \& Technology Law}, 20: 385.

\bibitem[{Gurumurthy and Bharthur(2019)}]{gurumurthy_taking_2019}
Gurumurthy, A.; and Bharthur, D. 2019.
\newblock Taking {Stock} of {AI} in {Indian} {Agriculture}.
\newblock Technical report, IT for Change.

\bibitem[{Gutierrez et~al.(2023)Gutierrez, Aguirre, Uuk, Boine, and Franklin}]{gutierrez_proposal_2023}
Gutierrez, C.~I.; Aguirre, A.; Uuk, R.; Boine, C.~C.; and Franklin, M. 2023.
\newblock A {Proposal} for a {Definition} of {General} {Purpose} {Artificial} {Intelligence} {Systems}.
\newblock \emph{Digital Society}, 2(3): 36.

\bibitem[{Haataja and Bryson(2022)}]{haataja_reflections_2022}
Haataja, M.; and Bryson, J. 2022.
\newblock Reflections on the {EU}'s {AI} {Act} and {How} {We} {Could} {Make} {It} {Even} {Better}.
\newblock \emph{TechREG Chronicle}, March 2022.

\bibitem[{Hacker(2023)}]{hacker_sustainable_2023}
Hacker, P. 2023.
\newblock Sustainable {AI} {Regulation}.
\newblock arXiv:2306.00292.

\bibitem[{Hacohen(2022)}]{hacohen_policy_2022}
Hacohen, U.~Y. 2022.
\newblock Policy {Implications} of {User}-{Generated} {Data} {Network} {Effects}.
\newblock \emph{Fordham Intellectual Property, Media \& Entertainment Law Journal}, 33: 340.

\bibitem[{Hadfield and Clark(2023)}]{hadfield_regulatory_2023}
Hadfield, G.~K.; and Clark, J. 2023.
\newblock Regulatory {Markets}: {The} {Future} of {AI} {Governance}.
\newblock arXiv:2304.04914.

\bibitem[{Hakenes and Schnabel(2014)}]{hakenes_regulatory_2014}
Hakenes, H.; and Schnabel, I. 2014.
\newblock Regulatory {Capture} by {Sophistication}.
\newblock SSRN:2501573.

\bibitem[{Hawking(2021)}]{hawking_rule_2021}
Hawking, C. 2021.
\newblock \emph{Rule of the algorithm: {Exploring} an ill-posed problem for democracy}.
\newblock Master's diss., Charles University.

\bibitem[{Heims and Moxon(2023)}]{heims_mechanisms_2023}
Heims, E.; and Moxon, S. 2023.
\newblock Mechanisms of regulatory capture: {Testing} claims of industry influence in the case of {Vioxx}.
\newblock \emph{Regulation \& Governance}, n/a(n/a).

\bibitem[{Hempling(2014)}]{hempling_regulatory_2014}
Hempling, S. 2014.
\newblock “{Regulatory} {Capture}”: {Sources} and {Solutions}.
\newblock \emph{Emory Corporate Governance and Accountability Review}, 1(1): 23.

\bibitem[{Hendrycks, Mazeika, and Woodside(2023)}]{hendrycks_overview_2023}
Hendrycks, D.; Mazeika, M.; and Woodside, T. 2023.
\newblock An {Overview} of {Catastrophic} {AI} {Risks}.
\newblock arXiv:2306.12001.

\bibitem[{Hennink and Kaiser(2022)}]{hennink_sample_2022}
Hennink, M.; and Kaiser, B.~N. 2022.
\newblock Sample sizes for saturation in qualitative research: {A} systematic review of empirical tests.
\newblock \emph{Social Science \& Medicine}, 292: 114523.

\bibitem[{Henshaw(2024)}]{henshaw_theres_2024}
Henshaw, W. 2024.
\newblock There’s an {AI} {Lobbying} {Frenzy} in {Washington}. {Big} {Tech} {Is} {Dominating}.
\newblock \textit{Time}. \url{https://perma.cc/DL64-V4P7}.
\newblock Accessed: 2024-05-01.

\bibitem[{Hermstrüwer and Langenbach(2023)}]{hermstruwer_fair_2023}
Hermstrüwer, Y.; and Langenbach, P. 2023.
\newblock Fair governance with humans and machines.
\newblock \emph{Psychology, Public Policy, and Law}, 29(4): 525--548.

\bibitem[{Herrman(2023)}]{herrman_how_2023}
Herrman, J. 2023.
\newblock How {Big} {Tech} {Companies} {Really} {Think} {About} {AI}.
\newblock \textit{New York Magazine}. \url{https://perma.cc/27RF-QTUQ}.
\newblock Accessed: 2023-12-16.

\bibitem[{Hicock(n.d.)}]{hicock_ai_nodate}
Hicock, M. n.d.
\newblock {AI} {Organizations}.
\newblock \url{https://perma.cc/8A2E-H6M2}.
\newblock Accessed: 2024-01-26.

\bibitem[{Hildén(2019)}]{hilden_politics_2019}
Hildén, J. 2019.
\newblock \emph{The {Politics} of {Datafication} : {The} influence of lobbyists on the {EU}’s data protection reform and its consequences for the legitimacy of the {General} {Data} {Protection} {Regulation}}.
\newblock Ph.{D}. diss., University of Helsinki.

\bibitem[{Hilty, Hoffmann, and Scheuerer(2020)}]{hilty_intellectual_2020}
Hilty, R.; Hoffmann, J.; and Scheuerer, S. 2020.
\newblock Intellectual {Property} {Justification} for {Artificial} {Intelligence}.
\newblock SSRN:3539406.

\bibitem[{Himmelreich(2023)}]{himmelreich_against_2023}
Himmelreich, J. 2023.
\newblock Against “{Democratizing} {AI}”.
\newblock \emph{AI \& Society}, 38(4): 1333--1346.

\bibitem[{Himmelreich and Lim(2022)}]{himmelreich_artificial_2022}
Himmelreich, J.; and Lim, D. 2022.
\newblock Artificial {Intelligence} and {Structural} {Injustice}: {Foundations} for {Equity}, {Values}, and {Responsibility}.
\newblock arXiv:2205.02389.

\bibitem[{Hirsch and Shotts(2018)}]{hirsch_policy-development_2018}
Hirsch, A.~V.; and Shotts, K.~W. 2018.
\newblock Policy-{Development} {Monopolies}: {Adverse} {Consequences} and {Institutional} {Responses}.
\newblock \emph{The Journal of Politics}, 80(4): 1339--1354.

\bibitem[{Ho, Marcus, and Ray(2021)}]{ho_quality_2021}
Ho, D.~E.; Marcus, D.; and Ray, G.~K. 2021.
\newblock Quality {Assurance} {Systems} in {Agency} {Adjudication}: {Emerging} {Practices} and {Insights}.
\newblock Technical report, Administrative Conference of the United States.

\bibitem[{Horowitz(2023)}]{horowitz_politics_2023}
Horowitz, B. 2023.
\newblock Politics and the {Future}.
\newblock \textit{Andreessen Horowitz}. \url{https://perma.cc/8J2N-3MR3}.

\bibitem[{Hu(2021)}]{hu_tech_2021}
Hu, L. 2021.
\newblock Tech {Ethics}: {Speaking} {Ethics} to {Power}, or {Power} {Speaking} {Ethics}?
\newblock \emph{Journal of Social Computing}, 2(3): 238--248.

\bibitem[{Hua and Belfield(2023)}]{hua_effective_2023}
Hua, S.-S.; and Belfield, H. 2023.
\newblock Effective {Enforceability} of {EU} {Competition} {Law} {Under} {AI} {Development} {Scenarios}: a {Framework} for {Anticipatory} {Governance}.
\newblock In \emph{Proceedings of the 2023 {AAAI}/{ACM} {Conference} on {AI}, {Ethics}, and {Society}}, {AIES} '23, 596--605. New York, NY, USA: Association for Computing Machinery.
\newblock ISBN 9798400702310.

\bibitem[{Huang and Ma(2023)}]{huang_legal_2023}
Huang, K.; and Ma, W. 2023.
\newblock Legal and {Ethics} {Responsibility} of {ChatGPT}.
\newblock In Huang, K.; Wang, Y.; Zhu, F.; Chen, X.; and Xing, C., eds., \emph{Beyond {AI}: {ChatGPT}, {Web3}, and the {Business} {Landscape} of {Tomorrow}}, Future of {Business} and {Finance}, 329--353. Cham: Springer Nature Switzerland.
\newblock ISBN 978-3-031-45282-6.

\bibitem[{{IAPP}(2024)}]{iapp_global_2024}
{IAPP}. 2024.
\newblock Global {AI} {Law} and {Policy} {Tracker}.
\newblock \textit{International Association of Privacy Professionals (IAPP)}. \url{https://perma.cc/FZ55-GUJ5}.
\newblock Accessed: 2024-04-26.

\bibitem[{Iliadis and Ford(2023)}]{iliadis_fast_2023}
Iliadis, A.; and Ford, H. 2023.
\newblock Fast {Facts}: {Platforms} {From} {Personalization} to {Centralization}.
\newblock \emph{Social Media + Society}, 9(3): 20563051231195546.

\bibitem[{Ilie and Welch(2014)}]{ilie_online_2014}
Ilie, A.; and Welch, G. 2014.
\newblock Online control of active camera networks for computer vision tasks.
\newblock \emph{ACM Transactions on Sensor Networks}, 10(2): 25:1--25:40.

\bibitem[{Jiang et~al.(2023)Jiang, Brown, Cheng, Khan, Gupta, Workman, Hanna, Flowers, and Gebru}]{jiang_ai_2023}
Jiang, H.~H.; Brown, L.; Cheng, J.; Khan, M.; Gupta, A.; Workman, D.; Hanna, A.; Flowers, J.; and Gebru, T. 2023.
\newblock {AI} {Art} and its {Impact} on {Artists}.
\newblock In \emph{Proceedings of the 2023 {AAAI}/{ACM} {Conference} on {AI}, {Ethics}, and {Society}}, {AIES} '23, 363--374. New York, NY, USA: Association for Computing Machinery.
\newblock ISBN 9798400702310.

\bibitem[{Jing, Berger, and Becerra~Sandoval(2023)}]{jing_towards_2023}
Jing, F.~S.; Berger, S.~E.; and Becerra~Sandoval, J.~C. 2023.
\newblock Towards {Labor} {Transparency} in {Situated} {Computational} {Systems} {Impact} {Research}.
\newblock In \emph{2023 {ACM} {Conference} on {Fairness}, {Accountability}, and {Transparency}}, 1026--1037. Chicago IL USA: ACM.
\newblock ISBN 9798400701924.

\bibitem[{Johnson(2020)}]{johnson_access_2020}
Johnson, K. 2020.
\newblock Access {Now} resigns from {Partnership} on {AI} due to lack of change among tech companies.
\newblock \textit{VentureBeat}. \url{https://perma.cc/4BGU-HTK3}.
\newblock Accessed: 2024-04-27.

\bibitem[{Kaminski(2022)}]{kaminski_regulating_2022}
Kaminski, M.~E. 2022.
\newblock Regulating the {Risks} of {AI}.
\newblock SSRN:4195066.

\bibitem[{Kaplan(2008)}]{kaplan_framing_2008}
Kaplan, S. 2008.
\newblock Framing {Contests}: {Strategy} {Making} {Under} {Uncertainty}.
\newblock \emph{Organization Science}, 19(5): 729--752.

\bibitem[{Katyal(2022)}]{katyal_democracy_2022}
Katyal, S.~K. 2022.
\newblock Democracy \& {Distrust} in an {Era} of {Artificial} {Intelligence}.
\newblock \emph{Daedalus}, 151(2): 322--334.

\bibitem[{Keller and Magalhães(2023)}]{keller_regulating_2023}
Keller, C.~I.; and Magalhães, J.~C. 2023.
\newblock Regulating {AI} in democratic erosion: context, imaginaries and voices in the {Brazilian} debate.
\newblock In \emph{Elgar {Companion} to {Regulating} {AI} and {Big} {Data} in {Emerging} {Economies}}, 183--200. Edward Elgar Publishing.
\newblock ISBN 978-1-78536-240-8.

\bibitem[{Khan(2023)}]{khan_regulating_2023}
Khan, F. 2023.
\newblock Regulating the {Revolution}: {A} {Legal} {Roadmap} to {Optimizing} {AI} in {Healthcare}.
\newblock SSRN:4562473.

\bibitem[{Khanal, Zhang, and Taeihagh(2024)}]{khanal_why_2024}
Khanal, S.; Zhang, H.; and Taeihagh, A. 2024.
\newblock Why and how is the power of {Big} {Tech} increasing in the policy process? {The} case of generative {AI}.
\newblock \emph{Policy and Society}, puae012.

\bibitem[{Killian(2021)}]{killian_taming_2021}
Killian, L.~J. 2021.
\newblock Taming the {Dark} {Side} of the {New} {Globalization}.
\newblock In Park, S.~H.; Gonzalez-Perez, M.~A.; and Floriani, D.~E., eds., \emph{The {Palgrave} {Handbook} of {Corporate} {Sustainability} in the {Digital} {Era}}, 355--376. Cham: Springer International Publishing.
\newblock ISBN 978-3-030-42412-1.

\bibitem[{Killion(2023)}]{killion_free_2023}
Killion, V.~L. 2023.
\newblock Free {Speech}: {When} and {Why} {Content}-{Based} {Laws} {Are} {Presumptively} {Unconstitutional}.
\newblock Technical Report IF12308, Congressional Research Service.

\bibitem[{King and Hayes(2018)}]{king_effects_2018}
King, D.~K.; and Hayes, J. 2018.
\newblock The effects of power relationships: knowledge, practice and a new form of regulatory capture.
\newblock \emph{Journal of Risk Research}, 21(9): 1104--1116.

\bibitem[{Klaessig(2021)}]{klaessig_traversing_2021}
Klaessig, F. 2021.
\newblock Traversing {Technology} {Trajectories}.
\newblock \emph{NanoEthics}, 15(2): 149--168.

\bibitem[{Kleizen(2020)}]{kleizen_big_2020}
Kleizen, B. 2020.
\newblock Big data, {AI}-based governance and trust in government: developing an analytical framework and a conceptual model.
\newblock Technical report, DIGI4FED.

\bibitem[{Knaack(2022)}]{knaack_global_2022}
Knaack, P. 2022.
\newblock \emph{Global {Financial} {Networked} {Governance}: {The} {Power} of the {Financial} {Stability} {Board} and its {Limits}}.
\newblock Taylor \& Francis.
\newblock ISBN 978-1-00-082963-1.

\bibitem[{Koene et~al.(2019)Koene, Clifton, Hatada, Webb, and Richardson}]{koene_governance_2019}
Koene, A.; Clifton, C.; Hatada, Y.; Webb, H.; and Richardson, R. 2019.
\newblock A governance framework for algorithmic accountability and transparency.
\newblock Technical Report PE 624.262, European Parliamentary Research Service.

\bibitem[{Kolt(2023)}]{kolt_algorithmic_2023}
Kolt, N. 2023.
\newblock Algorithmic {Black} {Swans}.
\newblock SSRN:4370566.

\bibitem[{Konya et~al.(2023)Konya, Turan, Ovadya, Qui, Masood, Devine, Schirch, Roberts, and Forum}]{konya_deliberative_2023}
Konya, A.; Turan, D.; Ovadya, A.; Qui, L.; Masood, D.; Devine, F.; Schirch, L.; Roberts, I.; and Forum, D.~A. 2023.
\newblock Deliberative {Technology} for {Alignment}.
\newblock arXiv:2312.03893.

\bibitem[{Kuntze and Mertins(2023)}]{kuntze_lobbying_2023}
Kuntze, M.~A.; and Mertins, V. 2023.
\newblock Lobbying through {Gifts}.
\newblock In Mause, K.; and Polk, A., eds., \emph{The {Political} {Economy} of {Lobbying}: {Channels} of {Influence} and their {Regulation}}, Studies in {Public} {Choice}, 201--219. Cham: Springer International Publishing.
\newblock ISBN 978-3-031-44393-0.

\bibitem[{Kuźniacki et~al.(2022)Kuźniacki, Almada, Tyliński, Górski, Winogradska, and Zeldenrust}]{kuzniacki_towards_2022}
Kuźniacki, B.; Almada, M.; Tyliński, K.; Górski, L.; Winogradska, B.; and Zeldenrust, R. 2022.
\newblock Towards {eXplainable} {Artificial} {Intelligence} ({XAI}) in tax law : the need for a minimum legal standard.
\newblock \emph{World tax journal}, 14: 573--616.

\bibitem[{Kwak(2013)}]{kwak_cultural_2013}
Kwak, J. 2013.
\newblock Cultural {Capture} and the {Financial} {Crisis}.
\newblock In Carpenter, D.; and Moss, D.~A., eds., \emph{Preventing {Regulatory} {Capture}: {Special} {Interest} {Influence} and {How} to {Limit} it}, 71--98. Cambridge: Cambridge University Press.
\newblock ISBN 978-1-107-03608-6.

\bibitem[{Lall(2009)}]{lall_why_2009}
Lall, R. 2009.
\newblock Why {Basel} {II} failed and why any {Basel} {III} is doomed.
\newblock \url{https://perma.cc/57YL-AKF7}.

\bibitem[{Lam et~al.(2024)Lam, Lange, Blili-Hamelin, Davidovic, Brown, and Hasan}]{lam_framework_2024}
Lam, K.; Lange, B.; Blili-Hamelin, B.; Davidovic, J.; Brown, S.; and Hasan, A. 2024.
\newblock A {Framework} for {Assurance} {Audits} of {Algorithmic} {Systems}.
\newblock arXiv:2401.14908.

\bibitem[{Larsen(2022)}]{larsen_governing_2022}
Larsen, B.~C. 2022.
\newblock \emph{Governing {Artificial} {Intelligence}: {Lessons} from the {United} {States} and {China}}.
\newblock Ph.{D}. diss., Copenhagen Business School.

\bibitem[{Laux, Wachter, and Mittelstadt(2021)}]{laux_taming_2021}
Laux, J.; Wachter, S.; and Mittelstadt, B. 2021.
\newblock Taming the few: {Platform} regulation, independent audits, and the risks of capture created by the {DMA} and {DSA}.
\newblock \emph{Computer Law \& Security Review}, 43: 105613.

\bibitem[{Lawrence et~al.(2023)Lawrence, Innes, Hogarth, and Brooks}]{lawrence_large_2023}
Lawrence, N.; Innes, J.; Hogarth, I.; and Brooks, B. 2023.
\newblock Large {Language} {Models} {Evidence} {Session} {No}. 1.
\newblock \url{https://perma.cc/QV64-VNHP}.

\bibitem[{LeCun(2023)}]{lecun_tegmark_2023}
LeCun, Y. 2023.
\newblock @tegmark @{RishiSunak} @vonderleyen {Altman}, {Hassabis}, and {Amodei} are the ones doing massive corporate lobbying at the moment. {They} are the ones who are attempting to perform a regulatory capture of the {AI} industry. {You}, {Geoff}, and {Yoshua} are giving ammunition to those who are lobbying for a ban on open {AI} {R}\&{D}. {If} . . .
\newblock \textit{Twitter}. \url{https://perma.cc/WX78-92QH}.

\bibitem[{Lee, Hilty, and Liu(2021)}]{lee_artificial_2021}
Lee, J.-A.; Hilty, R.; and Liu, K.-C. 2021.
\newblock \emph{Artificial {Intelligence} and {Intellectual} {Property}}.
\newblock Oxford University Press.
\newblock ISBN 978-0-19-264429-9.

\bibitem[{Leslie et~al.(2022)Leslie, Katell, Aitken, Singh, Briggs, Powell, Rincón, Chengeta, Birhane, Perini, Jayadeva, and Mazumder}]{leslie_advancing_2022}
Leslie, D.; Katell, M.; Aitken, M.; Singh, J.; Briggs, M.; Powell, R.; Rincón, C.; Chengeta, T.; Birhane, A.; Perini, A.; Jayadeva, S.; and Mazumder, A. 2022.
\newblock Advancing {Data} {Justice} {Research} and {Practice}: {An} {Integrated} {Literature} {Review}.
\newblock Technical report, The Alan Turing Institute.

\bibitem[{{Lexis}(2024)}]{lexis_artificial_2024}
{Lexis}. 2024.
\newblock Artificial {Intelligence} {Legislation} {Tracker} (2024).
\newblock \url{https://perma.cc/ME9Q-2SSW}.
\newblock Accessed: 2024-04-26.

\bibitem[{Li(2023)}]{li_regulatory_2023}
Li, W.~Y. 2023.
\newblock Regulatory capture’s third face of power.
\newblock \emph{Socio-Economic Review}, 21(2): 1217--1245.

\bibitem[{Lie(2023)}]{lie_trustworthy_2023}
Lie, S. 2023.
\newblock \emph{Trustworthy {AI} in {Radiology}: {Evolving} {AI} policy to navigate trust, transparency, and human-machine interactions}.
\newblock Master's diss., Tilburg Institute for Law, Technology, and Society.

\bibitem[{Liesenfeld, Lopez, and Dingemanse(2023)}]{liesenfeld_opening_2023}
Liesenfeld, A.; Lopez, A.; and Dingemanse, M. 2023.
\newblock Opening up {ChatGPT}: {Tracking} openness, transparency, and accountability in instruction-tuned text generators.
\newblock In \emph{Proceedings of the 5th {International} {Conference} on {Conversational} {User} {Interfaces}}, {CUI} '23, 1--6. New York, NY, USA: Association for Computing Machinery.
\newblock ISBN 9798400700149.

\bibitem[{Lin and Jackson(2023)}]{lin_bias_2023}
Lin, C.~K.; and Jackson, S.~J. 2023.
\newblock From {Bias} to {Repair}: {Error} as a {Site} of {Collaboration} and {Negotiation} in {Applied} {Data} {Science} {Work}.
\newblock \emph{Proceedings of the ACM on Human-Computer Interaction}, 7(CSCW1): 131:1--131:32.

\bibitem[{Livermore and Revesz(2013)}]{livermore_can_2013}
Livermore, M.~A.; and Revesz, R.~L. 2013.
\newblock Can {Executive} {Review} {Help} {Prevent} {Capture}?
\newblock In Carpenter, D.; and Moss, D.~A., eds., \emph{Preventing {Regulatory} {Capture}: {Special} {Interest} {Influence} and {How} to {Limit} it}, 420--450. Cambridge: Cambridge University Press.
\newblock ISBN 978-1-107-03608-6.

\bibitem[{Lomas(2023)}]{lomas_frances_2023}
Lomas, N. 2023.
\newblock France's {Mistral} dials up call for {EU} {AI} {Act} to fix rules for apps, not model makers.
\newblock \textit{TechCrunch}. \url{https://perma.cc/6V4D-NCX3}.
\newblock Accessed: 2023-12-16.

\bibitem[{London and Danks(2018)}]{london_regulating_2018}
London, A.~J.; and Danks, D. 2018.
\newblock Regulating {Autonomous} {Vehicles}: {A} {Policy} {Proposal}.
\newblock In \emph{Proceedings of the 2018 {AAAI}/{ACM} {Conference} on {AI}, {Ethics}, and {Society}}, {AIES} '18, 216--221. New York, NY, USA: Association for Computing Machinery.
\newblock ISBN 978-1-4503-6012-8.

\bibitem[{Luetz(2023)}]{luetz_gender_2023}
Luetz, F. 2023.
\newblock Gender {Equality} and {Algorithmic} {Discrimination}: the contribution of the {EU} standardisation request on {AI}.
\newblock \emph{ex/ante}, 2023(2): 4--15.

\bibitem[{Lupo(2023)}]{lupo_risky_2023}
Lupo, G. 2023.
\newblock Risky {Artificial} {Intelligence}: {The} {Role} of {Incidents} in the {Path} to {AI} {Regulation}.
\newblock \emph{Law, Technology and Humans}, 5(1): 133--152.

\bibitem[{Lévesque(2021)}]{levesque_scoping_2021}
Lévesque, M. 2021.
\newblock Scoping {AI} {Governance}: {A} {Smarter} {Tool} {Kit} for {Beneficial} {Applications}.
\newblock Technical Report 260, Centre for International Governnace Innovation.

\bibitem[{Magill(2013)}]{magill_courts_2013}
Magill, M.~E. 2013.
\newblock Courts and {Regulatory} {Capture}.
\newblock In Carpenter, D.; and Moss, D.~A., eds., \emph{Preventing {Regulatory} {Capture}: {Special} {Interest} {Influence} and {How} to {Limit} it}, 397--419. Cambridge: Cambridge University Press.
\newblock ISBN 978-1-107-03608-6.

\bibitem[{Manheim et~al.(2024)Manheim, Martin, Bailey, Samin, and Greutzmacher}]{manheim_necessity_2024}
Manheim, D.; Martin, S.; Bailey, M.; Samin, M.; and Greutzmacher, R. 2024.
\newblock The {Necessity} of {AI} {Audit} {Standards} {Boards}.
\newblock arXiv:2404.13060.

\bibitem[{Marcus(2023{\natexlab{a}})}]{marcus_controlling_2023}
Marcus, G. 2023{\natexlab{a}}.
\newblock Controlling {AI}.
\newblock \emph{Communications of the ACM}, 66(10): 6--7.

\bibitem[{Marcus(2023{\natexlab{b}})}]{marcus_mustafasuleymn_2023}
Marcus, G. 2023{\natexlab{b}}.
\newblock @mustafasuleymn {No}. {It}’s textbook regulatory capture, no mention whatsoever of independent scientists, who are critically needed if we want to have a good outcome for humanity. {Global} {AI} summit good; the framing of who is to be invited is worrisome.
\newblock \textit{Twitter}. \url{https://perma.cc/JEN8-XUEG}.

\bibitem[{Margulies(2023)}]{margulies_adjudicating_2023}
Margulies, P. 2023.
\newblock Adjudicating {Algorithms}: {Accountability} in {Regulation} of {Surveillance}, {Privacy}, and {Discrimination}.
\newblock SSRN:4375080.

\bibitem[{Mariniello, Neven, and Padilla(2015)}]{mariniello_antitrust_2015}
Mariniello, M.; Neven, D.; and Padilla, A.~J. 2015.
\newblock Antitrust, regulatory capture and economic integration.
\newblock Research {Report} 2015/11, Bruegel Policy Contribution.

\bibitem[{McDonald, Schoenebeck, and Forte(2019)}]{mcdonald_reliability_2019}
McDonald, N.; Schoenebeck, S.; and Forte, A. 2019.
\newblock Reliability and {Inter}-rater {Reliability} in {Qualitative} {Research}: {Norms} and {Guidelines} for {CSCW} and {HCI} {Practice}.
\newblock \emph{Proc. ACM Hum.-Comput. Interact.}, 3(CSCW): 72:1--72:23.

\bibitem[{McGraw and Mandl(2021)}]{mcgraw_privacy_2021}
McGraw, D.; and Mandl, K.~D. 2021.
\newblock Privacy protections to encourage use of health-relevant digital data in a learning health system.
\newblock \emph{npj Digital Medicine}, 4(1): 1--11.

\bibitem[{McInerney and Drage(2024)}]{mcinerney_good_2024}
McInerney, K.; and Drage, E., eds. 2024.
\newblock \emph{The {Good} {Robot} : {Why} {Technology} {Needs} {Feminism}}.
\newblock Bloomsbury Publishing.
\newblock ISBN 978-1-350-39995-2.

\bibitem[{Meghani(2021)}]{meghani_regulations_2021}
Meghani, Z. 2021.
\newblock Regulations {Matter}: {Epistemic} {Monopoly}, {Domination}, {Patents}, and the {Public} {Interest}.
\newblock \emph{Philosophy \& Technology}, 34(4): 1449--1474.

\bibitem[{Mehmood, Naseer, and Chen(2024)}]{mehmood_ai_2024}
Mehmood, S.; Naseer, S.; and Chen, D. 2024.
\newblock {AI} {Education} as {State} {Capacity}: {Experimental} {Evidence} from {Pakistan}.
\newblock \url{https://perma.cc/W52C-UJZ6}.

\bibitem[{Meßmer and Degeling(2023)}]{mesmer_auditing_2023}
Meßmer, A.-K.; and Degeling, M. 2023.
\newblock Auditing {Recommender} {Systems} -- {Putting} the {DSA} into practice with a risk-scenario-based approach.
\newblock arXiv:2302.04556.

\bibitem[{Miller(2023)}]{miller_us_2023}
Miller, G. 2023.
\newblock {US} {Senate} {AI} ‘{Insight} {Forum}’ {Tracker}.
\newblock \textit{Tech Policy Press}. \url{https://perma.cc/RT5Q-W6Y4}.
\newblock Accessed: 2024-04-24.

\bibitem[{Mitnick(2011)}]{mitnick_capturing_2011}
Mitnick, B.~M. 2011.
\newblock Capturing “capture”: definition and mechanisms.
\newblock In \emph{Handbook on the {Politics} of {Regulation}}. Edward Elgar Publishing.
\newblock ISBN 978-0-85793-611-0.

\bibitem[{Mitnick(2015)}]{mitnick_developing_2015}
Mitnick, B.~M. 2015.
\newblock Developing a {Normative} {Theory} of {Fiducial} {Regulation}.
\newblock In \emph{Jerusalem {Papers} in {Regulation} \& {Governance}}.

\bibitem[{Moberg and Gill-Pedro(2024)}]{moberg_law_2024}
Moberg, A.; and Gill-Pedro, E. 2024.
\newblock Law and the {Governance} of {Artificial} {Intelligence}.
\newblock In \emph{{YSEC} {Yearbook} of {Socio}-{Economic} {Constitutions} 2023}, {YSEC} {Yearbook} of {Socio}-{Economic} {Constitutions}, 1--13. Cham: Springer International Publishing.

\bibitem[{Moitra et~al.(2022)Moitra, Wagenaar, Kalirai, Ahmed, and Soden}]{moitra_ai_2022}
Moitra, A.; Wagenaar, D.; Kalirai, M.; Ahmed, S.~I.; and Soden, R. 2022.
\newblock {AI} and {Disaster} {Risk}: {A} {Practitioner} {Perspective}.
\newblock \emph{Proceedings of the ACM on Human-Computer Interaction}, 6(CSCW2): 272:1--272:20.

\bibitem[{Mollman(2023)}]{mollman_bill_2023}
Mollman, S. 2023.
\newblock Bill {Gurley} rips regulatory capture in {AI}, credits {Silicon} {Valley}’s success to being ‘so f--king far away from’ {D}.{C}.
\newblock \textit{Fortune}. \url{https://perma.cc/9XQH-TKDT}.
\newblock Accessed: 2023-12-16.

\bibitem[{Mügge(2023)}]{mugge_securitization_2023}
Mügge, D. 2023.
\newblock The securitization of the {EU}’s digital tech regulation.
\newblock \emph{Journal of European Public Policy}, 30(7): 1431--1446.

\bibitem[{Narayanan and Tan(2023)}]{narayanan_attitudinal_2023}
Narayanan, D.; and Tan, Z.~M. 2023.
\newblock Attitudinal {Tensions} in the {Joint} {Pursuit} of {Explainable} and {Trusted} {AI}.
\newblock \emph{Minds and Machines}, 33(1): 55--82.

\bibitem[{Neerven(2023)}]{neerven_will_2023}
Neerven, D. B.~K., Lennart~van. 2023.
\newblock Will {Disagreement} {Over} {Foundation} {Models} {Put} the {EU} {AI} {Act} at {Risk}?
\newblock \textit{Tech Policy Press}. \url{https://perma.cc/7EXE-9CKH}.
\newblock Accessed: 2023-12-16.

\bibitem[{Nemitz(2023)}]{nemitz_democracy_2023}
Nemitz, P. 2023.
\newblock Democracy through law {The} {Transatlantic} {Reflection} {Group} and its manifesto in defence of democracy and the rule of law in the age of “artificial intelligence”.
\newblock \emph{European Law Journal}, 29(1-2): 237--248.

\bibitem[{Neudert(2023)}]{neudert_regulatory_2023}
Neudert, L.-M. 2023.
\newblock Regulatory capacity capture: {The} {United} {Kingdom}’s online safety regime.
\newblock \emph{Internet Policy Review}, 12(4).

\bibitem[{Neumann(2018)}]{neumann_risks_2018}
Neumann, P.~G. 2018.
\newblock Risks to the {Public}.
\newblock \emph{ACM SIGSOFT Software Engineering Notes}, 43(2): 8--11.

\bibitem[{Neves(2023)}]{neves_freedom_2023}
Neves, I. 2023.
\newblock The freedom to conduct a business as a driver for {AI} governance.
\newblock \emph{UNIO – EU Law Journal}, 9(2): 13--56.

\bibitem[{{NIST}(2023)}]{nist_ai_2023}
{NIST}. 2023.
\newblock {AI} {Risk} {Management} {Framework}: {AI} {RMF} (1.0).
\newblock \url{https://nvlpubs.nist.gov/nistpubs/ai/NIST.AI.100-1.pdf}.

\bibitem[{{NquiringMinds}(2023)}]{nquiringminds_written_2023}
{NquiringMinds}. 2023.
\newblock Written evidence ({LLM0073}) to the {House} of {Lords} {Communications} and {Digital} {Select} {Committee} inquiry: {Large} language models.
\newblock \textit{House of Lords Communications and Digital Select Committee}. \url{https://perma.cc/NW48-5345}.

\bibitem[{Ochigame(2019)}]{ochigame_how_2019}
Ochigame, R. 2019.
\newblock How {Big} {Tech} {Manipulates} {Academia} to {Avoid} {Regulation}.
\newblock \textit{The Intercept}. \url{https://perma.cc/WCC3-MKML}.
\newblock Accessed: 2024-06-19.

\bibitem[{{OECD}(n.d.)}]{oecd_oecd_nodate}
{OECD}. n.d.
\newblock {OECD} {Working} {Party} and {Network} of {Experts} on {AI}.
\newblock \url{https://perma.cc/84GT-69UN}.
\newblock Accessed: 2024-01-26.

\bibitem[{Olteanu et~al.(2023)Olteanu, Ekstrand, Castillo, and Suh}]{olteanu_responsible_2023}
Olteanu, A.; Ekstrand, M.; Castillo, C.; and Suh, J. 2023.
\newblock Responsible {AI} {Research} {Needs} {Impact} {Statements} {Too}.
\newblock arXiv:2311.11776.

\bibitem[{Ong(2024)}]{ong_privacy_2024}
Ong, R. 2024.
\newblock Privacy and personal information protection in {China}’s all-seeing state.
\newblock \emph{International Journal of Law and Information Technology}, eaae003.

\bibitem[{{OpenUK}(2023)}]{openuk_written_2023}
{OpenUK}. 2023.
\newblock Written evidence ({LLM0015}) to the {House} of {Lords} {Communications} and {Digital} {Select} {Committee} inquiry: {Large} language models.
\newblock \textit{House of Lords Communications and Digital Select Committee}. \url{https://perma.cc/Z9YE-QRLS}.

\bibitem[{Opoku(2019)}]{opoku_regulation_2019}
Opoku, V. 2019.
\newblock \emph{Regulation of {Artificial} {Intelligence} in the {EU}}.
\newblock Master's diss., University of Hamburg.

\bibitem[{Oprysko(2023)}]{oprysko_openai_2023}
Oprysko, C. 2023.
\newblock {OpenAI} registers to lobby.
\newblock \url{https://perma.cc/U8UC-LQ6E}.
\newblock Accessed: 2023-12-16.

\bibitem[{Oprysko(2024)}]{oprysko_ai_2024}
Oprysko, C. 2024.
\newblock {AI} and a jam-packed agenda fueled another strong quarter on {K} {Street}.
\newblock \url{https://perma.cc/TRZ9-AJZK}.
\newblock Accessed: 2024-01-29.

\bibitem[{Outeda and Cacheda(2023)}]{outeda_artificial_2023}
Outeda, C.~C.; and Cacheda, B.~G. 2023.
\newblock Artificial {Intelligence}: {A} {Reading} from {European} {Politics}.
\newblock In Ramiro~Troitiño, D.; Kerikmäe, T.; and Hamuľák, O., eds., \emph{Digital {Development} of the {European} {Union}: {An} {Interdisciplinary} {Perspective}}, 363--381. Cham: Springer International Publishing.
\newblock ISBN 978-3-031-27312-4.

\bibitem[{O’Shaughnessy et~al.(2023)O’Shaughnessy, Schiff, Varshney, Rozell, and Davenport}]{oshaughnessy_what_2023}
O’Shaughnessy, M.~R.; Schiff, D.~S.; Varshney, L.~R.; Rozell, C.~J.; and Davenport, M.~A. 2023.
\newblock What governs attitudes toward artificial intelligence adoption and governance?
\newblock \emph{Science and Public Policy}, 50(2): 161--176.

\bibitem[{Pacione and Teixeira(2023)}]{pacione_implications_2023}
Pacione, M.; and Teixeira, N. 2023.
\newblock \emph{Implications of {Artificial} {Intelligence} on {Leadership} in {Complex} {Organizations}: {An} {Exploration} of the {Near} {Future}}.
\newblock Master's diss., OCAD University.

\bibitem[{Papyshev and Yarime(2022)}]{papyshev_limitation_2022}
Papyshev, G.; and Yarime, M. 2022.
\newblock The limitation of ethics-based approaches to regulating artificial intelligence: regulatory gifting in the context of {Russia}.
\newblock \emph{AI \& Society}.

\bibitem[{Papyshev and Yarime(2023)}]{papyshev_challenges_2023}
Papyshev, G.; and Yarime, M. 2023.
\newblock The challenges of industry self-regulation of {AI} in emerging economies: implications of the case of {Russia} for public policy and institutional development.
\newblock In \emph{Elgar {Companion} to {Regulating} {AI} and {Big} {Data} in {Emerging} {Economies}}, 81--98. Edward Elgar Publishing.
\newblock ISBN 978-1-78536-240-8.

\bibitem[{Paul(2022)}]{paul_politics_2022}
Paul, R. 2022.
\newblock The {Politics} of {Regulating} {Artificial} {Intelligence} {Technologies}: {A} {Competition} {State} {Perspective}.
\newblock SSRN:4272867.

\bibitem[{Pavel et~al.(2022)Pavel, Kind, Strait, Reeve, Peppin, Szymielewicz, Veale, MacDonald, Lynskey, Coyle, and Nemitz}]{pavel_rethinking_2022}
Pavel, V.; Kind, C.; Strait, A.; Reeve, O.; Peppin, A.; Szymielewicz, K.; Veale, M.; MacDonald, R.; Lynskey, O.; Coyle, D.; and Nemitz, P. 2022.
\newblock Rethinking data and rebalancing digital power.
\newblock Report, Ada Lovelace Institute, London, UK.

\bibitem[{Pavlopoulou(2022)}]{pavlopoulou_exploration_2022}
Pavlopoulou, T. 2022.
\newblock \emph{An exploration of {AI} governance through the lens of sustainability}.
\newblock Master's diss., University of Iceland.

\bibitem[{Peng, Lin, and Streinz(2021)}]{peng_artificial_2021}
Peng, S.-y.; Lin, C.-F.; and Streinz, T. 2021.
\newblock Artificial {Intelligence} and {International} {Economic} {Law}: {A} {Research} and {Policy} {Agenda}.
\newblock SSRN:3877055.

\bibitem[{Perlman(2020)}]{perlman_for_2020}
Perlman, R.~L. 2020.
\newblock For {Safety} or {Profit}? {How} {Science} {Serves} the {Strategic} {Interests} of {Private} {Actors}.
\newblock \emph{American Journal of Political Science}, 64(2): 293--308.

\bibitem[{Perrigo(2023)}]{perrigo_exclusive_2023}
Perrigo, B. 2023.
\newblock Exclusive: {OpenAI} {Lobbied} {E}.{U}. to {Water} {Down} {AI} {Regulation}.
\newblock \textit{Time}. \url{https://perma.cc/UL9Y-D4TN}.
\newblock Accessed: 2023-12-16.

\bibitem[{Pizzi, Romanoff, and Engelhardt(2020)}]{pizzi_ai_2020}
Pizzi, M.; Romanoff, M.; and Engelhardt, T. 2020.
\newblock {AI} for humanitarian action: {Human} rights and ethics.
\newblock \emph{International Review of the Red Cross}, 102(913): 145--180.

\bibitem[{Plantinga et~al.(2024)Plantinga, Shilongo, Mudongo, Umubyeyi, Gastrow, and Razzano}]{plantinga_responsible_2024}
Plantinga, P.; Shilongo, K.; Mudongo, O.; Umubyeyi, A.; Gastrow, M.; and Razzano, G. 2024.
\newblock Responsible artificial intelligence in {Africa}: {Towards} policy learning.
\newblock \url{https://perma.cc/QPM2-D32T}.

\bibitem[{Png(2022)}]{png_at_2022}
Png, M.-T. 2022.
\newblock At the {Tensions} of {South} and {North}: {Critical} {Roles} of {Global} {South} {Stakeholders} in {AI} {Governance}.
\newblock In \emph{Proceedings of the 2022 {ACM} {Conference} on {Fairness}, {Accountability}, and {Transparency}}, {FAccT} '22, 1434--1445. New York, NY, USA: Association for Computing Machinery.
\newblock ISBN 978-1-4503-9352-2.

\bibitem[{Polishchuk(2023)}]{polishchuk_exploring_2023}
Polishchuk, A. 2023.
\newblock \emph{Exploring the challenges in the harmonization of clinical evaluation of medical device software across {EU} member states}.
\newblock Master's diss., Laurea University of Applied Sciences.

\bibitem[{{Politico}(2024)}]{politico_davos_2024}
{Politico}. 2024.
\newblock Davos {Party}.
\newblock \url{https://web.archive.org/web/20240130162550/https://www.politico.eu/event/davos-party/}.
\newblock Accessed: 2024-04-26.

\bibitem[{Rainie, Anderson, and Vogels(2021)}]{rainie_experts_2021}
Rainie, L.; Anderson, J.; and Vogels, E.~A. 2021.
\newblock Experts {Doubt} {Ethical} {AI} {Design} {Will} {Be} {Broadly} {Adopted} as the {Norm} {Within} the {Next} {Decade}.
\newblock Technical report, Pew Research Center.

\bibitem[{Raji, Costanza-Chock, and Buolamwini(2023)}]{raji_change_2023}
Raji, I.~D.; Costanza-Chock, S.; and Buolamwini, J. 2023.
\newblock Change from the {Outside}: {Towards} {Credible} {Third}-{Party} {Audits} of {AI} {Systems}.
\newblock In Prud’homme, B.; Régis, C.; and Farnadi, G., eds., \emph{Missing {Links} in {AI} {Governance}}, 5--26. Montréal, Canada: UNESCO/Mila – Québec Institute of Artificial Intelligence.

\bibitem[{Ramdas(2022)}]{ramdas_identifying_2022}
Ramdas, V. 2022.
\newblock Identifying an {Actionable} {Algorithmic} {Transparency} {Framework}: {A} {Comparative} {Analysis} of {Initiatives} to {Enhance} {Accountability} of {Social} {Media} {Platforms}.
\newblock \emph{NLUD Journal of Legal Studies}, 4: 74--90.

\bibitem[{Ratanpal(2024)}]{ratanpal_federal_2024}
Ratanpal, H. 2024.
\newblock Federal lobbying on artificial intelligence grows as legislative efforts stall.
\newblock \textit{OpenSecrets}. \url{https://perma.cc/W7B8-FXB5}.
\newblock Accessed: 2024-01-29.

\bibitem[{Rawat, Prerna, and Singh(2024)}]{rawat_ethics_2024}
Rawat, P.; Prerna; and Singh, P. 2024.
\newblock Ethics and {Regulations} of {AI} in {Society} 5.0.
\newblock In \emph{Artificial {Intelligence} and {Society} 5.0}. Chapman and Hall/CRC.
\newblock ISBN 978-1-00-339705-2.

\bibitem[{Ren(2022)}]{ren_centralization_2022}
Ren, Y. 2022.
\newblock Centralization or {Decentralization}? {The} {Future} of {Global} {Climate} {Governance}.
\newblock In \emph{Proceedings of the 2021 {International} {Conference} on {Social} {Development} and {Media} {Communication} ({SDMC} 2021)}, 868--873. Atlantis Press.
\newblock ISBN 978-94-6239-512-1.

\bibitem[{Renieris(2023)}]{renieris_concerned_2023}
Renieris, E. 2023.
\newblock Concerned about regulatory capture wrt \#{AI} regulation? {Let}'s learn from the past! {Here}'s an excerpt from {Chapter} 2 of my book {BEYOND} {DATA} where {I} describe a similar phenomenon around \#{DataProtection} regulations.
\newblock \textit{Twitter}. \url{https://perma.cc/76CU-MF2N}.

\bibitem[{Reuel et~al.(2024)Reuel, Soder, Bucknall, and Undheim}]{reuel_position_2024}
Reuel, A.; Soder, L.; Bucknall, B.; and Undheim, T.~A. 2024.
\newblock Position {Paper}: {Technical} {Research} and {Talent} is {Needed} for {Effective} {AI} {Governance}.
\newblock arXiv:2406.06987.

\bibitem[{Rex(2020)}]{rex_anatomy_2020}
Rex, J. 2020.
\newblock Anatomy of agency capture: {An} organizational typology for diagnosing and remedying capture.
\newblock \emph{Regulation \& Governance}, 14(2): 271--294.

\bibitem[{Rex(2022)}]{rex_agency_2022}
Rex, J. 2022.
\newblock Agency capture.
\newblock In \emph{Handbook of {Regulatory} {Authorities}}, 362--378. Edward Elgar Publishing.
\newblock ISBN 978-1-83910-899-0.

\bibitem[{Rikap(2024)}]{rikap_dynamics_2024}
Rikap, C. 2024.
\newblock Dynamics of {Corporate} {Governance} {Beyond} {Ownership} in {AI}.
\newblock Technical report, Common Wealth.

\bibitem[{Rilinger(2023)}]{rilinger_who_2023}
Rilinger, G. 2023.
\newblock Who captures whom? {Regulatory} misperceptions and the timing of cognitive capture.
\newblock \emph{Regulation \& Governance}, 17(1): 43--60.

\bibitem[{Roberts et~al.(2023)Roberts, Babuta, Morley, Thomas, Taddeo, and Floridi}]{roberts_artificial_2023}
Roberts, H.; Babuta, A.; Morley, J.; Thomas, C.; Taddeo, M.; and Floridi, L. 2023.
\newblock Artificial intelligence regulation in the {United} {Kingdom}: a path to good governance and global leadership?
\newblock \emph{Internet Policy Review}, 12(2).

\bibitem[{Roberts et~al.(2021)Roberts, Cowls, Hine, Mazzi, Tsamados, Taddeo, and Floridi}]{roberts_achieving_2021}
Roberts, H.; Cowls, J.; Hine, E.; Mazzi, F.; Tsamados, A.; Taddeo, M.; and Floridi, L. 2021.
\newblock Achieving a ‘{Good} {AI} {Society}’: {Comparing} the {Aims} and {Progress} of the {EU} and the {US}.
\newblock \emph{Science and Engineering Ethics}, 27(6): 68.

\bibitem[{Rubinstein~Reiss(2012)}]{rubinstein_reiss_benefits_2012}
Rubinstein~Reiss, D. 2012.
\newblock The {Benefits} of {Capture} {The} {Asymmetry} of {Administrative} {Law}: {The} {Lack} of {Public} {Participation} and the {Public} {Interest}.
\newblock \emph{Wake Forest Law Review}, 47(3): 569--610.

\bibitem[{Saltelli et~al.(2022)Saltelli, Dankel, Di~Fiore, Holland, and Pigeon}]{saltelli_science_2022}
Saltelli, A.; Dankel, D.~J.; Di~Fiore, M.; Holland, N.; and Pigeon, M. 2022.
\newblock Science, the endless frontier of regulatory capture.
\newblock \emph{Futures}, 135: 102860.

\bibitem[{Sanchez-Graells(2023{\natexlab{a}})}]{sanchez-graells_discharging_2023}
Sanchez-Graells, A. 2023{\natexlab{a}}.
\newblock Discharging {Procurement} of the {Digital} {Regulation} {Role}: {External} {Oversight} and {Mandatory} {Requirements} for {Public} {Sector} {Digital} {Technology} {Adoption}.
\newblock SSRN:4412359.

\bibitem[{Sanchez-Graells(2023{\natexlab{b}})}]{sanchez-graells_procurement_2023}
Sanchez-Graells, A. 2023{\natexlab{b}}.
\newblock Procurement {Tools} for {AI} {Regulation} by {Contract}. {Not} the {Sharpest} in the {Shed}.
\newblock SSRN:4369297.

\bibitem[{Sanchez-Graells(2023{\natexlab{c}})}]{sanchez-graells_responsibly_2023}
Sanchez-Graells, A. 2023{\natexlab{c}}.
\newblock Responsibly {Buying} {Artificial} {Intelligence}: {A} ‘{Regulatory} {Hallucination}’.
\newblock SSRN:4643273.

\bibitem[{Sanchez-Graells(2023{\natexlab{d}})}]{sanchez-graells_two_2023}
Sanchez-Graells, A. 2023{\natexlab{d}}.
\newblock The {Two} {Roles} of {Procurement} in the {Transition} {Towards} {Digital} {Public} {Governance}: {Procurement} as {Regulatory} {Gatekeeper} and as {Site} for {Public} {Sector} {Experimentation}.
\newblock SSRN:4384037.

\bibitem[{Sanchez-Graells(2024{\natexlab{a}})}]{sanchez-graells_digital_2024}
Sanchez-Graells, A. 2024{\natexlab{a}}.
\newblock \emph{Digital {Technologies} and {Public} {Procurement}: {Gatekeeping} and {Experimentation} in {Digital} {Public} {Governance}}.
\newblock Oxford University Press.
\newblock ISBN 978-0-19-263660-7.

\bibitem[{Sanchez-Graells(2024{\natexlab{b}})}]{sanchez-graells_governing_2024}
Sanchez-Graells, A. 2024{\natexlab{b}}.
\newblock Governing the {Assessment} and {Taking} of {Risks} in {Digital} {Procurement} {Governance}.
\newblock In Sanchez-Graells, A., ed., \emph{Digital {Technologies} and {Public} {Procurement}: {Gatekeeping} and {Experimentation} in {Digital} {Public} {Governance}}, 0. Oxford University Press.
\newblock ISBN 978-0-19-886677-0.

\bibitem[{Sandoval et~al.(2023)Sandoval, de~Santana, Berger, Quigley, and Hobson}]{sandoval_responsible_2023}
Sandoval, J. C.~B.; de~Santana, V.~F.; Berger, S.; Quigley, L.~T.; and Hobson, S. 2023.
\newblock Responsible and {Inclusive} {Technology} {Framework}: {A} {Formative} {Framework} to {Promote} {Societal} {Considerations} in {Information} {Technology} {Contexts}.
\newblock arXiv:2302.11565.

\bibitem[{Sarel(2023)}]{sarel_restraining_2023}
Sarel, R. 2023.
\newblock Restraining {ChatGPT}.
\newblock SSRN:4354486.

\bibitem[{Sastry et~al.(2024)Sastry, Heim, Belfield, Anderljung, Brundage, Hazell, O'Keefe, Hadfield, Ngo, Pilz, Gor, Bluemke, Shoker, Egan, Trager, Avin, Weller, Bengio, and Coyle}]{sastry_computing_2024}
Sastry, G.; Heim, L.; Belfield, H.; Anderljung, M.; Brundage, M.; Hazell, J.; O'Keefe, C.; Hadfield, G.~K.; Ngo, R.; Pilz, K.; Gor, G.; Bluemke, E.; Shoker, S.; Egan, J.; Trager, R.~F.; Avin, S.; Weller, A.; Bengio, Y.; and Coyle, D. 2024.
\newblock Computing {Power} and the {Governance} of {Artificial} {Intelligence}.
\newblock arXiv:2402.08797.

\bibitem[{Saunders, Kitzinger, and Kitzinger(2015)}]{saunders_anonymising_2015}
Saunders, B.; Kitzinger, J.; and Kitzinger, C. 2015.
\newblock Anonymising interview data: challenges and compromise in practice.
\newblock \emph{Qualitative Research}, 15(5): 616--632.

\bibitem[{Saunders et~al.(2018)Saunders, Sim, Kingstone, Baker, Waterfield, Bartlam, Burroughs, and Jinks}]{saunders_saturation_2018}
Saunders, B.; Sim, J.; Kingstone, T.; Baker, S.; Waterfield, J.; Bartlam, B.; Burroughs, H.; and Jinks, C. 2018.
\newblock Saturation in qualitative research: exploring its conceptualization and operationalization.
\newblock \emph{Quality \& Quantity}, 52(4): 1893--1907.

\bibitem[{Scarcella(2024)}]{scarcella_openai_2024}
Scarcella, M. 2024.
\newblock {OpenAI} expands lobbying efforts, hiring former {US} senator.
\newblock \textit{Reuters}. \url{https://perma.cc/Y6ZJ-9QTJ}.
\newblock Accessed: 2024-04-24.

\bibitem[{Scherer(2016)}]{scherer_regulating_2016}
Scherer, M.~U. 2016.
\newblock Regulating {Artificial} {Intelligence} {Systems}: {Risks}, {Challenges}, {Competencies}, and {Strategies}.
\newblock \emph{Harvard Journal of Law \& Technology}, 29(2).

\bibitem[{Scherz(2024)}]{scherz_ai_2024}
Scherz, P. 2024.
\newblock {AI} as {Person}, {Paradigm}, and {Structure}: {Notes} toward an {Ethics} of {AI}.
\newblock \emph{Theological Studies}, 85(1): 124--144.

\bibitem[{Schiffrin(2021)}]{schiffrin_media_2021}
Schiffrin, A. 2021.
\newblock Media {Capture}: {How} {Money}, {Digital} {Platforms}, and {Governments} {Control} the {News}.
\newblock In \emph{Media {Capture}}. Columbia University Press.
\newblock ISBN 978-0-231-54802-1.

\bibitem[{Schlanger(2014)}]{schlanger_offices_2014}
Schlanger, M. 2014.
\newblock Offices of {Goodness}: {Influence} {Without} {Authority} in {Federal} {Agencies}.
\newblock \emph{Cardozo Law Review}, 36(1): 53--117.

\bibitem[{Schwarcz(2013)}]{schwarcz_preventing_2013}
Schwarcz, D. 2013.
\newblock Preventing {Capture} {Through} {Consumer} {Empowerment} {Programs}: {Some} {Evidence} from {Insurance} {Regulation}.
\newblock In Carpenter, D.; and Moss, D.~A., eds., \emph{Preventing {Regulatory} {Capture}: {Special} {Interest} {Influence} and {How} to {Limit} it}, 365--396. Cambridge: Cambridge University Press.
\newblock ISBN 978-1-107-03608-6.

\bibitem[{Schäferling(2023)}]{schaferling_case_2023}
Schäferling, S. 2023.
\newblock The {Case} for a {Right} {Against} {Automated} {Decision}-{Making}.
\newblock In Schäferling, S., ed., \emph{Governmental {Automated} {Decision}-{Making} and {Human} {Rights}: {Reconciling} {Law} and {Intelligent} {Systems}}, Law, {Governance} and {Technology} {Series}, 231--283. Cham: Springer Nature Switzerland.
\newblock ISBN 978-3-031-48125-3.

\bibitem[{Seger et~al.(2023)Seger, Dreksler, Moulange, Dardaman, Schuett, Wei, Winter, Arnold, hÉigeartaigh, Korinek, Anderljung, Bucknall, Chan, Stafford, Koessler, Ovadya, Garfinkel, Bluemke, Aird, Levermore, Hazell, and Gupta}]{seger_open-sourcing_2023}
Seger, E.; Dreksler, N.; Moulange, R.; Dardaman, E.; Schuett, J.; Wei, K.; Winter, C.; Arnold, M.; hÉigeartaigh, S.~O.; Korinek, A.; Anderljung, M.; Bucknall, B.; Chan, A.; Stafford, E.; Koessler, L.; Ovadya, A.; Garfinkel, B.; Bluemke, E.; Aird, M.; Levermore, P.; Hazell, J.; and Gupta, A. 2023.
\newblock Open-{Sourcing} {Highly} {Capable} {Foundation} {Models}: {An} evaluation of risks, benefits, and alternative methods for pursuing open-source objectives.
\newblock arXiv:2311.09227.

\bibitem[{Selbst(2021)}]{selbst_institutional_2021}
Selbst, A.~D. 2021.
\newblock An {Institutional} {View} of {Algorithmic} {Impact} {Assessments}.
\newblock \emph{Harvard Journal of Law \& Technology}, 35(1).

\bibitem[{Sengupta(2022)}]{sengupta_towards_2022}
Sengupta, U. 2022.
\newblock \emph{Towards a {Values}-{Based} {Data} {Governance} {Theory} in the {Social} {Economy} in {Ontario}}.
\newblock Ph.{D}. diss., University of Toronto.

\bibitem[{Shapiro(2012)}]{shapiro_complexity_2012}
Shapiro, S.~A. 2012.
\newblock The {Complexity} of {Regulatory} {Capture}: {Diagnosis}, {Causality}, and {Remediation} {Blowout}: {Legal} {Legacy} of the {Deepwater} {Horizon} {Catastrophe}.
\newblock \emph{Roger Williams University Law Review}, 17(1): 221--257.

\bibitem[{Shneiderman(2020)}]{shneiderman_bridging_2020}
Shneiderman, B. 2020.
\newblock Bridging the {Gap} {Between} {Ethics} and {Practice}: {Guidelines} for {Reliable}, {Safe}, and {Trustworthy} {Human}-centered {AI} {Systems}.
\newblock \emph{ACM Transactions on Interactive Intelligent Systems}, 10(4): 26:1--26:31.

\bibitem[{Shrier and Pentland(2022)}]{shrier_global_2022}
Shrier, D.~L.; and Pentland, A. 2022.
\newblock \emph{Global {Fintech}: {Financial} {Innovation} in the {Connected} {World}}.
\newblock MIT Press.
\newblock ISBN 978-0-262-54366-8.

\bibitem[{Sifat(2023)}]{sifat_artificial_2023}
Sifat, I. 2023.
\newblock Artificial {Intelligence} ({AI}) and {Retail} {Investment}.
\newblock SSRN:4539625.

\bibitem[{Sitaraman and Eyre(2023)}]{sitaraman_building_2023}
Sitaraman, G.; and Eyre, R. 2023.
\newblock Building {Public} {Capacity} on {Artificial} {Intelligence}.
\newblock Technical report, Vanderbilt Policy Accelerator.

\bibitem[{Sivan-Sevilla and Sharvit(2021)}]{sivan-sevilla_when_2021}
Sivan-Sevilla, I.; and Sharvit, S. 2021.
\newblock When {Cybersecurity} {Meets} the {Regulatory} {State}: {Case}-{Study} {Analysis} of the {Israeli} {Cybersecurity} {Regulatory} {Regime}.
\newblock In Tevet, E.; Shiffer, V.; and Galnoor, I., eds., \emph{Regulation in {Israel}: {Values}, {Effectiveness}, {Methods}}, 173--193. Cham: Springer International Publishing.
\newblock ISBN 978-3-030-56247-2.

\bibitem[{Slayton and Clark-Ginsberg(2018)}]{slayton_beyond_2018}
Slayton, R.; and Clark-Ginsberg, A. 2018.
\newblock Beyond regulatory capture: {Coproducing} expertise for critical infrastructure protection.
\newblock \emph{Regulation \& Governance}, 12(1): 115--130.

\bibitem[{Soest(2023)}]{soest_why_2023}
Soest, C.~v. 2023.
\newblock Why {Do} {We} {Speak} to {Experts}? {Reviving} the {Strength} of the {Expert} {Interview} {Method}.
\newblock \emph{Perspectives on Politics}, 21(1): 277--287.

\bibitem[{Solow-Niederman(2019)}]{solow-niederman_administering_2019}
Solow-Niederman, A. 2019.
\newblock Administering {Artificial} {Intelligence}.
\newblock \emph{Southern California Law Review}, 93: 633--696.

\bibitem[{{Sourcelist}(n.d.)}]{sourcelist_sourcelist_nodate}
{Sourcelist}. n.d.
\newblock Sourcelist ({Women}+).
\newblock \url{https://perma.cc/VPM8-KE25}.
\newblock Accessed: 2024-01-26.

\bibitem[{Southgate(2021)}]{southgate_artificial_2021}
Southgate, E. 2021.
\newblock Artificial {Intelligence} and {Machine} {Learning}: {A} {Practical} and {Ethical} {Guide} for {Teachers}.
\newblock In \emph{Digital {Disruption} in {Teaching} and {Testing}}. Routledge.
\newblock ISBN 978-1-00-304579-3.

\bibitem[{Southgate et~al.(2019)Southgate, Blackmore, Pieschl, Grimes, McGuire, and Smithers}]{southgate_artificial_2019}
Southgate, E.; Blackmore, K.; Pieschl, S.; Grimes, S.; McGuire, J.; and Smithers, K. 2019.
\newblock Artificial intelligence and emerging technologies in schools: {Research} report.
\newblock Technical report, University of Newcastle, Newcastle, NSW.

\bibitem[{{SRI}(2023)}]{sri_discerning_2023}
{SRI}. 2023.
\newblock Discerning signal from noise: {The} state of global {AI} standardization and what it means for {Canada}.
\newblock Technical report, Schwartz Reisman Institute for Technology and Society, University of Toronto.

\bibitem[{Stahl et~al.(2022)Stahl, Rodrigues, Santiago, and Macnish}]{stahl_european_2022}
Stahl, B.~C.; Rodrigues, R.; Santiago, N.; and Macnish, K. 2022.
\newblock A {European} {Agency} for {Artificial} {Intelligence}: {Protecting} fundamental rights and ethical values.
\newblock \emph{Computer Law \& Security Review}, 45: 105661.

\bibitem[{Stam and Diaz(2023)}]{stam_qualitative_2023}
Stam, A.; and Diaz, P. 2023.
\newblock Qualitative data anonymisation: theoretical and practical considerations for anonymising interview transcripts.
\newblock Technical Report No. 20, Version 1.0, Swiss Centre of Expertise in the Social Sciences FORS., Lausanne.

\bibitem[{Steed and Acquisti(2024)}]{steed_adoption_2024}
Steed, R.; and Acquisti, A. 2024.
\newblock Adoption of '{Privacy}-{Preserving}' {Analytics}: {Drivers}, {Designs}, \& {Decoupling}.
\newblock SSRN:4718865.

\bibitem[{Stempeck(2015)}]{stempeck_are_2015}
Stempeck, M. 2015.
\newblock Are {Uber} and {Facebook} {Turning} {Users} into {Lobbyists}?
\newblock \emph{Harvard Business Review}.

\bibitem[{Stigler(1971)}]{stigler_theory_1971}
Stigler, G.~J. 1971.
\newblock The {Theory} of {Economic} {Regulation}.
\newblock \emph{The Bell Journal of Economics and Management Science}, 2(1): 3--21.

\bibitem[{Stix and Maas(2021)}]{stix_bridging_2021}
Stix, C.; and Maas, M.~M. 2021.
\newblock Bridging the gap: the case for an ‘{Incompletely} {Theorized} {Agreement}’ on {AI} policy.
\newblock \emph{AI and Ethics}, 1(3): 261--271.

\bibitem[{Stockbauer(2021)}]{stockbauer_how_2021}
Stockbauer, J. 2021.
\newblock \emph{How can we explain the development of {Project} {Maven} from agenda-setting theory, and to what extent did this {Project} affect the relationship between the military and industry?}
\newblock Bachelor's diss., University of Twente.

\bibitem[{Straub et~al.(2023)Straub, Morgan, Hashem, Francis, Esnaashari, and Bright}]{straub_multidomain_2023}
Straub, V.~J.; Morgan, D.; Hashem, Y.; Francis, J.; Esnaashari, S.; and Bright, J. 2023.
\newblock A multidomain relational framework to guide institutional {AI} research and adoption.
\newblock In \emph{Proceedings of the 2023 {AAAI}/{ACM} {Conference} on {AI}, {Ethics}, and {Society}}, {AIES} '23, 512--519. New York, NY, USA: Association for Computing Machinery.
\newblock ISBN 9798400702310.

\bibitem[{Stuurman and Lachaud(2022)}]{stuurman_regulating_2022}
Stuurman, K.; and Lachaud, E. 2022.
\newblock Regulating {AI}. {A} label to complete the proposed {Act} on {Artificial} {Intelligence}.
\newblock \emph{Computer Law \& Security Review}, 44: 105657.

\bibitem[{Sun and Guo(2013)}]{sun_unequal_2013}
Sun, W.; and Guo, Y. 2013.
\newblock \emph{Unequal {China}: {The} {Political} {Economy} and {Cultural} {Politics} of {Inequality}}.
\newblock Routledge.
\newblock ISBN 978-0-415-62910-2.

\bibitem[{Tabakovic and Wollmann(2018)}]{tabakovic_revolving_2018}
Tabakovic, H.; and Wollmann, T.~G. 2018.
\newblock From {Revolving} {Doors} to {Regulatory} {Capture}? {Evidence} from {Patent} {Examiners}.
\newblock \url{https://perma.cc/AZK8-4GVV}.

\bibitem[{Taeihagh(2021)}]{taeihagh_governance_2021}
Taeihagh, A. 2021.
\newblock Governance of artificial intelligence.
\newblock \emph{Policy and Society}, 40(2): 137--157.

\bibitem[{Tafani(2022)}]{tafani_whats_2022}
Tafani, D. 2022.
\newblock What's wrong with “{AI} ethics” narratives.
\newblock \emph{Bollettino telematico di filosofia politica}.

\bibitem[{Tambini(2021)}]{tambini_reconceptualizing_2021}
Tambini, D. 2021.
\newblock Reconceptualizing {Media} {Freedom}.
\newblock In Moore, M.; and Tambini, D., eds., \emph{Regulating {Big} {Tech}: {Policy} {Responses} to {Digital} {Dominance}}, 0. Oxford University Press.
\newblock ISBN 978-0-19-761609-3.

\bibitem[{Tambini(2023)}]{tambini_regulation_2023}
Tambini, D. 2023.
\newblock Regulation of election communication.
\newblock In \emph{Handbook of {Digital} {Politics}}, 401--431. Edward Elgar Publishing.
\newblock ISBN 978-1-80037-758-5.

\bibitem[{Tartaro(2023{\natexlab{a}})}]{tartaro_regulating_2023}
Tartaro, A. 2023{\natexlab{a}}.
\newblock Regulating by standards: current progress and main challenges in the standardisation of {Artificial} {Intelligence} in support of the {AI} {Act}.
\newblock \emph{European Journal of Privacy Law \& Technologies}, 2023(1).

\bibitem[{Tartaro(2023{\natexlab{b}})}]{tartaro_towards_2023}
Tartaro, A. 2023{\natexlab{b}}.
\newblock Towards {European} {Standards} {Supporting} the {AI} {Act}: {Alignment} {Challenges} on the {Path} to {Trustworthy} {AI}.
\newblock SSRN:4470766.

\bibitem[{Tartaro, Smith, and Shaw(2023)}]{tartaro_assessing_2023}
Tartaro, A.; Smith, A.~L.; and Shaw, P. 2023.
\newblock Assessing the impact of regulations and standards on innovation in the field of {AI}.
\newblock arXiv:2302.04110.

\bibitem[{Taylor(2021)}]{taylor_public_2021}
Taylor, L. 2021.
\newblock Public {Actors} {Without} {Public} {Values}: {Legitimacy}, {Domination} and the {Regulation} of the {Technology} {Sector}.
\newblock \emph{Philosophy \& Technology}, 34(4): 897--922.

\bibitem[{Taylor and Dencik(2020)}]{taylor_constructing_2020}
Taylor, L.; and Dencik, L. 2020.
\newblock Constructing {Commercial} {Data} {Ethics}.
\newblock \emph{Technology and Regulation}, 2020: 1--10.

\bibitem[{Teachout and Khan(2014)}]{teachout_market_2014}
Teachout, Z.; and Khan, L.~M. 2014.
\newblock Market {Structure} and {Political} {Law}: {A} {Taxonomy} of {Power}.
\newblock \emph{Duke Journal of Constitutional Law \& Public Policy}, 9: 37.

\bibitem[{{TechNet}(2024)}]{technet_technets_2024}
{TechNet}. 2024.
\newblock {TechNet}’s “{AI} for {America}” {Initiative} {Expands} {Nationally} with {New} {Ad}.
\newblock \url{https://perma.cc/M27P-MLZJ}.
\newblock Accessed: 2024-04-26.

\bibitem[{Tegmark(2023)}]{tegmark_classic_2023}
Tegmark, M. 2023.
\newblock Classic regulatory capture. {Now} we’re seeing {AI} companies trying to repeat the same trick.
\newblock \textit{Twitter}. \url{https://perma.cc/7WKC-RKYN}.

\bibitem[{Terzis, Veale, and Gaumann(2024)}]{terzis_law_2024}
Terzis, P.; Veale, M.; and Gaumann, N. 2024.
\newblock Law and the {Emerging} {Political} {Economy} of {Algorithmic} {Audits}.
\newblock In \emph{Proceedings of the 2024 {ACM} {Conference} on {Fairness}, {Accountability}, and {Transparency}}. ACM.
\newblock ISBN 979-8-4007-0450-5/24/06.

\bibitem[{Thaw(2014)}]{thaw_enlightened_2014}
Thaw, D. 2014.
\newblock Enlightened {Regulatory} {Capture}.
\newblock \emph{Washington Law Review}, 89: 329.

\bibitem[{{TheBridge}(2024)}]{thebridge_thebridge_2024}
{TheBridge}. 2024.
\newblock {TheBridge} {Leaders} {Directory}.
\newblock \url{https://perma.cc/ZR6X-ESE4}.
\newblock Accessed: 2024-01-26.

\bibitem[{Thierer and Chilson(2023)}]{thierer_problem_2023}
Thierer, A.; and Chilson, N. 2023.
\newblock The {Problem} with {AI} {Licensing} \& an “{FDA} for {Algorithms}”.
\newblock \url{https://perma.cc/98RJ-WWXL}.
\newblock Accessed: 2023-10-13.

\bibitem[{Thierer and Haaland(2021)}]{thierer_does_2021}
Thierer, A.~D.; and Haaland, C. 2021.
\newblock Does the {United} {States} {Need} a {More} {Targeted} {Industrial} {Policy} for {High} {Tech}?
\newblock SSRN:3965696.

\bibitem[{Thierer and Skorup(2013)}]{thierer_history_2013}
Thierer, A.~D.; and Skorup, B. 2013.
\newblock A {History} of {Cronyism} and {Capture} in the {Information} {Technology} {Sector}.
\newblock \emph{Journal of Technology Law \& Policy}, 18.

\bibitem[{Thompson(2022{\natexlab{a}})}]{thompson_ex-google_2022}
Thompson, A. 2022{\natexlab{a}}.
\newblock Ex-{Google} boss helps fund dozens of jobs in {Biden}’s administration.
\newblock \textit{Politico}. \url{https://perma.cc/R6XA-6YSW}.
\newblock Accessed: 2023-10-13.

\bibitem[{Thompson(2022{\natexlab{b}})}]{thompson_google_2022}
Thompson, A. 2022{\natexlab{b}}.
\newblock A {Google} billionaire's fingerprints are all over {Biden}'s science office.
\newblock \textit{Politico}. \url{https://perma.cc/9D2T-NH5R}.
\newblock Accessed: 2023-10-13.

\bibitem[{Thornhill(2023)}]{thornhill_ai_2023}
Thornhill, J. 2023.
\newblock {AI} will never threaten humans, says top {Meta} scientist.
\newblock \textit{Financial Times}. \url{https://perma.cc/PNQ8-3WZ4}.
\newblock Accessed: 2023-12-16.

\bibitem[{Thönnes et~al.(2023)Thönnes, Guild, Brouwer, and Salomon}]{thonnes_future_2023}
Thönnes, C.; Guild, E.; Brouwer, E.; and Salomon, S. 2023.
\newblock The {Future} of the {European} {Security} {Architecture} - a {Debate} {Series}.
\newblock SSRN:4547775.

\bibitem[{Timcke(2023)}]{timcke_considerations_2023}
Timcke, S. 2023.
\newblock Considerations for {African} {Democrats} about {AI}.
\newblock Technical report, Research ICT Africa.

\bibitem[{Timmers(2021)}]{timmers_ai_2021}
Timmers, P. 2021.
\newblock {AI} {Challenging} {Sovereignty} and {Democracy}.
\newblock \emph{Turkish Policy Quarterly}, 20(4): 45--55.

\bibitem[{Trager et~al.(2023)Trager, Harack, Reuel, Carnegie, Heim, Ho, Kreps, Lall, Larter, hÉigeartaigh, Staffell, and Villalobos}]{trager_international_2023}
Trager, R.; Harack, B.; Reuel, A.; Carnegie, A.; Heim, L.; Ho, L.; Kreps, S.; Lall, R.; Larter, O.; hÉigeartaigh, S.~O.; Staffell, S.; and Villalobos, J.~J. 2023.
\newblock International {Governance} of {Civilian} {AI}: {A} {Jurisdictional} {Certification} {Approach}.
\newblock arXiv:2308.15514.

\bibitem[{Triguero et~al.(2024)Triguero, Molina, Poyatos, Del~Ser, and Herrera}]{triguero_general_2024}
Triguero, I.; Molina, D.; Poyatos, J.; Del~Ser, J.; and Herrera, F. 2024.
\newblock General {Purpose} {Artificial} {Intelligence} {Systems} ({GPAIS}): {Properties}, definition, taxonomy, societal implications and responsible governance.
\newblock \emph{Information Fusion}, 103: 102135.

\bibitem[{Turner(2018)}]{turner_robot_2018}
Turner, J. 2018.
\newblock \emph{Robot {Rules}: {Regulating} {Artificial} {Intelligence}}.
\newblock Springer.
\newblock ISBN 978-3-319-96235-1.

\bibitem[{Turner(2019)}]{turner_building_2019}
Turner, J. 2019.
\newblock Building a {Regulator}.
\newblock In Turner, J., ed., \emph{Robot {Rules} : {Regulating} {Artificial} {Intelligence}}, 207--262. Cham: Springer International Publishing.
\newblock ISBN 978-3-319-96235-1.

\bibitem[{Vallor(2022)}]{vallor_social_2022}
Vallor, S. 2022.
\newblock Social {Networking} and {Ethics}.
\newblock \emph{Stanford Encyclopedia of Philosophy}.

\bibitem[{van~der Merwe and Al~Achkar(2022)}]{van_der_merwe_data_2022}
van~der Merwe, J.; and Al~Achkar, Z. 2022.
\newblock Data responsibility, corporate social responsibility, and corporate digital responsibility.
\newblock \emph{Data \& Policy}, 4: e12.

\bibitem[{Varoglu, Gokten, and Ozdogan(2021)}]{varoglu_digital_2021}
Varoglu, A.; Gokten, S.; and Ozdogan, B. 2021.
\newblock Digital {Corporate} {Governance}: {Inevitable} {Transformation}.
\newblock In Hacioglu, U.; and Aksoy, T., eds., \emph{Financial {Ecosystem} and {Strategy} in the {Digital} {Era}: {Global} {Approaches} and {New} {Opportunities}}, Contributions to {Finance} and {Accounting}, 219--236. Cham: Springer International Publishing.
\newblock ISBN 978-3-030-72624-9.

\bibitem[{Vasse'i(2019)}]{vassei_ethical_2019}
Vasse'i, R.~M. 2019.
\newblock The {Ethical} {Guidelines} for {Trustworthy} {AI} – {A} {Procrastination} of {Effective} {Law} {Enforcement}: {Weaknesses} of ethical principles in general and the {EU}’s approach in particular.
\newblock \emph{Computer Law Review International}, 20(5): 129--136.

\bibitem[{Veale(2020)}]{veale_critical_2020}
Veale, M. 2020.
\newblock A {Critical} {Take} on the {Policy} {Recommendations} of the {EU} {High}-{Level} {Expert} {Group} on {Artificial} {Intelligence}.
\newblock \emph{European Journal of Risk Regulation}, 11(1): e1.

\bibitem[{Veale, Matus, and Gorwa(2023)}]{veale_ai_2023}
Veale, M.; Matus, K.; and Gorwa, R. 2023.
\newblock {AI} and {Global} {Governance}: {Modalities}, {Rationales}, {Tensions}.
\newblock \emph{Annual Review of Law and Social Science}, 19: 255--275.

\bibitem[{Verma(2023)}]{verma_ai_2023}
Verma, M. 2023.
\newblock {AI} {Safety} and {Regulations}: {Navigating} the {Post}-{COVID} {Era}: {Aims}, {Opportunities}, and {Challenges}: {A} {ChatGPT} {Analysis}.
\newblock \emph{International Journal of Trend in Scientific Research and Development}, 7(6): 12--20.

\bibitem[{Vescent and Blakley(2018)}]{vescent_shifting_2018}
Vescent, H.; and Blakley, B. 2018.
\newblock Shifting {Paradigms}: {Using} {Strategic} {Foresight} to {Plan} for {Security} {Evolution}.
\newblock In \emph{Proceedings of the {New} {Security} {Paradigms} {Workshop}}, {NSPW} '18, 28--40. New York, NY, USA: Association for Computing Machinery.
\newblock ISBN 978-1-4503-6597-0.

\bibitem[{Viljoen(2021)}]{viljoen_promise_2021}
Viljoen, S. 2021.
\newblock The {Promise} and {Limits} of {Lawfulness}: {Inequality}, {Law}, and the {Techlash}.
\newblock \emph{Journal of Social Computing}, 2(3): 284--296.

\bibitem[{Vipra and Korinek(2023)}]{vipra_market_2023}
Vipra, J.; and Korinek, A. 2023.
\newblock Market {Concentration} {Implications} of {Foundation} {Models}.
\newblock arXiv:2311.01550.

\bibitem[{Volpicelli(2023)}]{volpicelli_power_2023}
Volpicelli, G. 2023.
\newblock Power grab by {France}, {Germany} and {Italy} threatens to kill {EU}’s {AI} bill.
\newblock \textit{Politico}. \url{https://perma.cc/8WAD-8BD2}.
\newblock Accessed: 2024-05-02.

\bibitem[{von Thun(2023)}]{von_thun_monopoly_2023}
von Thun, M. 2023.
\newblock Monopoly {Power} {Is} the {Elephant} in the {Room} in the {AI} {Debate}.
\newblock \textit{Tech Policy Press}. \url{https://perma.cc/T5ZK-9S3E}.
\newblock Accessed: 2023-12-16.

\bibitem[{Vries, Kanevskaia, and Jager(2023)}]{vries_internal_2023}
Vries, S.~d.; Kanevskaia, O.; and Jager, R.~d. 2023.
\newblock Internal {Market} 3.0: {The} {Old} “{New} {Approach}” for {Harmonising} {AI} {Regulation}.
\newblock \emph{European Papers - A Journal on Law and Integration}, 2023 8(2): 583--610.

\bibitem[{Wagner(2009)}]{wagner_administrative_2009}
Wagner, W.~E. 2009.
\newblock Administrative {Law}, {Filter} {Failure}, and {Information} {Capture}.
\newblock \emph{Duke Law Journal}, 59(7): 1321--1432.

\bibitem[{{WAIE}(2024)}]{waie_open_2024}
{WAIE}. 2024.
\newblock Open directory of {Women} in {AI} {Ethics}.
\newblock \url{https://perma.cc/7Q8J-3DYK}.
\newblock Accessed: 2024-01-26.

\bibitem[{Walters(2019)}]{walters_capturing_2019}
Walters, D.~E. 2019.
\newblock Capturing the {Regulatory} {Agenda}: {An} {Empirical} {Study} of {Agency} {Responsiveness} to {Rulemaking} {Petitions}.
\newblock \emph{Harvard Environmental Law Review}, 43: 175.

\bibitem[{Wang(2023)}]{wang_sponsored_2023}
Wang, A. 2023.
\newblock Sponsored {Content}: {The} {AI} {War} and {What} the {U}.{S}. {Must} {Do} to {Win} {It}.
\newblock \textit{Politico}. \url{https://perma.cc/B72C-LTLW}.
\newblock Accessed: 2024-04-26.

\bibitem[{Wansley(2015)}]{wansley_virtuous_2015}
Wansley, M. 2015.
\newblock Virtuous {Capture}.
\newblock \emph{Administrative Law Review}, 67: 419.

\bibitem[{Wedam(2023)}]{wedam_how_2023}
Wedam, J. 2023.
\newblock \emph{How to {Control} {AI}: {Changing} {Our} {Trajectory} and {Democratizing} {Technology}}.
\newblock Outskirts Press.
\newblock ISBN 978-1-977269-04-1.

\bibitem[{Weidinger et~al.(2021)Weidinger, Mellor, Rauh, Griffin, Uesato, Huang, Cheng, Glaese, Balle, Kasirzadeh, Kenton, Brown, Hawkins, Stepleton, Biles, Birhane, Haas, Rimell, Hendricks, Isaac, Legassick, Irving, and Gabriel}]{weidinger_ethical_2021}
Weidinger, L.; Mellor, J.; Rauh, M.; Griffin, C.; Uesato, J.; Huang, P.-S.; Cheng, M.; Glaese, M.; Balle, B.; Kasirzadeh, A.; Kenton, Z.; Brown, S.; Hawkins, W.; Stepleton, T.; Biles, C.; Birhane, A.; Haas, J.; Rimell, L.; Hendricks, L.~A.; Isaac, W.; Legassick, S.; Irving, G.; and Gabriel, I. 2021.
\newblock Ethical and social risks of harm from {Language} {Models}.
\newblock arXiv:2112.04359.

\bibitem[{Weidinger et~al.(2023)Weidinger, Rauh, Marchal, Manzini, Hendricks, Mateos-Garcia, Bergman, Kay, Griffin, Bariach, Gabriel, Rieser, and Isaac}]{weidinger_sociotechnical_2023}
Weidinger, L.; Rauh, M.; Marchal, N.; Manzini, A.; Hendricks, L.~A.; Mateos-Garcia, J.; Bergman, S.; Kay, J.; Griffin, C.; Bariach, B.; Gabriel, I.; Rieser, V.; and Isaac, W. 2023.
\newblock Sociotechnical {Safety} {Evaluation} of {Generative} {AI} {Systems}.
\newblock arXiv:2310.11986.

\bibitem[{Weil(2024)}]{weil_tort_2024}
Weil, G. 2024.
\newblock Tort {Law} as a {Tool} for {Mitigating} {Catastrophic} {Risk} from {Artificial} {Intelligence}.
\newblock SSRN:4694006.

\bibitem[{Weinkle(2019)}]{weinkle_experts_2019}
Weinkle, J. 2019.
\newblock Experts, regulatory capture, and the “governor's dilemma”: {The} politics of hurricane risk science and insurance.
\newblock \emph{Regulation \& Governance}, 14.

\bibitem[{Westgarth et~al.(2022)Westgarth, Chen, Hay, and Heath}]{westgarth_understanding_2022}
Westgarth, T.; Chen, W.; Hay, G.; and Heath, R. 2022.
\newblock Understanding {UK} {Artificial} {Intelligence} {R}\&{D} commercialisation and the role of standards.
\newblock Technical report, UK Department for Digital, Culture, Media and Sport, Office for Artificial Intelligence.

\bibitem[{Wexler(2011)}]{wexler_which_2011}
Wexler, M.~N. 2011.
\newblock Which {Fox} in {What} {Henhouse} and {When}? {Conjectures} on {Regulatory} {Capture}.
\newblock \emph{Business and Society Review}, 116(3): 277--302.

\bibitem[{{White House}(2023)}]{white_house_ensuring_2023}
{White House}. 2023.
\newblock Ensuring {Safe}, {Secure}, and {Trustworthy} {AI}.
\newblock \textit{White House}. \url{https://perma.cc/FQ5Y-MPXS}.

\bibitem[{Whittaker(2021)}]{whittaker_steep_2021}
Whittaker, M. 2021.
\newblock The steep cost of capture.
\newblock \emph{Interactions}, 28(6): 50--55.

\bibitem[{Whyman(2023)}]{whyman_ai_2023}
Whyman, B. 2023.
\newblock {AI} {Regulation} is {Coming}- {What} is the {Likely} {Outcome}?
\newblock Technical report, CSIS.

\bibitem[{Widder, West, and Whittaker(2023)}]{widder_open_2023}
Widder, D.~G.; West, S.; and Whittaker, M. 2023.
\newblock Open ({For} {Business}): {Big} {Tech}, {Concentrated} {Power}, and the {Political} {Economy} of {Open} {AI}.
\newblock SSRN:4543807.

\bibitem[{Widder et~al.(2023)Widder, Zhen, Dabbish, and Herbsleb}]{widder_its_2023}
Widder, D.~G.; Zhen, D.; Dabbish, L.; and Herbsleb, J. 2023.
\newblock It’s about power: {What} ethical concerns do software engineers have, and what do they (feel they can) do about them?
\newblock In \emph{2023 {ACM} {Conference} on {Fairness}, {Accountability}, and {Transparency}}, 467--479. Chicago IL USA: ACM.
\newblock ISBN 9798400701924.

\bibitem[{Williams(2023)}]{williams_amazon_2023}
Williams, Z. 2023.
\newblock Amazon, {Google} {Among} {Firms} {Focusing} on {AI} {Lobbying} in {States}.
\newblock \textit{Bloomberg Law}. \url{https://perma.cc/NER9-WYU9}.
\newblock Accessed: 2024-01-29.

\bibitem[{Won(2021)}]{won_missing_2021}
Won, D. 2021.
\newblock The {Missing} {Algorithm}: {Safeguarding} {Brady} against the {Rise} of {Trade} {Secrecy} in {Policing} {Note}.
\newblock \emph{Michigan Law Review}, 120(1): 157--194.

\bibitem[{Wong(2023)}]{wong_ai_2023}
Wong, M. 2023.
\newblock {AI} {Doomerism} {Is} a {Decoy}.
\newblock \textit{The Atlantic}. \url{https://perma.cc/TMW4-YA5G}.
\newblock Accessed: 2023-12-16.

\bibitem[{Wouters(2022)}]{wouters_corporations_2022}
Wouters, J. 2022.
\newblock Corporations and the {Making} of {Public} {Standards} in {International} {Law}.
\newblock \emph{Integration}, 683: 66.

\bibitem[{Wren-Lewis(2011)}]{wren-lewis_regulatory_2011}
Wren-Lewis, L. 2011.
\newblock Regulatory {Capture}: {Risks} and {Solutions}.
\newblock In Estache, A., ed., \emph{Emerging {Issues} in {Competition}, {Collusion}, and {Regulation} of {Network} {Industries}}. London: Centre for Economic Policy Research.
\newblock ISBN 978-1-907142-35-2.

\bibitem[{Wörsdörfer(2023)}]{worsdorfer_eus_2023}
Wörsdörfer, M. 2023.
\newblock The {E}.{U}.’s artificial intelligence act: an ordoliberal assessment.
\newblock \emph{AI and Ethics}.

\bibitem[{Xenidis(2024)}]{xenidis_beyond_2024}
Xenidis, R. 2024.
\newblock Beyond bias: algorithmic machines, discrimination law and the analogy trap.
\newblock \emph{Transnational Legal Theory}, 0(0): 1--35.

\bibitem[{Xia et~al.(2024)Xia, Lu, Zhu, and Xing}]{xia_ai_2024}
Xia, B.; Lu, Q.; Zhu, L.; and Xing, Z. 2024.
\newblock An {AI} {System} {Evaluation} {Framework} for {Advancing} {AI} {Safety}: {Terminology}, {Taxonomy}, {Lifecycle} {Mapping}.
\newblock In \emph{Proceedings of the 1st {ACM} {International} {Conference} on {AI}-{Powered} {Software}}, {AIware} 2024, 74--78. New York, NY, USA: Association for Computing Machinery.
\newblock ISBN 9798400706851.

\bibitem[{Yackee(2022)}]{yackee_regulatory_2022}
Yackee, S.~W. 2022.
\newblock Regulatory {Capture}'s {Self}-{Serving} {Application}.
\newblock \emph{Public Administration Review}, 82(5): 866--875.

\bibitem[{Yaghmai(2021)}]{yaghmai_critical_2021}
Yaghmai, R. 2021.
\newblock A {Critical} {Examination} of {How} {Artificial} {Intelligence} {Should} {Be} {Regulated} in the {United} {Kingdom}.
\newblock \emph{Bristol Law Review}, 2021: 59--83.

\bibitem[{Yang et~al.(2024)Yang, Wong, Jackson, Junginger, Hagan, Gilbert, and Zimmerman}]{yang_future_2024}
Yang, Q.; Wong, R.~Y.; Jackson, S.; Junginger, S.; Hagan, M.~D.; Gilbert, T.; and Zimmerman, J. 2024.
\newblock The {Future} of {HCI}-{Policy} {Collaboration}.
\newblock In \emph{Proceedings of the {CHI} {Conference} on {Human} {Factors} in {Computing} {Systems}}, {CHI} '24, 1--15. New York, NY, USA: Association for Computing Machinery.
\newblock ISBN 9798400703300.

\bibitem[{Yeoh(2019)}]{yeoh_capture_2019}
Yeoh, P. 2019.
\newblock Capture of {Regulatory} {Agencies}: {A} {Time} for {Reflection} {Again}.
\newblock \emph{Business Law Review}, 40(4).

\bibitem[{Young, Katell, and Krafft(2022)}]{young_confronting_2022}
Young, M.; Katell, M.; and Krafft, P. 2022.
\newblock Confronting {Power} and {Corporate} {Capture} at the {FAccT} {Conference}.
\newblock In \emph{Proceedings of the 2022 {ACM} {Conference} on {Fairness}, {Accountability}, and {Transparency}}, {FAccT} '22, 1375--1386. New York, NY, USA: Association for Computing Machinery.
\newblock ISBN 978-1-4503-9352-2.

\bibitem[{Zhang, Ong, and Findlay(2023)}]{zhang_digital_2023}
Zhang, W.; Ong, L.~M.; and Findlay, M. 2023.
\newblock Digital self-determination: an alternative paradigm for emerging economies.
\newblock In \emph{Elgar {Companion} to {Regulating} {AI} and {Big} {Data} in {Emerging} {Economies}}, 158--179. Edward Elgar Publishing.
\newblock ISBN 978-1-78536-240-8.

\bibitem[{Zingales(2013)}]{zingales_preventing_2013}
Zingales, L. 2013.
\newblock Preventing {Economists}' {Capture}.
\newblock In Carpenter, D.; and Moss, D.~A., eds., \emph{Preventing {Regulatory} {Capture}: {Special} {Interest} {Influence} and {How} to {Limit} it}, 124--151. Cambridge: Cambridge University Press.
\newblock ISBN 978-1-107-03608-6.

\bibitem[{Zumbansen(2022)}]{zumbansen_corporate_2022}
Zumbansen, P.~C. 2022.
\newblock Corporate {Governance} {Choices} and the {Actual} {Stakes} of {Stakeholder} {Governance}.
\newblock SSRN:4092148.

\bibitem[{Zumbansen(2023)}]{zumbansen_corporation_2023}
Zumbansen, P.~C. 2023.
\newblock The {Corporation} in an {Age} of {Divisiveness}.
\newblock SSRN:4374323.

\end{thebibliography}

\appendix

\clearpage

\section{Appendix: Interview Protocol} \label{sec:Appendix_Interview_Protocol}

This appendix contains our interview instrument. All interviews were conducted by the first author.

We first set forth a definition of ``AI policy'' and ``general-purpose AI systems''\footnote{See also \citetalias[Art. 3(63)]{AIA}; \citet{gutierrez_proposal_2023, triguero_general_2024, xia_ai_2024} for additional commentary and definitions of ``general-purpose AI.''} for all interviewees:

\begin{quote}
    ``One quick note before we begin: In this interview, we will be discussing AI policy. We are primarily interested in general-purpose AI systems that have a wide variety of use cases, rather than narrow or domain-specific AI systems. By `AI policy,' we mean government policies intended to regulate, restrict, or promote the development and deployment of general-purpose AI systems. Does that make sense to you?''
\end{quote}

Our interview instrument then contained four components: one question on the interviewee's background, questions to help identify outcomes of industry influence in AI policy, questions on mechanisms used by industry actors to influence AI policy, and one open-ended question. The interviews were semi-structured, so follow-ups were asked as necessary. 

We began with one question concerning interviewees' backgrounds:

\begin{enumerate}
    \item Can you tell me what your job and job description is? What areas of policy are you responsible for?
\end{enumerate}

We then ask a set of questions concerning the industry goals in AI policy:

\begin{enumerate}
    \setcounter{enumi}{1}
    \item Briefly, how would you define the public interest goals of regulation of AI?
    \item Briefly, what actors or types of actors from industry currently contribute to the design, development, or enforcement of AI policy?
    \item In general, what goals do industry actors have when they attempt to shape AI policy? In particular, what types of changes do they propose for regulation, and do these changes tend to interact with the public interest goals of [goals identified in Q2]?
\end{enumerate}

We then presented interviewees with a version of Table \ref{tab:Appendix_Mechanisms_Definitions}.\footnote{This table was adapted throughout our interview process with new examples, clarifications, and mechanisms based on previous interviewee questions and suggestions.} We asked a set of questions about mechanisms of industry influence in AI policy:

\begin{enumerate}
    \setcounter{enumi}{5}
    \item In this table, we have identified a list of different mechanisms for how industry can influence policymaking in different sectors. Do you have any questions about any of the mechanisms, or can we provide any examples to make these more clear for you?
    \begin{enumerate}
        \item Please describe any mechanisms for industry influence in policy that are missing from this table.
    \end{enumerate}
    \item Which of the mechanisms listed in Table [\ref{tab:Appendix_Mechanisms_Definitions}] are currently most relevant to AI policy, to your knowledge? These could be mechanisms that industry is currently using or likely to use in the future to influence policy.
    \item {[For each of the mechanisms identified in Q6]} For [mechanism], do industry actors currently use this mechanism to influence policy to your knowledge, or is it not currently used but likely to be used in the future?
    \item {[For mechanisms currently used to influence policy]}
    \begin{enumerate}
        \item Have you personally seen [mechanism] in action? 
        \item Can you give us a few examples of when you have seen this occur?
        \item What procedures or mechanisms are currently in place that would prevent industry influence on policymakers through [mechanism]?
        \item To the best of your knowledge, are these preventative procedures are currently effective? Why or why not?
    \end{enumerate}
    \item {[For mechanisms likely to be used to influence policy in the future]}
    \begin{enumerate}
        \item What features of the AI industry or of current AI policymaking make [mechanism] likely to be used in the future?
        \item Are there any industries in which you have seen [mechanism] occur that inform why [mechanism] may occur in AI?
        \item What procedures or mechanisms are currently in place that would prevent industry influence on policymakers through [mechanism]?
        \item To the best of your knowledge, are these preventative procedures currently effective? Why or why not?
    \end{enumerate}
\end{enumerate}

Finally, we concluded with an open-ended question:

\begin{enumerate}
    \setcounter{enumi}{10}
    \item Is there anything that we did not discuss that you would like to mention?
\end{enumerate}

\clearpage

\section{Appendix: Literature Review} \label{sec:Appendix_Review}

\subsection{Methods} \label{subsec:Appendix_Review_Methods}
We conducted a limited scoping review of the academic literature that discussed regulatory capture in the context of AI policy. Our review was used to give us an overview of the literature, to inform Section \ref{sec:Definition}, to develop examples and talking points for interviews, and to corroborate interview findings where applicable. It is not our main contribution and is not as rigorous as might be expected from a systematic review.

Using the search terms in Table \ref{tab:Review_Keywords}, we queried the ACM Digital Library, IEEE Xplore, arXiv,\footnote{Although preprints on arXiv are not peer reviewed, a not-insignificant proportion of academic literature appears on arXiv before formal publication, as do various articles in the gray literature, e.g., reports by think tanks or advocacy institutions.} and Google Scholar for articles containing terms in Table \ref{tab:Review_Keywords}. Searches were conducted in October 2023 and again in January--February 2024. Both articles in the academic and gray literature were included.

The search string was constructed to return any article that contained an exact match of (any) one term from \textit{both} columns in Table \ref{tab:Review_Keywords}. In other words, articles needed to contain one term related to AI policy and one term related to regulatory capture. All searches were full text, so the terms could be contained anywhere in the text of the article or in any metadata field (e.g., title, abstract).  

\begin{table}[!hbtp]
    \centering
    \small
    \begin{tabularx}{\linewidth}{ X X }
        \toprule
        
        \textbf{AI Governance Keywords} & \textbf{Capture Keywords} 
        \\ 
        
        \midrule
        
        ``artificial intelligence'' & ``regulatory capture'' 
        \\
        
        ``AI governance'' & ``industry capture'' 
        \\
        
        ``AI policy'' & ``agency capture'' 
        \\

        ``AI ethics'' & ``corporate capture'' 
        \\

        \bottomrule
    \end{tabularx}
    \caption{Search terms for scoping review}
    \label{tab:Review_Keywords}
\end{table}

The search returned $n = 255$ unique articles in the English language. The second author filtered these results by reading the titles and abstracts of all articles. Filtering was conducted using the inclusion and exclusion criteria in Tables \ref{tab:Review_Inclusion} and \ref{tab:Review_Exclusion}. 

\begin{table}[!htb]
    \centering
    \small
    \begin{tabularx}{\linewidth}{ X }
        \toprule
        
        \textbf{Inclusion Criteria} 
        \\ 
        
        \midrule
        
        The article describes influence related to AI policy or regulation 
        \\ 
        
        \midrule
        
        The article describes influence exerted on policymakers, or on an entity that may influence policymakers 
        \\
        
        \bottomrule
    \end{tabularx}
\caption{Inclusion criteria for scoping review}
\label{tab:Review_Inclusion}
\end{table}

\begin{table}[!htb]
    \centering
    \small
    \begin{tabularx}{\linewidth}{ X }
        \toprule
        
        \textbf{Exclusion Criteria} 
        \\ 
        
        \midrule
        
        The article only discusses influence by non-industry or non-corporate actors 
        \\ 
        
        \midrule
        
        The article includes neither a mechanism nor a concrete outcome of capture 
        \\

        \midrule
        
        The article is a thesis or dissertation
        \\
        
        \bottomrule
    \end{tabularx}
\caption{Exclusion criteria for scoping review}
\label{tab:Review_Exclusion}
\end{table}

A list of the articles remaining after filtering based on the inclusion and exclusion criteria may be found in Table \ref{tab:Review_Article_List}. Note that our method led us to be over-inclusive about which articles remained in the final review. As long as \textit{any part of the text} satisfied the inclusion criteria, we included the article in our final review. Oftentimes, the included articles contained only a few sentences to a paragraph of relevant discussion.

Based on the article text, the second author then coded each article based on the mechanism of influence discussed in that article (either explicitly or---more often---inferred from the text).\footnote{Since the primary research questions of the vast majority of these articles were not about regulatory capture or industry influence, we generally screened only the relevant parts of the text instead of the full text of every article (see Section \ref{subsec:Appendix_Review_Results}).}

\subsection{Results} \label{subsec:Appendix_Review_Results}

A brief overview of our results is below. Generally, most articles' research questions were not centered on industry influence or regulatory capture. Most often, the relevant portions of the text were extremely brief and situated as warnings against capture in the context of broader policy proposals or discussions; most articles did not discuss specific mechanisms of influence in depth. Thus, we do not believe that the number of articles in our review or our results below are particularly indicative of the maturity of the research into industry influence in AI policy.

Of the included articles ($n=120$), the most frequently discussed types of mechanisms are information capture ($n=57$), indirect mechanisms ($n=46$), and personal engagement ($n=25$). Discussion of information capture was divided mostly between information management ($n=29$) and agenda-setting ($n=27$). Discussion of indirect mechanisms for capture were limited to private regulator capture ($n=29$) and academic capture ($n=17$). Most articles coded as personal engagement discussed advocacy ($n=25$). 

Incentive shaping ($n=12$) and cultural capture ($n=11$) are discussed relatively less frequently. The majority of articles about incentive shaping discussed the revolving door phenomenon ($n=8$), with only half as many articles discussing donations and gifts ($n=4$). Discussion of cultural capture was split mostly between group identity ($n=5$) and relationship networks ($n=5$), with only one article about status ($n=1$).

\begin{table*}[!htbp]
    \centering
    \small
    \begin{tabularx}{\textwidth}{s X}
        \toprule

        &
        \textbf{Articles}
        \\

        \midrule

        Included ($n = 120$) &
        \citet{abdu_empirical_2023, abebe_adversarial_2022, ai_governance_alliance_generative_2024, alaga_coordinated_2023, allen_regulating_2019, almada_eu_2023, anderljung_frontier_2023, attard-frost_ethics_2023, badran_thoughts_2021, bajohr_whoever_2023, bannerman_platforms_2020, bender_power_2024, berman_scoping_2024, bova_both_2024, brandusescu_artificial_2021, broughel_rules_2023, browne_tech_2024, brynjolfsson_big_2023, bryson_is_2021, bryson_artificial_2020, casper_black-box_2024, cath_your_2022, chan_visibility_2024, chan_reclaiming_2023, chesterman_weapons_2021, chilson_public_2024, chomanski_missing_2021, cihon_fragmentation_2020, cihon_should_2020, clarke_survey_2022, cui_governing_2024, de_laat_companies_2021, vries_internal_2023, dempsey_transnational_2024, derczynski_assessing_2023, dickens_right_2021, ebers_standardizing_2022, egan_oversight_2023, erman_democratization_2024, evans_truthful_2021, fagleman_ai_2023, frazier_administrative_2023, friedman_policing_2022, gaske_regulation_2023, gazendam_mind_2023, gilbert_choices_2022, giraudo_competing_2023, goanta_regulation_2023, gornet_european_2023, greenleaf_does_2019, guha_ai_2024, guihot_nudging_2017, haataja_reflections_2022, hacker_sustainable_2023, hadfield_regulatory_2023, himmelreich_artificial_2022, hu_tech_2021, hua_effective_2023, jiang_ai_2023, jing_towards_2023, katyal_democracy_2022, khan_regulating_2023, koene_governance_2019, kolt_algorithmic_2023, lam_framework_2024, laux_taming_2021, leslie_advancing_2022, levesque_scoping_2021, liesenfeld_opening_2023, luetz_gender_2023, lupo_risky_2023, marcus_controlling_2023, margulies_adjudicating_2023, mugge_securitization_2023, narayanan_attitudinal_2023, nemitz_democracy_2023, oshaughnessy_what_2023, ochigame_how_2019, papyshev_limitation_2022, paul_politics_2022, peng_artificial_2021, pizzi_ai_2020, png_at_2022, raji_change_2023, ramdas_identifying_2022, roberts_achieving_2021, roberts_artificial_2023, sanchez-graells_governing_2024, sanchez-graells_two_2023, sanchez-graells_responsibly_2023, sanchez-graells_discharging_2023, sanchez-graells_procurement_2023, sarel_restraining_2023, schaferling_case_2023, sri_discerning_2023, seger_open-sourcing_2023, selbst_institutional_2021, solow-niederman_administering_2019, straub_multidomain_2023, stuurman_regulating_2022, taeihagh_governance_2021, tafani_whats_2022, tartaro_regulating_2023, tartaro_towards_2023, taylor_constructing_2020, thonnes_future_2023, timcke_considerations_2023, trager_international_2023, veale_critical_2020, veale_ai_2023, vipra_market_2023, weil_tort_2024, westgarth_understanding_2022, whittaker_steep_2021, widder_open_2023, won_missing_2021, worsdorfer_eus_2023, wouters_corporations_2022, vallor_social_2022, young_confronting_2022}
        \\

        \midrule

        Excluded ($n = 135$) &
        \citet{abdalla_elephant_2023, abou-zeid_au_2022, ajena_agroecology_2022, barabas_uninventing_2023, baumberger_unveiling_2023, bechara_impact_2021, bedford_post-capitalocentric_2022, bennett_transmuting_2023, bietti_genealogy_2023, boffel_influence_2023, brandusescu_canadas_2023, bremmer_ai_2023, brownsword_law_2019, carlizzi_artificial_2023, carter_machine_2023, cebulla_future_2023, chan_balancing_2022, charisi_operationalizing_2024, charlesworth_regulating_2021, charlesworth_response_2023, chauhan_what_2023, chesterman_we_2021, chesterman_tragedy_2023, chinen_international_2023, cohen_rights_2019, correa_global_2023, couldry_data_2019, critch_tasra_2023, cuellar_artificially_2022, dancy_integrating_2023, edwards_regulating_2022, eliot_culling_2022, fahey_introduction_2022, fahey_eu_2022, fenwick_banking_2020, fenwick_fintech_2020, fenwick_future_2021, findlay_data_2020, findlay_regulatory_2022, findlay_ai_2023, ford_embracing_2021, fraser_acceptable_2023, gans_how_2024, gantzias_dynamics_2021, gaske_operational_2023, gegenhuber_institutional_2022, geiger_making_2023, gilbert_modes_2021, goodlad_editors_2023, goodman_beyond_2019, gottardo_algorithmic_2023, gurumurthy_taking_2019, hacohen_policy_2022, hawking_rule_2021, hermstruwer_fair_2023, hilty_intellectual_2020, himmelreich_against_2023, ho_quality_2021, huang_legal_2023, iliadis_fast_2023, ilie_online_2014, kaplan_framing_2008, keller_regulating_2023, killian_taming_2021, klaessig_traversing_2021, kleizen_big_2020, knaack_global_2022, konya_deliberative_2023, kuzniacki_towards_2022, larsen_governing_2022, lee_artificial_2021, lie_trustworthy_2023, lin_bias_2023, london_regulating_2018, mcgraw_privacy_2021, mcinerney_good_2024, meghani_regulations_2021, mehmood_ai_2024, mesmer_auditing_2023, moberg_law_2024, moitra_ai_2022, neudert_regulatory_2023, neumann_risks_2018, neves_freedom_2023, olteanu_responsible_2023, ong_privacy_2024, opoku_regulation_2019, outeda_artificial_2023, pacione_implications_2023, papyshev_challenges_2023, pavel_rethinking_2022, pavlopoulou_exploration_2022, plantinga_responsible_2024, polishchuk_exploring_2023, rainie_experts_2021, rawat_ethics_2024, ren_centralization_2022, roberts_artificial_2023, sanchez-graells_digital_2024, sandoval_responsible_2023, sastry_computing_2024, sengupta_towards_2022, scherz_ai_2024, shneiderman_bridging_2020, shrier_global_2022, sifat_artificial_2023, sitaraman_building_2023, sivan-sevilla_when_2021, southgate_artificial_2019, southgate_artificial_2021, stahl_european_2022, steed_adoption_2024, stix_bridging_2021, stockbauer_how_2021, sun_unequal_2013, tambini_reconceptualizing_2021, tambini_regulation_2023, taylor_public_2021, thierer_does_2021, timmers_ai_2021, turner_building_2019, turner_robot_2018, van_der_merwe_data_2022, varoglu_digital_2021, verma_ai_2023, vescent_shifting_2018, viljoen_promise_2021, wedam_how_2023, widder_its_2023, xenidis_beyond_2024, yaghmai_critical_2021, yang_future_2024, zhang_digital_2023, zumbansen_corporate_2022, zumbansen_corporation_2023}
        \\

        \bottomrule
    \end{tabularx}
    \caption{A complete list of the 255 articles resulting from our search, including the final 120 articles in our review}
    \label{tab:Review_Article_List}
\end{table*}

\end{document}